\documentclass{emulateapj}
\usepackage{apjfonts}
\usepackage{multirow}
\usepackage{array}
\usepackage[usenames]{color}
\begin{document}
\submitted{ApJ Accepted}
\newcommand{\comment}[1]{}
\newcommand{\risa}[1]{\textcolor{red}{(\bf #1)}}
\newcommand{\michael}[1]{\textcolor{blue}{(\bf #1)}}
\definecolor{purple}{RGB}{160,32,240}
\newcommand{\peter}[1]{\textcolor{purple}{(\bf #1)}}
\newcommand{\macc}{M_\mathrm{acc}}
\newcommand{\mpeak}{M_\mathrm{peak}}
\newcommand{\mnow}{M_\mathrm{now}}
\newcommand{\vacc}{v^\mathrm{acc}_\mathrm{max}}
\newcommand{\vpeak}{v^\mathrm{peak}_\mathrm{max}}
\newcommand{\vnow}{v^\mathrm{now}_\mathrm{max}}

\newcommand{\hinv}{h^{-1}}
\newcommand{\mpc}{\rm{Mpc}}
\newcommand{\hmpc}{$\hinv\mpc$}

\shortauthors{BEHROOZI, WECHSLER, \& WU}
\shorttitle{The \textsc{rockstar} Halo Finder}

\title{The \textsc{rockstar} Phase-Space Temporal Halo Finder and the Velocity Offsets of Cluster Cores}

\author{Peter S. Behroozi, Risa H. Wechsler, Hao-Yi Wu}
\affil{Kavli Institute for Particle Astrophysics and Cosmology, 
Department of Physics, Stanford University; \\
Department of Particle Physics and Astrophysics,  SLAC National  Accelerator Laboratory; 
Stanford, CA 94305} 

\begin{abstract}
We present a new algorithm for identifying dark matter halos, substructure, and tidal features.  The approach is based on adaptive hierarchical refinement of friends-of-friends groups in six phase-space dimensions and one time dimension, which allows for robust (grid-independent, shape-independent, and noise-resilient) tracking of substructure; as such, it is named \textsc{rockstar} (Robust Overdensity Calculation using K-Space Topologically Adaptive Refinement).  Our method is massively parallel (up to $10^5$ CPUs) and runs on the largest current simulations (>$10^{10}$ particles) with high efficiency (10 CPU hours and 60 gigabytes of memory required per billion particles analyzed).  A previous paper \citep{Knebe11} has shown \textsc{rockstar} to have excellent recovery of halo properties; we expand on these comparisons with more tests and higher-resolution simulations.  We show a significant improvement in substructure recovery compared to several other halo finders and discuss the theoretical and practical limits of simulations in this regard.  Finally, we present results which demonstrate conclusively that dark matter halo cores are not at rest relative to the halo bulk or substructure average velocities and have coherent velocity offsets across a wide range of halo masses and redshifts.  For massive clusters, these offsets can be up to 350 km s$^{-1}$ at $z=0$ and even higher at high redshifts.  Our implementation is publicly available at {\tt http://code.google.com/p/rockstar} \hspace{-0.6ex}.
\end{abstract}
\keywords{dark matter --- galaxies: abundances --- galaxies:
  evolution --- methods: N-body simulations}

\newcommand{\Mnfw}{M_\mathrm{NFW}}
\newcommand{\Msun}{M_{\odot}}
\newcommand{\mvir}{M_\mathrm{vir}}
\newcommand{\rvir}{R_\mathrm{vir}}
\newcommand{\vmax}{v_\mathrm{max}}
\newcommand{\vmac}{v_\mathrm{max}^\mathrm{acc}}
\newcommand{\mvac}{M_\mathrm{vir}^\mathrm{acc}}
\newcommand{\sfr}{\mathrm{SFR}}
\newcommand{\plotgrace}[1]{\includegraphics[width=\columnwidth,type=pdf,ext=.pdf,read=.pdf]{#1}}
\newcommand{\plotgraceflip}[1]{\includegraphics[width=\columnwidth,type=pdf,ext=.pdf,read=.pdf]{#1}}
\newcommand{\plotlargegrace}[1]{\includegraphics[width=2\columnwidth,type=pdf,ext=.pdf,read=.pdf]{#1}}
\newcommand{\plotlargegraceflip}[1]{\includegraphics[width=2\columnwidth,type=pdf,ext=.pdf,read=.pdf]{#1}}
\newcommand{\plotminigrace}[1]{\includegraphics[width=0.5\columnwidth,type=pdf,ext=.pdf,read=.pdf]{#1}}
\newcommand{\plotmicrograce}[1]{\includegraphics[width=0.25\columnwidth,type=pdf,ext=.pdf,read=.pdf]{#1}}
\newcommand{\plotsmallgrace}[1]{\includegraphics[width=0.7\columnwidth,type=pdf,ext=.pdf,read=.pdf]{#1}}

\section{Introduction}

In the paradigm of Lambda Cold Dark Matter ($\Lambda$CDM), the majority of the matter density of the universe does not couple to electromagnetic fields, leaving it detectable only through its gravitational and possibly weak force interactions.  Nonetheless, the effects of dark matter on the visible universe are spectacular, as the steep gravitational potentials of bound dark matter halos channel baryons together, forming the birthplaces of visible galaxies.  In this model, the locations, sizes, and merging histories for galaxies are thus intricately connected to the growth of bound dark matter structures.

Testing this model in detail requires extensive computer simulations, as the complicated nonlinear evolution of structure growth cannot be fully evaluated by hand.  The simulations generally follow the evolution of a set of dark matter particles and output the positions and velocities of the particles at several discrete timesteps.  These outputs must then be postprocessed to determine the locations and properties of bound dark matter structures also known as ``halos''---namely, the locations and properties which influence the formation of visible galaxies.  This postprocessing (``halo-finding'') necessarily involves both ambiguity and imprecision---ambiguity in the definitions (e.g., the center of a bound halo) and imprecision in determining halo properties due to limited information (e.g., for halos consisting of only a few dark matter particles, or for determining particle membership in two overlapping halos).  Currently, as a consequence, essential statistical measures (e.g., the number density of halos with a given mass) are known to at best 5-10\% even for a specific cosmology \citep{tinker-umf}.  Moreover, halo properties are rarely checked for consistency over multiple timesteps, which can be a serious problem for robust modeling of galaxy formation theories \citep[see however \citealt{BehrooziTree}; this has also been addressed in specific contexts when creating merger trees, 
for example improving subhalo tracking and smoother mass accretion histories, see e.g.][]{Wechsler02, Springel05Mill,Faltenbacher05, Allgood06, Harker06, Tweed09, Wetzel09}.

This level of uncertainty is unacceptable for current and future surveys, including data expected to come from the Baryon Oscillation Spectroscopic Survey (BOSS), Dark Energy Survey (DES), BigBOSS, Panoramic Survey Telescope and Rapid Response System (Pan-STARRS), Extended Roentgen Survey with an Imaging Telescope Array (eROSITA), Herschel, Planck, James Webb Space Telescope (JWST), and Large Synoptic Survey Telescope (LSST) \citep{BOSS,DES,BigBOSS,Pan-STARRS,eROSITA,Herschel,Planck,JWST,LSST}, in order to fully realize the constraining power of these observations.  Indeed, derived quantities such as the halo mass function and autocorrelation functions must be understood at the one percent level in order for theoretical uncertainties to be at the same level as statistical uncertainties in constraining, e.g., dark energy \citep[for example]{Wu10,Cunha10}.   Although there may be a limit to the accuracy possible with dark matter simulations alone, given the impact of baryons on dark matter halo profiles \citep{Stanek09} and the 
power spectrum \citep{Rudd08}, different halo finders running on the same simulation demonstrate that a large fraction of this uncertainty is still due to the process of halo finding itself \citep{Knebe11}.

As previously mentioned, some of these uncertainties are due to limited use of information.  Considerable progress has been made since the first generation of position-space halo finders \citep{Davis85,Cole94}, which used only particle locations to determine bound halos.  Currently, the most advanced algorithms are adaptive phase-space finders \citep[e.g.,][]{Maciejewski09, Ascasibar10}, which make use of the full six dimensions of particle positions and velocities.  At the same time, an often-overlooked (with the possible exception of \citealt{Tweed09}) aspect of halo finding is the extra information stored in the \textit{temporal} evolution of bound particles.  In this paper, we detail an advanced adaptive phase-space and temporal finder which is designed to maximize consistency of halo properties across timesteps, rather than just within a single simulation timestep.  Together with a companion paper \citep{BehrooziTree}, which details the process of comparing and validating halo catalogs across timesteps to create gravitationally consistent merger trees, our combined approach is the first to use particle information in \textit{seven} dimensions to determine halo catalogs, allowing unprecedented accuracy and precision in determining halo properties.

Furthermore, in contrast to previous grid-based adaptive phase-space approaches, ours is the first grid-independent and orientation-independent approach; it is also the first publicly-available adaptive phase-space code designed to run on massively parallel systems on the very large simulations ($>10^{10}$ particles) which are necessary to constrain structure formation across the range of scales probed by current and future galaxy observations.  Finally, we remark that our halo finder is the first which is successfully able to probe substructure masses down to the very centers of host halos (see also \citealt{Knebe11}), which is essential for a full modeling of galaxy statistics and will enable future studies of the expected breakdown between halo positions and galaxy positions due to the effects of baryon interactions at the very centers of galaxies.

Throughout, we have paid careful attention not only to the basic task of assigning particles to halos, but also to the process of estimating useful properties from them to compare with observations.  While in many cases galaxy surveys are not able to probe halo properties to the same precision as halo finders in simulations, one significant counterexample exists.  It is a common practice especially for halo finders based on the friends-of-friends algorithm (e.g., \citealt{Davis85,Springel01,Habib09,Rasera10,Gardner07}; see also \S \ref{s:previous_finders}) to calculate halo velocities by averaging all halo particle velocities together to find a bulk velocity.  Examination of the difference between velocities in the inner regions of halos and the bulk averaged velocity suggests that the bulk average velocity may be offset by several hundred km s$^{-1}$ from the velocity at the radius where galaxies are expected to reside; differences at this scale are easily detectable in cluster redshift surveys and may also factor in interpreting observations of the kinetic Sunyaev-Zel'dovich effect.  As this difference has an important impact on the usefulness of derived halo properties, we additionally perform an investigation of the core-bulk velocity difference in halos across a wide range of redshifts and masses.

We begin this paper with a survey of previous work in halo finding as well as previous limitations in \S \ref{s:previous}.  We discuss our improved methodology in \S \ref{s:methodology} and conduct detailed tests of our approach in \S \ref{s:tests}.  
We present an analysis of the theoretical and practical limitations of simulations in terms of tracking substructure in \S \ref{s:substructure}.  Finally, our results concerning the velocity offsets of cluster cores are presented in \S \ref{s:cores}; we summarize our conclusions in \S \ref{s:conclusions}.  Multiple simulations including slight variations of cosmological parameters are considered in this paper; all simulations model a flat $\Lambda$CDM universe and we always take the Hubble constant $H_0$ to be 70 km s$^{-1}$ Mpc$^{-1}$; equivalently, $h = 0.7$.

\section{Previous Halo Finders}

\label{s:previous}

\subsection{Summary of Previous Approaches}

\label{s:previous_finders}

Previously published approaches to halo finding may be classified, with few exceptions, into two large groups.  Spherical overdensity (SO) finders, such as Amiga's Halo Finder (AHF; \citealt{Knollmann09}), Adaptive Spherical Overdensity Halo Finder (ASOHF; \citealt{Planelles10}), Bound Density Maxima (BDM; \citealt{Klypin99}), SO \citep{Cole94,Jenkins01, Evrard02}, parallel SO (pSO; \citealt{Sutter10}), Voronoi Bound Zones (VOBOZ; \citealt{Neyrinck05}), and Spline Kernel Interpolative Denmax (SKID; \citealt{Stadel01}) proceed by identifying density peaks in the particle distribution and then adding particles in spheres of increasing size around the peaks until the enclosed mass falls below a predetermined density threshold (a top-down approach).  Friends-of-friends (FOF) and HOP-based \citep{HOP} halo finders, such as FOF, SUBFIND, the LANL halo finder, parallel FOF (pFOF), Ntropy-fofsv, and AdaptaHOP \citep{Davis85,Springel01,Habib09,Rasera10,Gardner07,Tweed09}, collect particles together which fall above a certain density threshold and then, if so designed, search for substructure inside these particle collections (a bottom-up approach).  Phase-space finders typically extend these two approaches to include particle velocity information, either by calculating a phase-space density, such as the Hierarchical Structure Finder (HSF; \citealt{Maciejewski09}) and the Six-Dimensional Hierarchical Overdensity Tree (HOT6D; \citealt{Ascasibar10}) or by using a phase-space linking length, as does Six-Dimensional Friends-of-Friends (6DFOF; \citealt{Diemand06}).

There are three notable exceptions to these algorithms, namely the ORIGAMI halo finder (discussed in \citealt{Knebe11}), the Hierarchical Bound-Tracing algorithm (HBT; \citealt{Han11}), and SURV \citep{Giocoli10}.  ORIGAMI operates by examining phase-space shell crossings for the current particle distribution as compared to the initial particle distribution; shells which have crossed along three dimensions are considered to be halos (as opposed to shells which have crossed along one or two dimensions, which would be considered as sheets and filaments, respectively).  HBT uses a friends-of-friends approach to find distinct halos and uses particle lists from distinct halos at previous timesteps to test for the presence of subhalos.  SURV is a very similar algorithm, except that distinct halos are identified using spherical overdensities.  These algorithms all rely heavily on temporal information in their approach to halo finding.

\subsection{Limitations of Previous Algorithms}

\label{s:limitations}

In order to develop an improved halo finder, it is important to understand some of the shortcomings of previous approaches.  For the vast majority of halos, even the most basic of algorithms (FOF and SO) do an acceptable job of determining halo properties to $10\%$ accuracy \citep{Knebe11}.  However, recent interest in the detailed properties and histories of halos --- e.g., precision mass functions and merger trees and the shape of tidal structures --- requires improvements to older approaches; this has resulted in a proliferation of new codes in the past decade \citep[summarized in][]{Knebe11}.

The most significant improvements to halo finding have come from using the information from six-dimensional (position and velocity) phase space.  Two traditional weak points for 3D (position-space) halo finders have been major mergers and subhalos close to the centers of their host halos.  In both cases, the density contrast alone is not enough to distinguish which particles belong to which halo: when two halos are close enough, the assignment of particles to halos becomes essentially random in the overlap region.  However, as long as the two halos have relative motion, six-dimensional halo finders can use particle velocity space information to very effectively determine particle-halo membership.  This, coupled with the ability of 6D halo finders to find tidal remnants (which are condensed in phase space but not in position space), means that phase-space capabilities are required for the most accurate and interesting studies of dark matter halos.

At the same time, phase space presents a unique challenge.  While position space and velocity space have well-defined distance metrics, there is not a single, unique way to combine position and velocity distance into a single metric.  For a useful definition of phase-space distance, one needs to be able to decide, e.g., whether an object just passing by the origin at 1 km s$^{-1}$ is closer or farther than an object at rest 1 kpc away.  One approach, used by 6DFOF, is to choose in advance a static conversion between velocity and position space.  While simple, this approach seems somewhat self-limiting: if too short a linking-length is chosen, the full extent of substructures cannot be found; if too large a linking-length is chosen, then otherwise distinct substructures will be merged together.

A demonstrably superior approach, at least in terms of recovering halo properties \citep{Knebe11}, is to \textit{adaptively} define a phase-space metric.  Both HSF and HOT6D subdivide the simulation space into six-dimensional hyperboxes containing (at the maximum refinement level) as little as a single particle each.  For a given particle, the enclosing hyperbox gives a local estimate of the phase-space metric, based on the relative sizes of the hyperbox's dimensions in position and velocity space.  The usefulness of this estimate depends heavily on the method for partitioning space into hyperboxes; HSF uses, for example, an entropy-based approach to determine whether more information is contained in the particle locations for position space or velocity space.

These algorithms all give excellent results for identifying halo centers at a single timestep.  However, consistent halo catalogs across timesteps are often compromised by a fundamental ambiguity in the definition of a host halo.  For major mergers, it is often unclear which halo center represents the larger ``host'' or central halo, and which represents the subhalo.  Phase-space halo finding helps when the two halo centers are relatively far apart (i.e., weakly interacting), because there exists a strong correlation between the velocities of particles in the halo outskirts and halo centers.  However, when the centers come close enough to interact strongly, this correlation is weakened, and it becomes much more difficult to accurately assign particles to the halos.  As a result, it is much more difficult to determine which of the halo centers should be considered the host halo.  Since the definition of halo mass often includes the mass of subhalos, this problem can result in large mass fluctuations across timesteps for merging halos.

A number of solutions to this problem have been proposed and examined with the AdaptaHOP halo finder \citep{Tweed09}.  \cite{Tweed09} found that a temporal approach (examining the host vs. subhalo assignment at earlier timesteps) was most successful at fixing this problem.  Other methods, such as choosing the densest halo center to be the host halo, have inherent instabilities because of the spread in halo concentrations at a given halo mass.  At the same time, while \cite{Tweed09} successfully resolves this problem, it nonetheless only finds halos in position space, and thus has the same weaknesses in identifying subhalo and major merger properties.

The HBT and SURV algorithms \citep{Han11,Giocoli10} use the ingenious approach of tracing subhalos by using the particles found in previously distinct halos, which could potentially also solve many of these problems.  Yet,  they both also include assumptions about accretion onto subhalos (e.g., that subhalos cannot accrete background matter from the host) which are untrue in HBT's case for large linking-lengths (as halos will be identified as satellites far from the actual virial radius of the host) and for both halo finders with major mergers (where satellite and host are more ambiguous; they can in any case easily trade particles with each other).  These assumptions vastly simplify the code at some expense to the completeness and accuracy of the mass function.  More seriously, the design of the algorithms \textit{requires} temporal information to find subhalos; in cases where simulation outputs are spaced very far apart or when only one timestep is available, they cannot effectively find substructure.  These issues are in principle fixable: future versions of the algorithms could easily combine advanced single-timestep substructure finding with checks against previous timesteps' particle lists.

\section{An Improved Approach: \textsc{rockstar}}

\label{s:methodology}

Our primary motivation in developing a halo finder was to improve the accuracy of halo merger trees that are required for an understanding of galaxy evolution.  The design of our halo finder was thus motivated by a requirement for consistent accuracy \textit{across multiple timesteps}.  This interest led to the development of a unique, adaptive phase-space temporal algorithm which is provably independent of halo orientation and velocity relative to the simulation axes, and which also attempts to be highly preserving of particle-halo and halo-subhalo membership across timesteps.  In addition, we paid special attention to the algorithm's efficiency and parallelizability, to allow it to run on the largest current datasets and so that it could easily scale to the next generation of simulations.  Thus far, we have run the halo finder (and in many cases the partner merger tree code) on the Bolshoi \citep[$\sim10^{10}$ particles,][]{Bolshoi} and LasDamas simulations (200 boxes of $1-4\times10^{9}$ particles each, McBride et al. in preparation), on several 2048$^3$ simulations run to create Dark Energy Survey (DES) simulated sky catalogs, on $\sim$ one hundred high resolution halos simulated as part of the RHAPSODY project \citep{Wu12b,Wu12a}, and on halos A-1 through A-5 of the Aquarius Simulation \citep[up to $1.4\times 10^9$ particles in a single halo;][]{Aquarius}.

\begin{figure}
\begin{center}
\large
\begin{tabular}{m{70pt}@{\hspace{0ex}}m{173pt}}
\raisebox{8.5ex}{\multirow{6}{*}{\hspace{-3ex}\includegraphics[width=70pt]{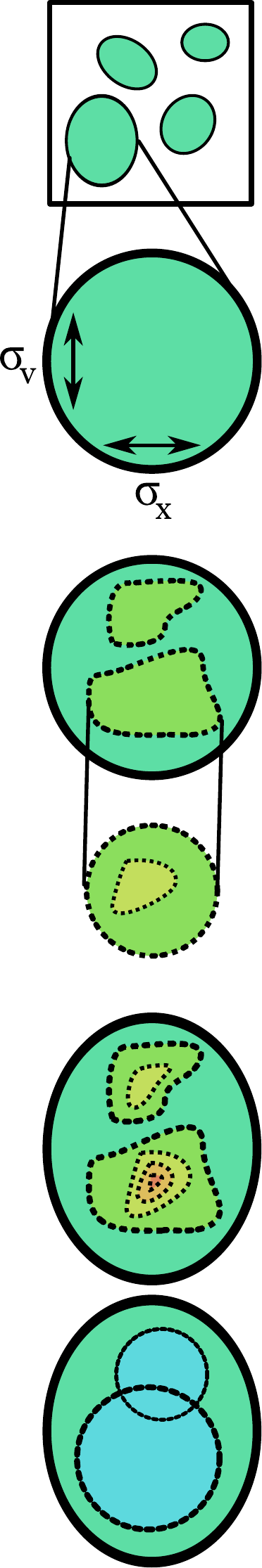}}} & \textbf{1.}\ The simulation volume is divided into 3D Friends-of-Friends groups for easy parallelization.\\[3ex]
& \textbf{2.}\ For each group, particle positions and velocities are divided (normalized) by the group position and velocity dispersions, giving a natural phase-space metric.\\[2ex]
& \textbf{3.}\ A phase-space linking length is adaptively chosen such that 70\% of the group's particles are linked together in subgroups.\\[1.5ex]
& \textbf{4.}\ The process repeats for each subgroup: renormalization, a new linking-length, and a new level of substructure calculated.\\[1.5ex]
& \textbf{5.}\ Once all levels of substructure are found, seed halos are placed at the lowest substructure levels and particles are assigned hierarchically to the closest seed halo in phase space.\\[1.5ex]
& \textbf{6.}\ Once particles have been assigned to halos, unbound particles are removed and halo properties (positions, velocities, etc.) are calculated.\\
\end{tabular}
\setlength{\extrarowheight}{0ex}
\normalsize
\caption{A visual summary of the particle-halo assignment algorithm.}
\label{f:summary}
\end{center}
\end{figure}

As a first step, our algorithm performs a rapid variant of the 3D friends-of-friends (FOF) method to find overdense regions which are then distributed among processors for analysis (\S\ref{s:fof}).  Then, it builds a hierarchy of FOF subgroups in phase space by progressively and adaptively reducing the 6D linking length, so that a tunable fraction of particles are captured at each subgroup as compared to the immediate parent group (\S\ref{s:subfof}).  Next, it converts this hierarchy of FOF groups into a list of particle memberships for halos (\S\ref{s:halos}).  It then computes the host halo/subhalo relationships among halos, using information from the previous timestep if available (\S\ref{s:hostsub}).  Finally, it removes unbound particles from halos and calculates halo properties, before automatically generating particle-based merger trees (\S\ref{s:halo_props}).  A visual summary of these steps is shown in Fig.\ \ref{f:summary}.

\subsection{Efficient, Parallel FOF Group Calculation}

\label{s:fof}

The 3D friends-of-friends (FOF) algorithm has existed since at least 1985 \citep{Davis85}.  In principle, implementation of the algorithm is straightforward: one attaches two particles to the same group if they are within a prespecified linking length.  Typically, this linking length is chosen in terms of a fraction $b$ of the mean interparticle distance (in our code, as for others, the cube root of the mean particle volume); common values for generating halo catalogs range from $b=0.15$ to $b=0.2$ \citep{More11}.

In practice, this means that one must determine neighbors for every particle within a sphere of radius equal to the linking length.  Even with an efficient tree code (we use a custom binary space partitioning tree), this represents a great deal of wasted computation, especially in dense cluster cores.  In such cases, particles might have tens of thousands of neighbors within a linking length, all of which will eventually end up in the same FOF group.

As the 3D FOF groups are used in our method only to divide up the simulation volume into manageable work units, we instead make use of a modified algorithm which is an order of magnitude faster.  As is usual, neighboring particles are assigned to be in the same group if their distance is within the linking length.  However, if a particle has more than a certain number of neighbors within the linking length (16, in our version), then the neighbor-finding process for those neighboring particles is skipped.  Instead, neighbors for the original particle out to \textit{twice} the original linking length are calculated.  If any of those particles belong to another FOF group, that corresponding \textit{group} is joined with that of the original particle.  Thus, although the neighbor-finding process has been skipped for the nearest particles, groups which would have been linked through those intermediate neighbors are still joined together.  

This process therefore links together at minimum the same particles as in the standard FOF algorithm---fine for our desired purpose---but does so much faster: neighbors must be calculated over a larger distance, but many fewer of those calculations must be performed.  Indeed, rather than slowing down as the linking length is increased, our variation of FOF becomes faster.  Because of this, we are free to choose an exceptionally large value for the linking length.
A value of $b=0.2$ is too small, as it does not include all particles out to the virial radius \citep{More11}; after evaluating different choices for $b$ (see \S \ref{s:def_param}), we chose $b=0.28$, which guarantees that virial spherical overdensities can be determined for even the most ellipsoidal halos.

The most important parallelization work occurs at this stage.  Separate reader tasks load particles from snapshot files.  Depending on the number of available CPUs for analysis, a master process divides the simulation region into rectangular boundaries, and it directs the reader tasks to send particles within those boundaries to the appropriate analysis tasks.

Each analysis task first calculates 3D FOFs in its assigned analysis region, and FOFs which span processor boundaries are automatically stitched together.  The FOF groups are then distributed for further phase-space analysis according to individual processor load.  The load-balancing procedure is described in further detail in Appendix \ref{a:load_balancing}.  Currently, single 3D FOF groups are analyzed by at most one processor.  Also, in the current implementation, there is no support for multiple particle masses, although support could easily be added by varying the linking length depending on particle mass.  Provided enough interest, support for multiprocessor analysis of single large halos as well as support for multiple particle masses may be added in a future version of \textsc{rockstar}.

\subsection{The Phase-Space FOF Hierarchy}

\label{s:subfof}

For each 3D FOF group which is created in the previous step, the algorithm proceeds by building a hierarchy of FOF subgroups in phase space.  Deeper levels of subgroups have a tighter linking-length criterion in phase space, which means that deeper levels correspond to increasingly tighter isodensity contours around peaks in the phase-space density distribution.  This enables an easy way to distinguish separate substructures --- above some threshold phase-space density, their particle distributions must be distinct in phase space; otherwise, it would be difficult to justify the separation into different structures.\footnote{This would not be true if, e.g., halos had Plummer profiles or otherwise flat density profiles in their centers.}

Beginning with a base FOF group, \textsc{rockstar} adaptively chooses a phase-space linking length based on the standard deviations of the particle distribution in position and velocity space.  That is, for two particles $p_1$ and $p_2$ in the base group, the phase-space distance metric is defined as:
\begin{equation}
d(p_1, p_2) = \left(\frac{|\vec{x}_1-\vec{x}_2|^2}{\sigma_x^2} + \frac{|\vec{v}_1-\vec{v}_2|^2}{\sigma_v^2}\right)^{1/2},
\end{equation}
where $\sigma_x$ and $\sigma_v$ are the particle position and velocity dispersions for the given FOF group; this is identical to the metric of \cite{Stefan98}.  For each particle, the distance to the nearest neighbor is computed; the phase-space linking length is then chosen such that a constant fraction $f$ of the particles are linked together with at least one other particle.  In large groups (>10,000 particles), where computing the nearest neighbor for all particles can be very costly, the nearest neighbors are only calculated for a random 10,000-particle subset of the group, as this is sufficient to determine the linking length to reasonable precision.

The proper choice of $f$ is constrained by two considerations.  If one chooses too large a value ($f>0.9$), the algorithm will take much longer, and it can also find spurious (not statistically significant) subgroups.  If one chooses too low of a value ($f<0.5$), the algorithm may not find smaller substructures.  As such, we use an intermediate value ($f=0.7$); with the recommended minimum threshold for halo particles ($20$).  In our tests of the mock NFW \citep{NFW97} halos described in \cite{Knebe11}, this results in a false positive rate of 10\% for 20-particle groups compared to a cosmological subhalo distribution, which declines exponentially for larger group sizes.  These false positives are easily removed by the significance and boundedness tests described in \S \ref{s:halos} and \S \ref{s:unbound}.

Once subgroups have been found in the base FOF group, this process is repeated.  For each subgroup, the phase-space metric is recalculated, and a new linking-length is selected such that a fraction $f$ of the subgroup's particles are linked together into sub-subgroups.  Group finding proceeds hierarchically in phase space until a predetermined minimum number of particles remain at the deepest level of the hierarchy.  Here we set this minimum number to 10 particles, although halo properties are not robust approaching this minimum.

\subsection{Converting FOF Subgroups to Halos}

\label{s:halos}

For each of the subgroups at the deepest level of the FOF hierarchy (corresponding to the local phase-space density maxima), a seed halo is generated.  The algorithm then recursively analyzes higher levels of the hierarchy to assign particles to these seed halos until all particles in the original FOF group have been assigned.  To prevent cases where noise gives rise to duplicated seed halos, we automatically calculate the Poisson uncertainty in seed halo positions and velocities, and merge the two seed halos if their positions and velocities are within 10$\sigma$ of the uncertainties. Specifically, the uncertainties are calculated as $\mu_x = \sigma_x / \sqrt{n}$ and $\mu_v = \sigma_v / \sqrt{n}$, where $\sigma_x$ and $\sigma_v$ are the position and velocity dispersions, and $n$ is the number of particles, all for the smaller of the two seed halos.  The two halos are merged if
\begin{equation}
\sqrt{(x_1-x_2)^2\mu_x^{-2} + (v_1-v_2)^2\mu_v^{-2}} < 10\sqrt{2}.
\end{equation}
In our tests, this threshold yields a near-featureless halo autocorrelation function; lower values result in a spurious upturn in the autocorrelation function close to the simulation force resolution.

For a parent group which contains only a single seed halo, all the particles in the group are assigned to the single seed.  For a parent group which contains multiple seed halos, however, particles in the group are assigned to the closest seed halo in phase space.  In this case, the phase-space metric is set by the seed halo properties, so that the distance between a halo $h$ and a particle $p$ is defined as:
\begin{eqnarray}
\label{e:membership}
d(h, p) & = & \left(\frac{|\vec{x}_h-\vec{x}_p|^2}{r_{\mathrm{dyn,vir}}^2} + \frac{|\vec{v}_h-\vec{v}_p|^2}{\sigma_v^2}\right)^{1/2}\\
r_{\mathrm{dyn,vir}} & = & \vmax t_\mathrm{dyn,vir} = \frac{\vmax}{\sqrt{\frac{4}{3}\pi G \rho_{vir}}} 
\end{eqnarray}
where $\sigma_v$ is the seed halo's current velocity dispersion, $\vmax$ is its current maximum circular velocity (see \S\ref{s:halo_props}), and ``$\mathrm{vir}$'' specifies the virial overdensity, using the definition of $\rho_{vir}$ from \cite{mvir_conv}, which corresponds to 360 times the background density at $z=0$, however, other choices of this density can easily be applied.

Using the radius $r_{\mathrm{dyn,vir}}$ as the position-space distance normalization may seem unusual at first, but the natural alternative (using $\sigma_x$) gives unstable and nonintuitive results.  At fixed phase-space density, subhalos and tidal streams (which have lower velocity dispersions than the host halo) will have larger position-space dispersions than the host halo.  Thus, if $\sigma_x$ were used, particles in the outskirts of a halo could be easily mis-assigned to a subhalo instead of the host halo.  Using $r_{\mathrm{den,vir}}$, on the other hand, prevents this problem by ensuring that particles assigned to subhalos cannot be too far from the main density peak even if they are close in velocity space.\footnote{For determination of tidal streams, this ``problem'' becomes a ``feature,'' and use of $\sigma_x$ may be preferable to $r_{\mathrm{vir}}$.}  Intuitively, the largest effect of using $r_\mathrm{vir}$ is that velocity-space information becomes the dominant method of distinguishing particle membership when two halos are within each other's virial radii.\footnote{An alternate radius (e.g., $r_{200b}$ or $r_{500c}$) could be used instead, but it would have an effect only on a small fraction of particles in a small fraction of halos (major mergers).}

This process of particle assignment assures that substructure masses are calculated correctly independently of the choice of $f$, the fraction of particles present in each subgroup relative to its parent group.  In addition, for a subhalo close to the center of its host halo, it assures that host particles are not mis-assigned to the subhalo --- the central particles of the host will naturally be closer in phase space to the true host center than they are to the subhalo's center.

\subsection{Calculating Substructure Membership}

\label{s:hostsub}

In addition to calculating particle-halo membership, it is also necessary to determine which halos are substructures of other halos.  The most common definition of substructure is a bound halo contained within another, larger halo.  Yet, as halo masses are commonly defined to include substructure, the question of which of two halos is the largest (and thus, which should be called a satellite of the other) can change depending on which substructures have been assigned to them.  This is one of the largest sources of ambiguity between spherical overdensity halo finders, even those which limit themselves to distinct halos.

We break the self-circularity by assigning satellite membership based on phase-space distances before calculating halo masses.  Treating each halo center like a particle, we use the same metric as Eq.\ \ref{e:membership} and calculate the distance to all other halos with larger numbers of assigned particles.  The satellite halo in question is then assigned to be a subhalo of the closest larger halo within the same 3D friends-of-friends group, if one exists.  If the halo catalog at an earlier timestep is available, this assignment is modified to include temporal information.  Halo cores at the current timestep are associated with halos at the previous timestep by finding the halo at the previous timestep with the largest contribution to the current halo core's particle membership.  Then, host-subhalo relationships are checked against the previous timestep; if necessary, the choice of which halo is the host may be switched so as to preserve the host-subhalo relationship of the previous timestep.

As explained above, these host-subhalo relationships are only used internally for calculating masses: particles assigned to the host are not counted within the mass of the subhalo, but particles within the subhalo are counted as part of the mass of the host halo.  This choice assures that the mass of a halo won't suddenly change as it crosses the virial radius of a larger halo, and it provides more stable mass definitions in major mergers.  Once halo masses have been calculated, the subhalo membership is recalulated according to the standard definition (subhalos are within $r_\Delta$ of more massive host halos) when the merger trees are constructed.

For clarity, it should be noted that every density peak within the original FOF analysis group will correspond to either a host halo or a subhalo in the final catalog.  It has been observed that FOF groups will ``bridge'' or ``premerge'' long before their corresponding SO halo counterparts \citep[e.g.,][]{Bolshoi}.  However, as we calculate full SO properties associated with each density peak, a single FOF group is naturally allowed to contain multiple SO host halos; thus the bridging or premerging of FOF groups does not affect the final halo catalogs.

\subsection{Calculating Halo Properties and Merger Trees}

\label{s:halo_props}

Typically, several properties of interest are generated for halo catalogs.  Regardless of the quality of particle assignment in the halo finder, careful attention to halo property calculation is essential for consistent, unbiased results.

\subsubsection{Halo Positions and Velocities}

\label{s:posvel}

For positions, \cite{Knebe11} demonstrated that halo finders which calculated halo locations based on the maximum density peak were more accurate than FOF-based halo finders which use the averaged location of all halo particles (see also \citealt{Gao06}).  The reason for this may be simply understood: as particle density rapidly drops in the outer reaches of a halo, the corresponding dispersion of particle positions climbs precipitously.  Consequently, rather than increasing the statistical accuracy of the halo center calculation, including the particles at the halo boundary actually reduces it.  The highest accuracy is instead achieved when the expected Poisson error ($\sigma_x / \sqrt{N}$) is minimized.  As our halo finder has access (via the hierarchy of FOF subgroups) to the inner regions of the halo density distribution, an accurate calculation of the center is possible by averaging the particle locations for the inner subgroup which best minimizes the Poisson error.  Typically, for a $10^6$ particle halo, this estimator ends up averaging the positions of the innermost $10^3$ particles.

The picture for halo velocities is not quite as simple.  As noted in \S \ref{s:cores}, the halo core can have a substantial velocity offset from the halo bulk.  Since the galaxy hosted by the halo will presumably best track the halo core, we calculate the main velocity for the halo using the average particle velocity within the innermost 10\% of the halo radius.  For calculating the bound/unbound mass of the halo (see \ref{s:hmass}), however, we use the more appropriate averaged halo bulk velocity including substructure.

\subsubsection{Halo Masses}

\label{s:hmass}

For halo masses, \textsc{rockstar} calculates spherical overdensities according to multiple user-specified density thresholds: e.g., the virial threshold, from \cite{mvir_conv}, or a density threshold relative to the background or to the critical density.  As is usual, these overdensities are calculated using all the particles for all the substructure contained in a halo.  On the other hand, subhalo masses have traditionally been a major point of ambiguity (for density-space halo finders).  With a phase-space halo finder, such as \textsc{rockstar}, the particles belonging to the subhalo can be more reliably isolated from the host, and thus less ambiguity exists: the same method of calculating spherical overdensities may be applied to just the particles belonging to the subhalo.  In terms of the definition of where the subhalo ``ends,'' Eq.\ \ref{e:membership} implies that the subhalo edge is effectively where the distribution of its particles in phase space becomes equidistant from the subhalo and its host halo.  If alternate mass definitions are necessary, the halo finder can output the full phase-space particle-halo assignments; these may then be post-processed by the user to obtain the desired masses.

\subsubsection{Unbinding Particles}

\label{s:unbound}

By default, \textsc{rockstar} performs an unbinding procedure before calculating halo mass and $v_{max}$, although this may be switched off for studies of e.g., tidal remnants.  Because the algorithm operates in phase space, the vast majority of halo particles assigned to central halos are actually bound.  We find typical boundedness values of 98\% at $z=0$; see \S \ref{s:def_param} and \cite{BehrooziUnbound}.  Even for substructure, unbound particles typically correspond to tidal streams at the outskirts of the subhalo, making a complicated unbinding algorithm unnecessary.  For this reason, as well as to improve consistency of halo masses across timesteps, we perform only a single-pass unbinding procedure using a modified Barnes-Hut method to accurately calculate particle potentials (see Appendix \ref{a:unbinding} for details).\footnote{Provided enough interest, we may add the option of a multi-pass unbinding procedure in future versions of the halo finder.}  Since many users will be interested in classical halo finding only as opposed to recovering tidal streams, the code by default does not output halos where fewer than 50\% of the particles are bound; this threshold is user-adjustable, but changing it does not produce statistically significant effects on the clustering or mass function until halos with a bound fraction of less than 15\% are included (see \S \ref{s:def_param}).

We note that, in major mergers, a more careful approach to unbinding must be used.  In many cases where merging halos initially have large velocity offsets, particles on the outskirts of the halos can mix in phase space before the halo cores themselves merge.  This results in many particles being unbound with respect to either of the two halo cores, even though they are bound to the overall merging system.  As such, a naive unbinding of the particles would lead to the merging halos' masses dropping sharply for several timesteps prior to the final merger.\footnote{With the \textsc{bdm}  halo finder \citep{Bolshoi}, for example, we have observed halo masses which drop by a factor of three on account of this affect.}  To counter this effect in major mergers, \textsc{rockstar} calculates the gravitational potential of the combined merging system, rather than for individual halos, when determining whether to unbind particles.  This behavior is triggered by default when two halos have a merger ratio of 1:3 or larger; this value is user-adjustable, but has little effect on the recovered mass function or clustering (see \S \ref{s:def_param}).

\subsubsection{Additional Halo Properties and Merger Trees}

Two more common outputs of halo finders are $v_{max}$, the maximum circular velocity and $R_s$, the scale radius.  $v_{max}$ is simply taken as the maximum of the quantity $\sqrt{GM(r)r^{-1}}$; it should be noted that this quantity is robust even for the smallest halos because of the extremely shallow dependence of $v_{max}$ on radius.  For $R_s$, we divide halo particles into up to 50 radial equal-mass bins (with a minimum of 15 particles per bin) and directly fit an NFW \citep{NFW97} profile to determine the maximum-likelihood fit.

We also calculate the Klypin scale radius for comparison \citep{Bolshoi}, which uses $v_{max}$ and $M_{vir}$ to calculate $R_s$ under the assumption of an NFW \citep{NFW97} profile.  In particular, for NFW profiles, the radius $R_{max}$ at which $v_{max}$ is measured is a constant multiple $b$ of the radius $R_s$, and is given by:
\begin{equation}
\frac{d}{db}\left[b^{-1}\ln(1+b) - \frac{1}{1+b}\right] = 0
\end{equation}
This may be solved numerically, with the result that
\begin{equation}
R_{max} = 2.1626 R_s
\end{equation}
Instead of using $R_{max}$ directly (which is ill-determined for small halos), we make use of the ratio of $R_{max}/R_s$ to relate the mass enclosed within $2.1626 R_s$ to $\vmax$, $M_\mathrm{vir}$, and the concentration $c = R_{vir} / R_s$:
\begin{eqnarray}
\label{e:concentration}
\frac{c}{f(c)} & = & v_{max}^2 \frac{R_{vir}}{G M_{vir}} \frac{2.1626}{f(2.1626)}\\
f(x) & \equiv & \ln(1+x) - \frac{x}{1+x}
\end{eqnarray}
Thus, by numerically inverting the function on the left-hand side of Eq.\ \ref{e:concentration}, $c$ may be found as a function of $M_{vir}$ and $v_{max}$, and the Klypin scale radius $R_s$ can be derived.  The Klypin scale radius is more robust than the fitted scale radius for halos with less than 100 particles; this is due not only to shot noise, but also due to the fact that halo profiles are not well-determined at distances comparable to the simulation force resolution.

We additionally calculate the angular momentum of the halo (using bound particles out to the desired halo radius) and the halo spin parameter ($\lambda$), as introduced by \cite{Peebles69}:
\begin{equation}
\lambda = \frac{J\sqrt{|E|}}{G M_\mathrm{vir}^{2.5}}
\end{equation}
where $J$ is the magnitude of the halo angular momentum and $E$ is the total energy of the halo (potential and kinetic).  For comparison, we also calculate the Bullock spin parameter \citep{Bullock01}, defined as
\begin{equation}
\lambda_B = \frac{J}{\sqrt{2} M_\mathrm{vir} V_\mathrm{vir} R_\mathrm{vir}} = \frac{J}{\sqrt{2 G M_\mathrm{vir}^3 R_\mathrm{vir}}}
\end{equation}

To help with cluster studies, we calculate several halo relaxation parameters.  These include the central position offset ($X_\mathrm{off}$, defined as the distance between the halo density peak and the halo center-of-mass), the central velocity offset ($V_\mathrm{off}$, defined as the difference between the halo core velocity and bulk velocity), and the ratio of kinetic to potential energy ($\frac{T}{|U|}$) for particles within the halo radius.  We refer interested readers to \cite{Neto07} for a description and comparison of methods for determining halo relaxedness.

We also calculate ellipsoidal shape parameters for halos.  Following the recommendations of \cite{Zemp11}, we calculate the mass distribution tensor for particles within the halo radius, excluding substructure:
\begin{equation}
M_{ij} = \frac{1}{N} \sum_{N} x_i x_j
\end{equation}
The sorted eigenvalues of this matrix correspond to the squares of the principal ellipsoid axes ($a^2 > b^2 > c^2$).  We output both the axis ratios ($\frac{b}{a}$ and $\frac{c}{a}$) as well as the largest ellipsoid axis vector, $\vec{A}$.

Finally, we mention that our halo finder automatically creates particle-based merger trees.  For a given halo, its descendant is assigned as the halo in the next timestep which has the maximum number of particles in common (excluding particles from subhalos).  While it is possible to use these merger trees directly, we recommend instead to use the advanced merger tree algorithm discussed in \cite{BehrooziTree}.  This algorithm detects and corrects inconsistencies across timesteps (e.g., halos which disappear and reappear as they cross the detection threshold) to further improve the temporal consistency of the merger trees.

\begin{figure*}[h!]
\begin{center}
\vspace{-12ex}
\includegraphics[width=1.5\columnwidth]{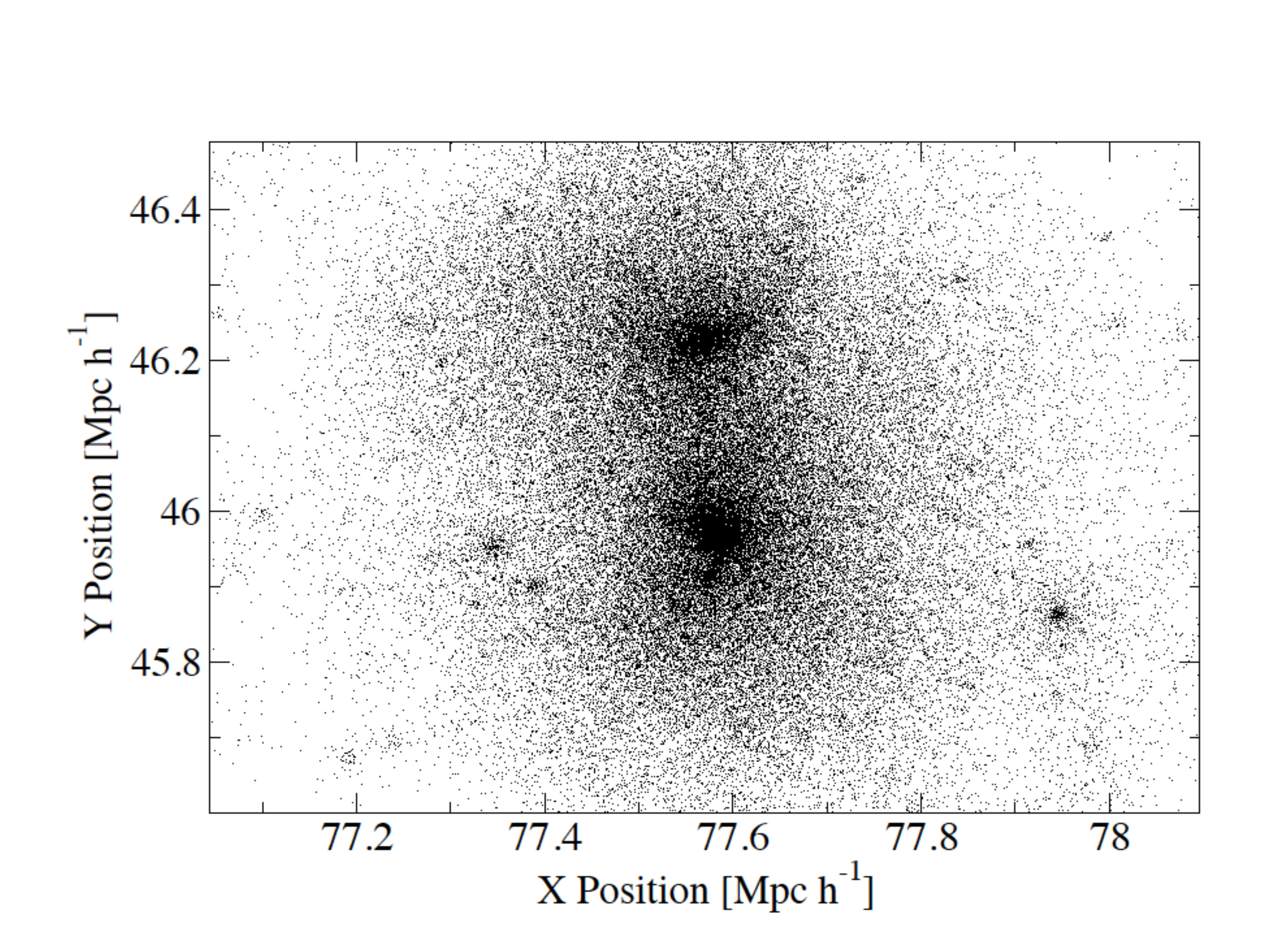}\\[-3ex] \includegraphics[width=1.02\columnwidth]{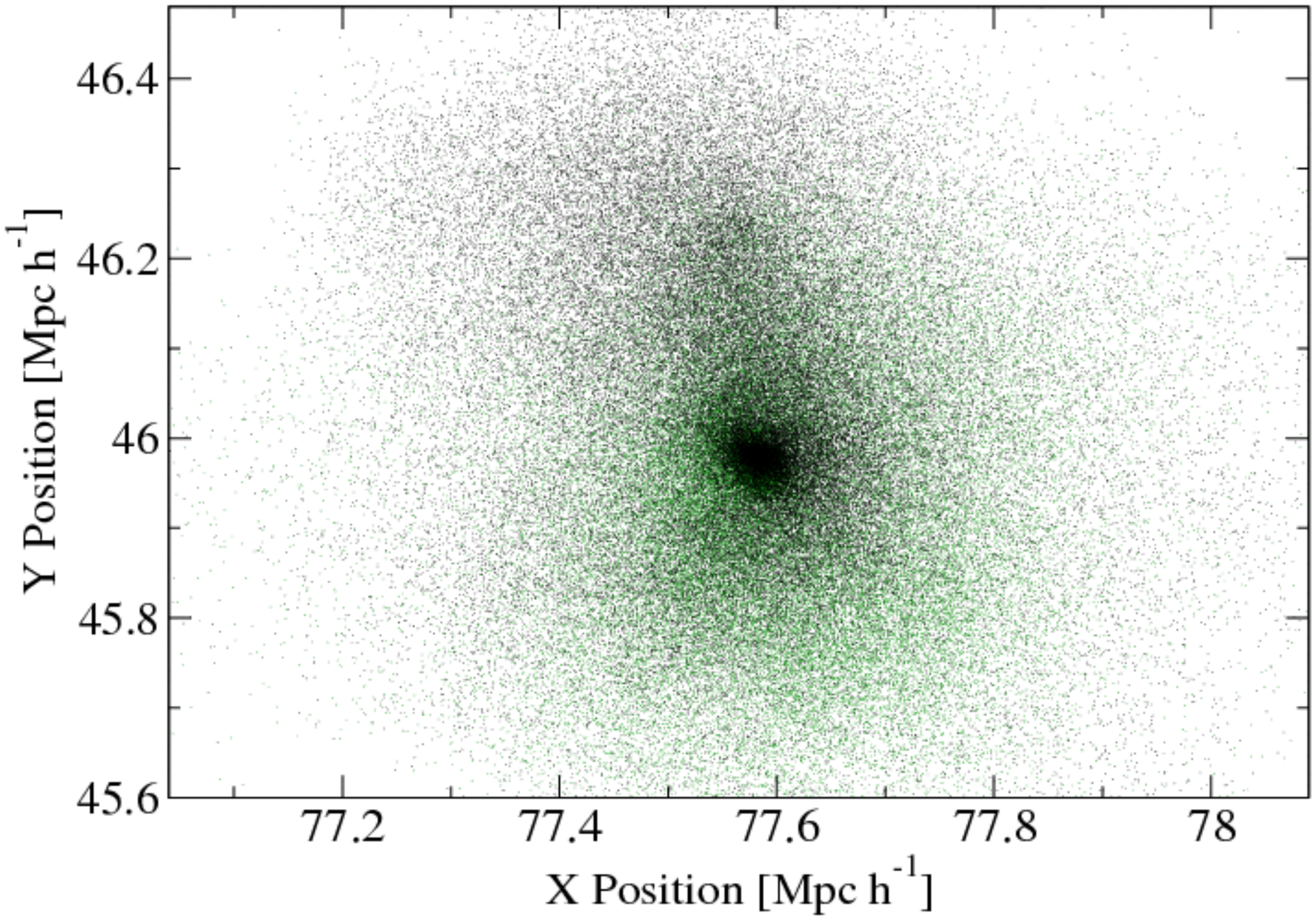}\hspace{-3ex} \includegraphics[width=1.02\columnwidth]{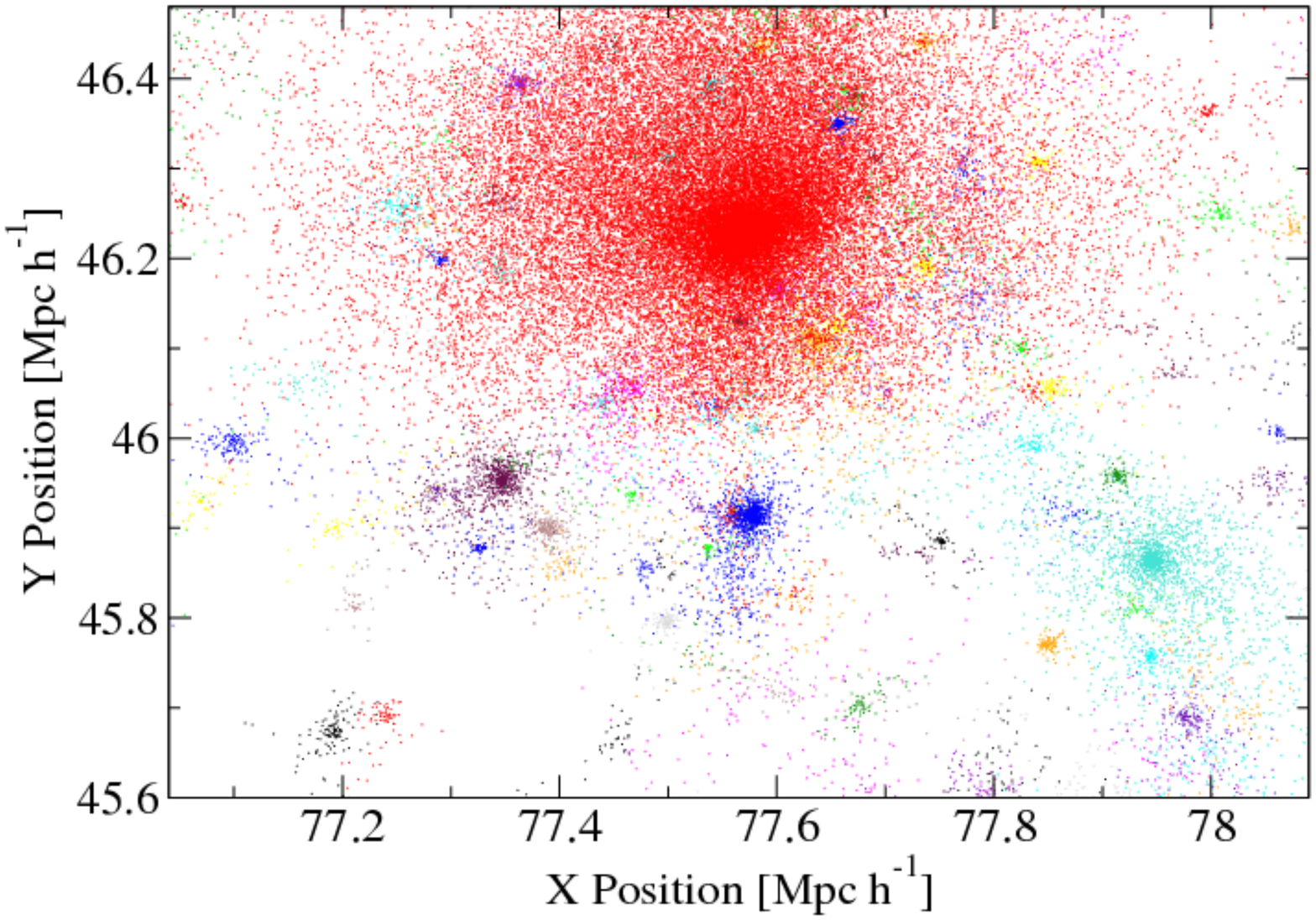}\\
\includegraphics[width=0.95\columnwidth]{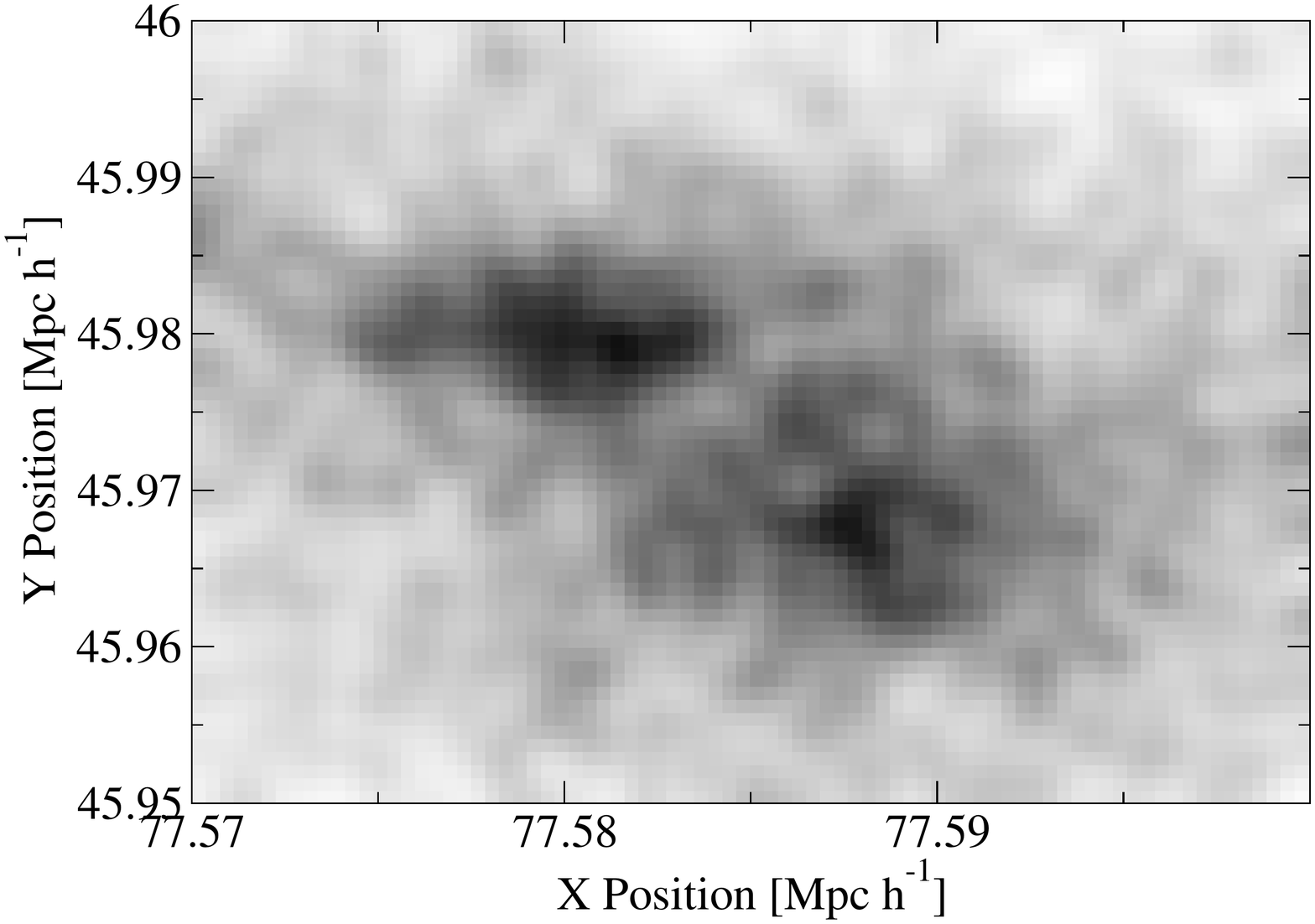} \includegraphics[width=0.95\columnwidth]{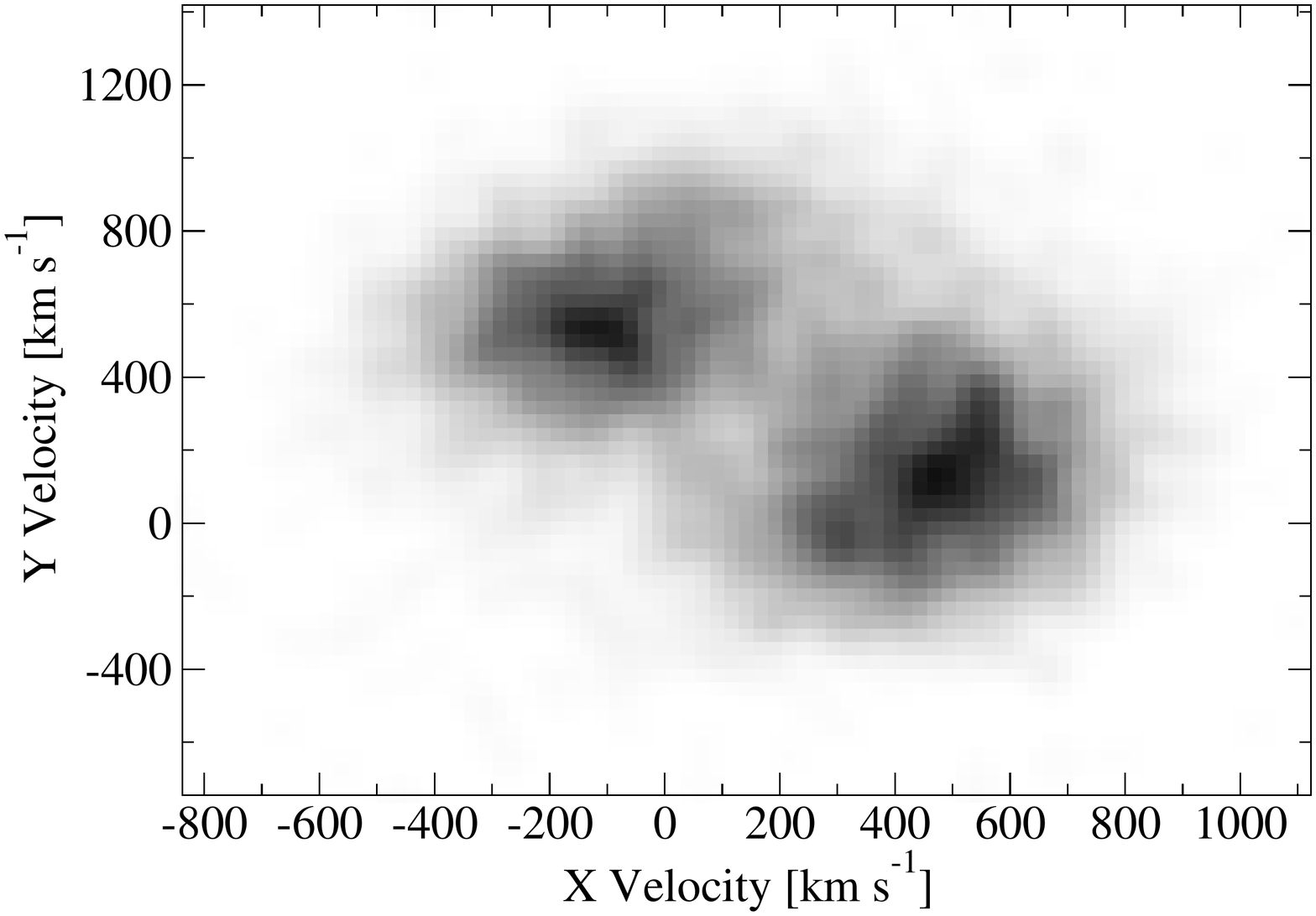}
\end{center}
\caption{\textsc{rockstar} allows recovery of even very close major mergers.  This figure shows an example of a major merger involving $10^{13} \Msun$ halos from the Bolshoi simulation \citep{Bolshoi}.  The \textbf{top} panel shows the complete particle distribution around the merging halos.  In the \textbf{second row}, the left panel shows the host particle distribution, and the right panel shows the subhalo particle distribution, with particles colored according to subhalo membership.  (The particle plotting size has been increased to show more clearly the extent of the small substructures in the right-hand panel). The two different colors in the left-hand panel hint at the fact that there are indeed \textit{three} halos involved in the major merger, two of which are extremely close to merging.  The uniform subhalo shapes in the right-hand panel suggest that subhalo particles can be distinguished without bias despite extreme variations in the host particle density between the subhalo centers and the subhalo outskirts.  The \textbf{bottom row} shows more clearly the extremely close major merger.  The bottom left-hand panel shows the full particle distribution in position space in a small region close to the merging halo cores; here, the bimodal distribution is evident, but distinguishing particle membership is impossible beyond the immediate vicinity of the cores.  On the other hand, the bottom right-hand panel shows the same particles in velocity space, where the bimodality from the two cores shows a clear velocity separation, allowing particles to be reasonably assigned even in the halo bulk.}
\label{f:vis_demo}
\end{figure*}

\begin{figure*} 
  \plotgrace{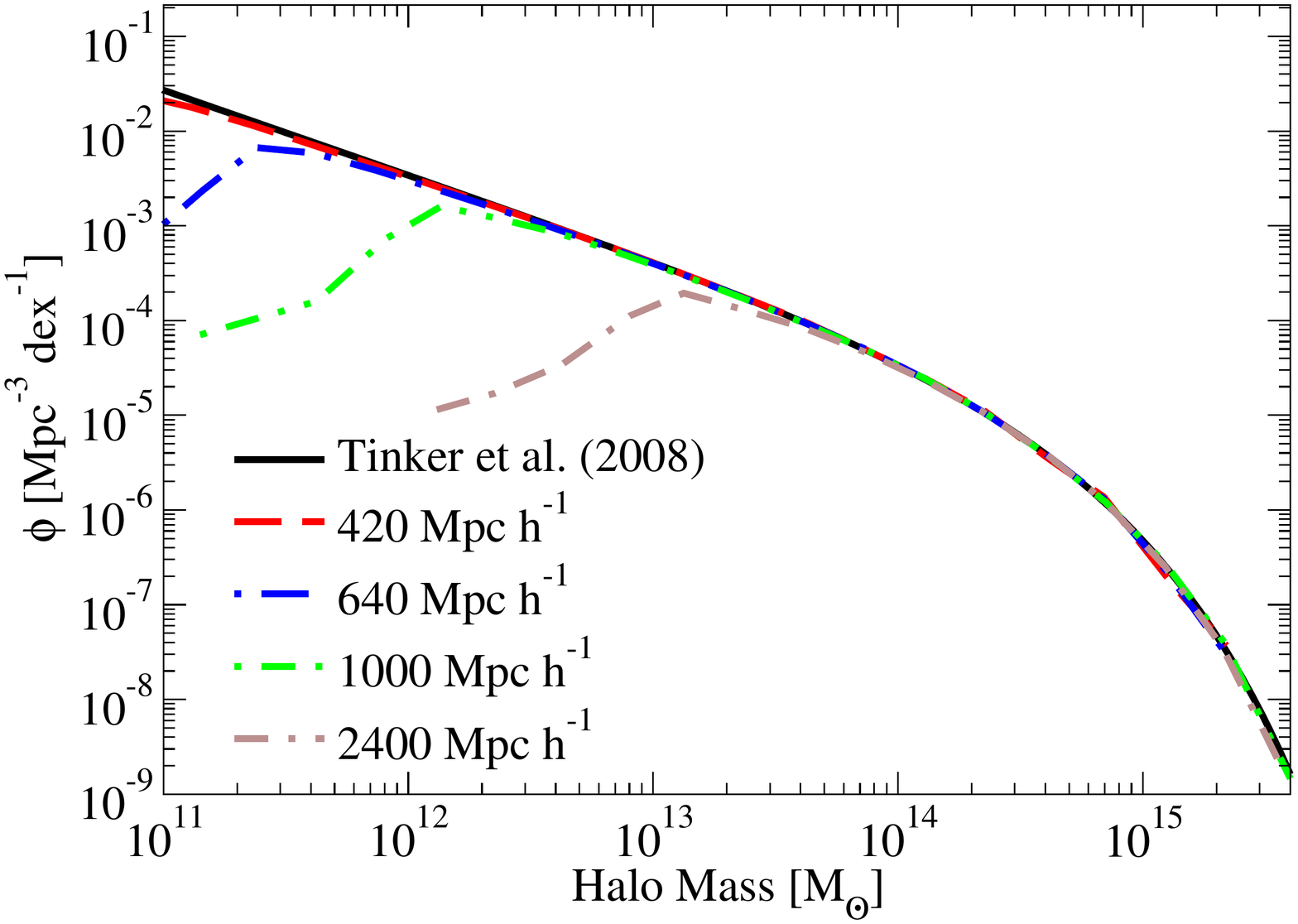} \hspace{2ex} \plotgrace{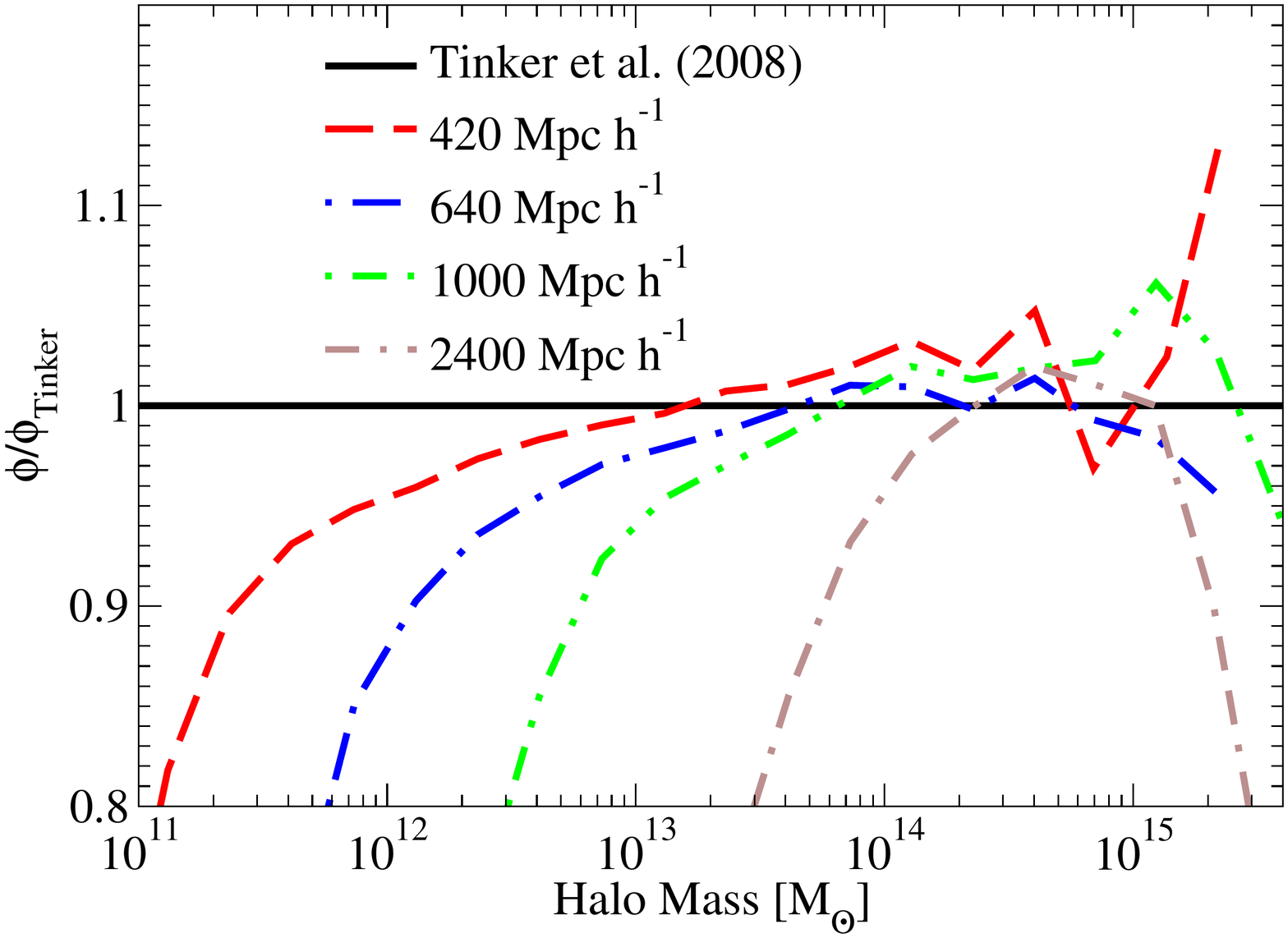}\\[-2ex]
  \plotgrace{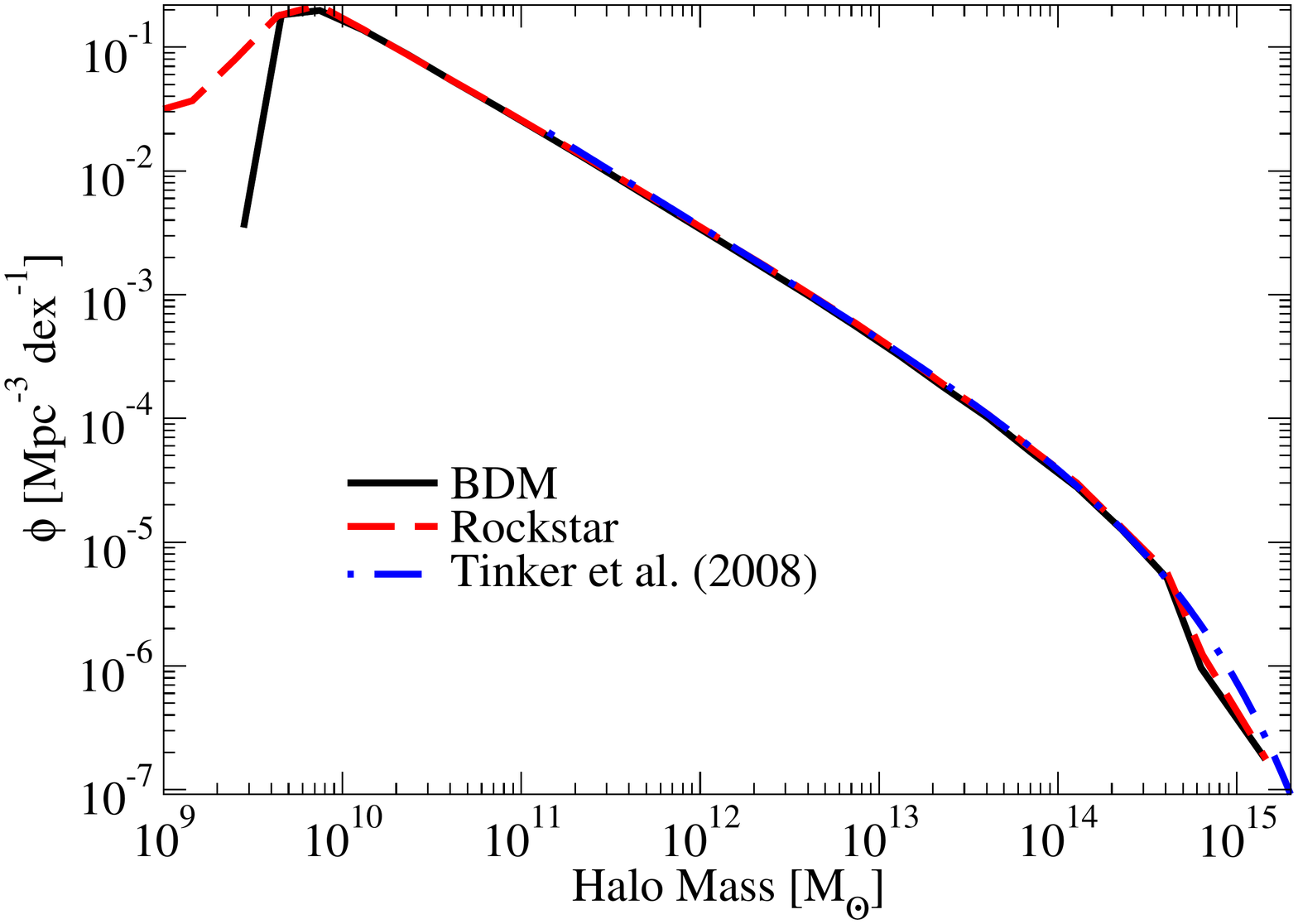} \hspace{2ex} \plotgrace{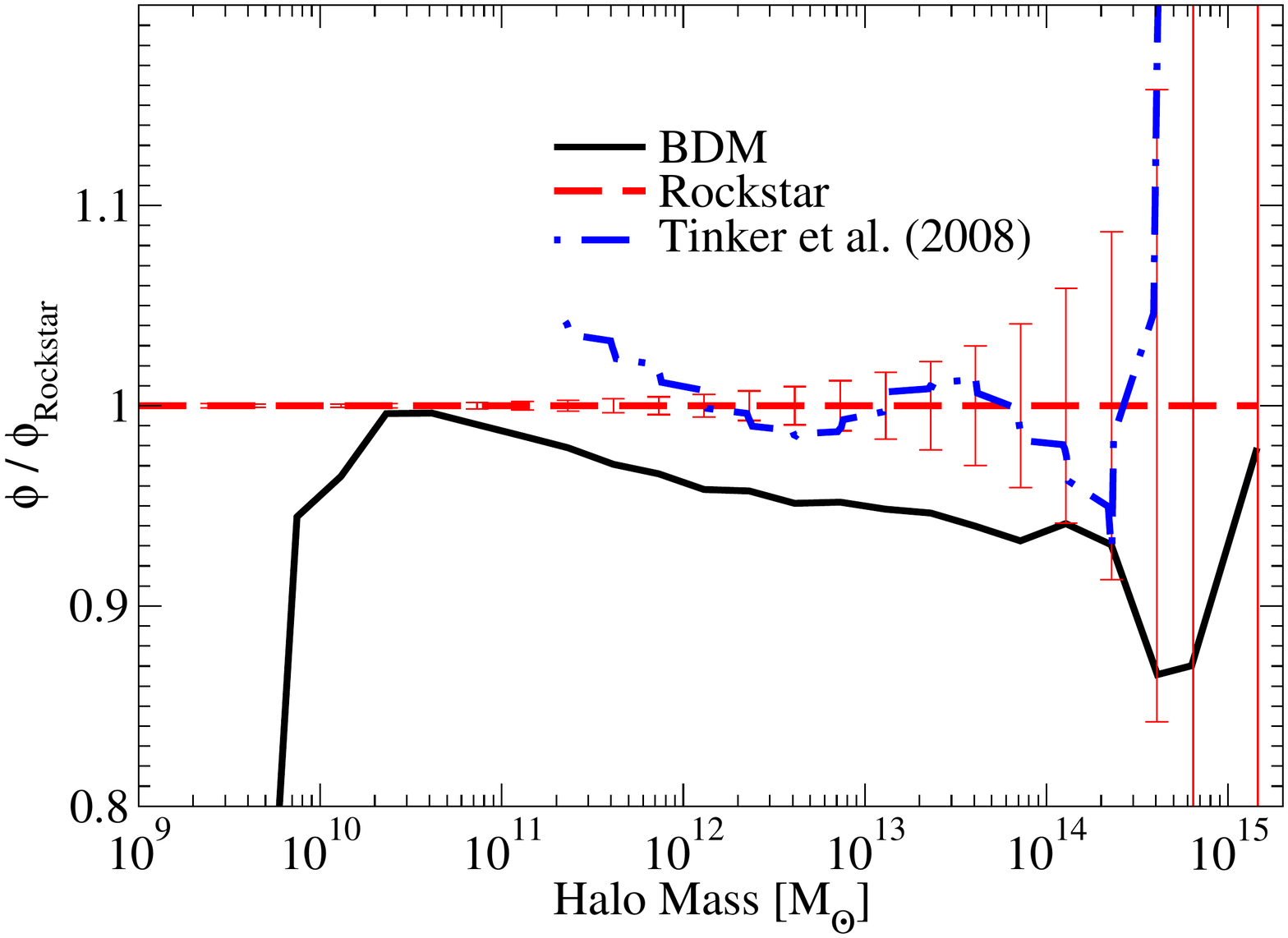}\\
  \caption{The halo mass function for distinct halos is very similar to previously published results.  The \textbf{top row} shows comparisons to the \cite{tinker-umf} central halo mass function for each of the four LasDamas boxes (see Table \ref{t:sim_data}).  Good agreement is seen above 100 particles.  The \textbf{bottom row} shows a comparison between the \textsc{rockstar} and \textsc{bdm}  \citep{Bolshoi} halo finders on the Bolshoi simulation (2048$^3$ particles, 250 Mpc $h^{-1}$).  The \textbf{left-hand} plots show the full mass functions, and the \textbf{right-hand} plots show the residuals, with Poisson errors shown for the Bolshoi simulation.  As noted in \cite{tinker-umf}, the calibrated mass range does not extend below $10^{11}$ h$^{-1} \Msun$; furthermore, the authors state ``the behavior of the fitting function at lower masses is arbitrary.''  We therefore do not extrapolate the \cite{tinker-umf} mass function in our comparisons.}
\label{f:mf_comp}
\end{figure*}

\begin{figure*}
\plotgrace{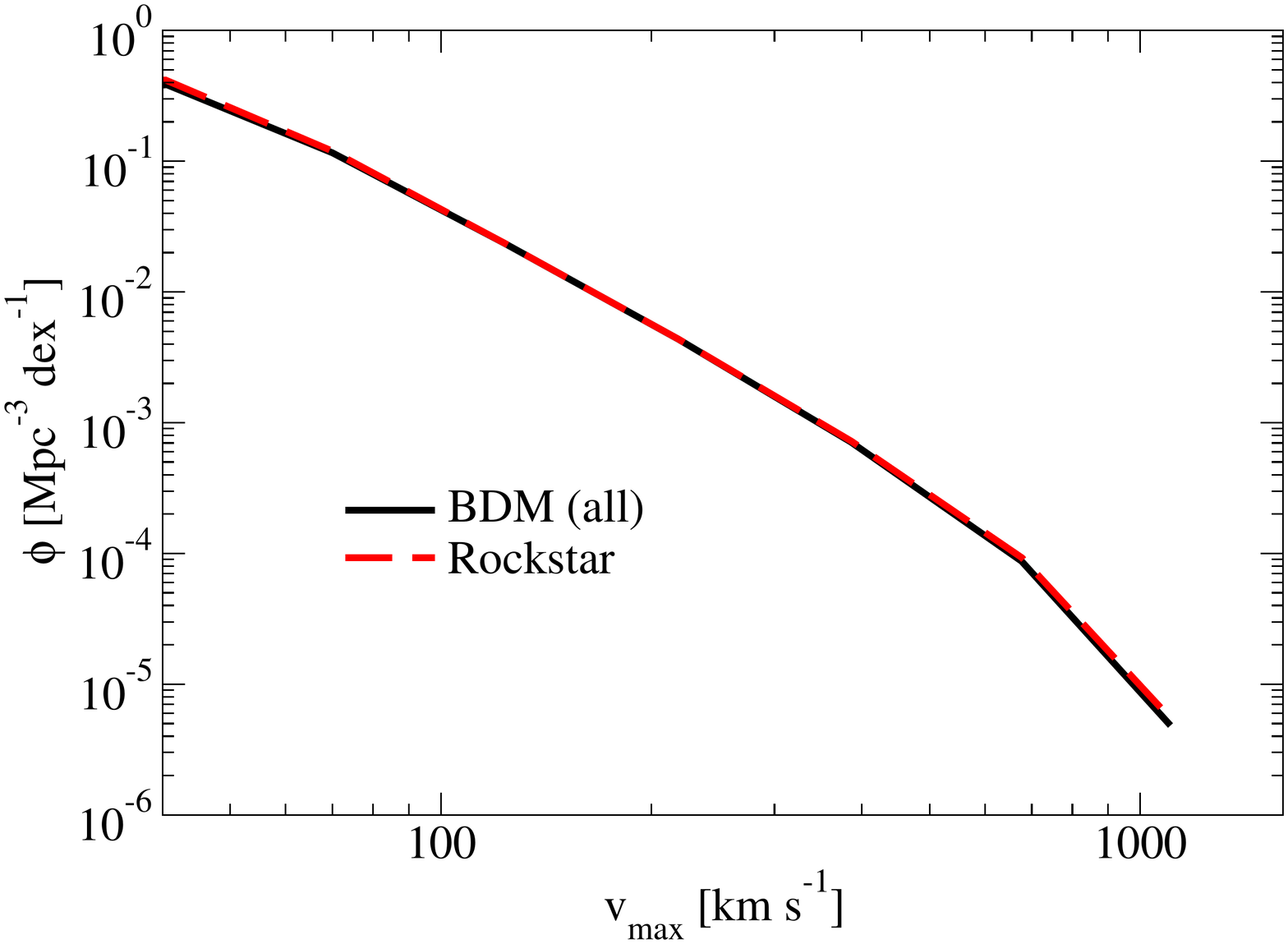} \hspace{2ex} \plotgrace{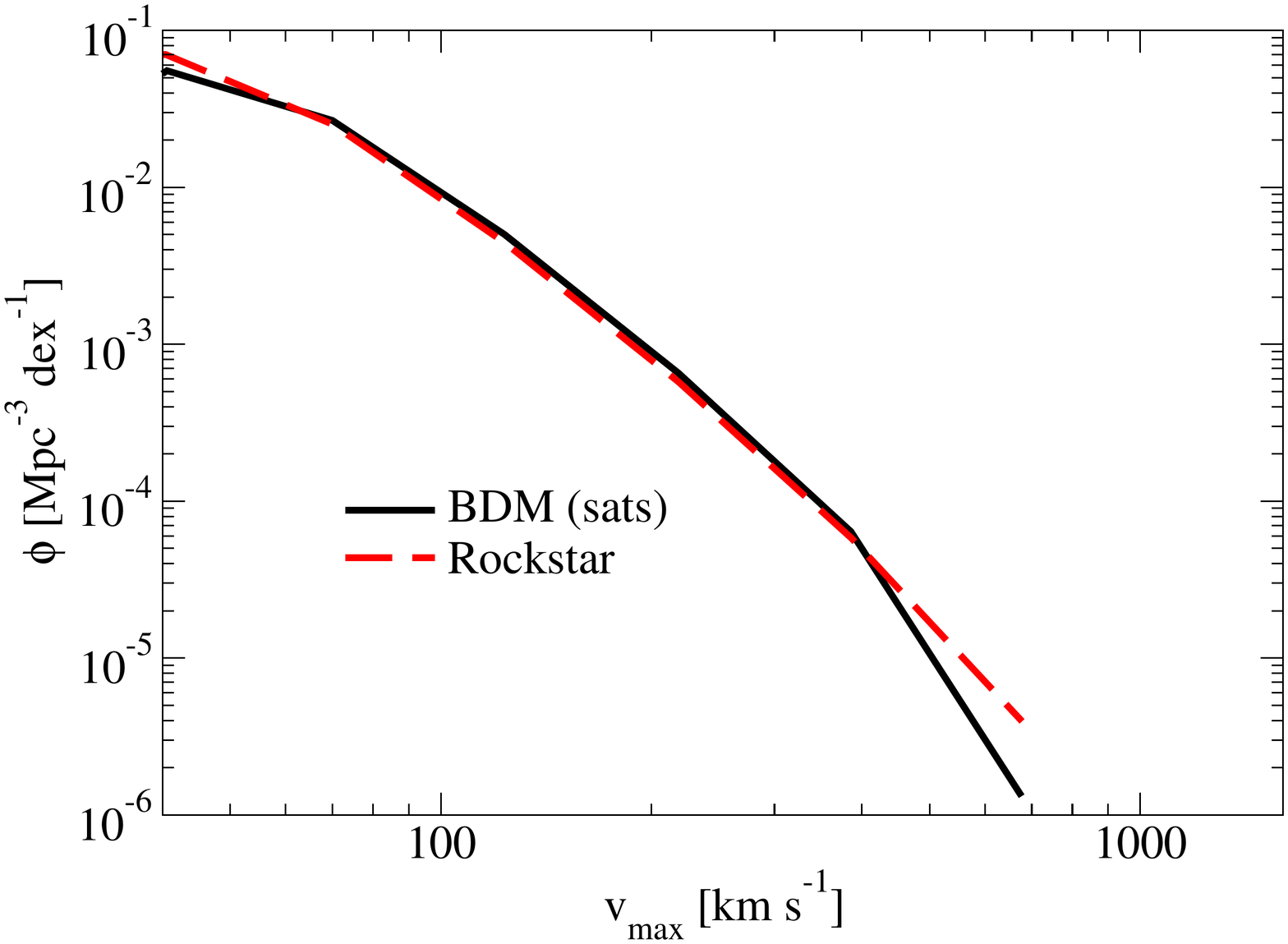}
\caption{The halo velocity function is also very similar to previously published results.  This figure shows comparisons between velocity ($v_\mathrm{max}$) functions for the \textsc{rockstar} halo finder and \textsc{bdm}  on the Bolshoi simulation (2048$^3$ particles, 250 Mpc $h^{-1}$).  The \textbf{left-hand} plot shows all halos; the \textbf{right-hand} plot shows satellite halos only.}
\label{f:vf_comp}
\end{figure*}

\section{Tests \& Comparisons}

\label{s:tests}

The \textsc{rockstar} algorithm has already undergone extensive testing and comparison to other halo finders in \cite{Knebe11}.  In tests with generated mock halos, \textsc{rockstar} successfully recovered halo properties for halos down to 20 particles, in many cases (e.g., for halo mass and bulk velocity) better than all seventeen other participating halo finders.  In cases where it did not perform best, it was often only marginally behind one of the other phase-space halo finders.  Notably, out of all the halo finders, it was the only one to fully successfully recover all halo properties (mass, location, position, velocity, and $v_{max}$) for a subhalo coinciding with the center of its host halo.  In addition, \cite{Knebe11} compared mass functions, $v_{max}$ functions, correlation functions (for $r > 2$ Mpc), and halo bulk velocities for a cosmological simulation with $1024^3$ particles; \textsc{rockstar} had results comparable to other halo finders in all these results, although only the other phase-space halo finders shared its low mass completeness limit ($\sim$ 25 particles for $M_{200c}$).

We thus focus on comparisons beyond those already covered in \cite{Knebe11}.  Our comparisons cover results for several different dark matter simulations, briefly summarized in \S \ref{s:sims}.  We provide a visual demonstration of the algorithm's performance in \S \ref{s:vis_demo}, a detailed comparison with the mass and correlation functions for other halo finders in \S \ref{s:mf_comp}, an evaluation of the dynamical accuracy of halo properties in \S \ref{s:dyn_perf}, and we present justification for our choice of the default parameters in \S \ref{s:def_param}.  Finally, we show figures demonstrating the excellent performance of the halo finder in \S \ref{s:performance}.

\subsection{Simulation Parameters}

\begin{table*}
\begin{center}
\caption{Simulation Parameters}
\label{t:sim_data}
\begin{tabular}{rcccccccccc}
\hline
\hline
Name & Particles & Size & Particle Mass & Softening & $\Omega_m$ & $\Omega_\Lambda$ & $h$ & $\sigma_8$ & $n_s$ & Code\\
\hline
Bolshoi & $2048^3$ & 250 $h^{-1}$ Mpc & $1.36\times 10^8 h^{-1}\Msun$ & 1 $h^{-1}$ kpc & 0.27 & 0.73 & 0.7 & 0.82 & 0.95 & ART\\
Consuelo & $1400^3$ & 420 $h^{-1}$ Mpc & $1.87\times 10^9 h^{-1}\Msun$ & 8 $h^{-1}$ kpc & 0.25 & 0.75 & 0.7 & 0.8 & 1.0 & GADGET-2\\
Esmeralda & $1250^3$ & 640 $h^{-1}$ Mpc & $9.31\times 10^9 h^{-1}\Msun$ & 15 $h^{-1}$ kpc & 0.25 & 0.75 & 0.7 & 0.8 & 1.0 & GADGET-2\\
Carmen & $1120^3$ & 1000 $h^{-1}$ Mpc & $4.94\times 10^{10} h^{-1}\Msun$ & 25 $h^{-1}$ kpc & 0.25 & 0.75 & 0.7 & 0.8 & 1.0 & GADGET-2\\
Oriana & $1280^3$ & 2400 $h^{-1}$ Mpc & $4.57\times 10^{11} h^{-1}\Msun$ & 53 $h^{-1}$ kpc & 0.25 & 0.75 & 0.7 & 0.8 & 1.0 & GADGET-2\\
\hline
\end{tabular}
\end{center}
\tablecomments{Descriptions are in \S \ref{s:sims}.}
\end{table*}

\label{s:sims}

We use five sets of simulations for this paper.  The first of these, Bolshoi, has already been extensively detailed in \cite{Bolshoi}.  To summarize the relevant parameters, it is a 2048$^3$ particle simulation of comoving side length 250 Mpc $h^{-1}$, run using the ART simulation code \citep{kravtsov_etal:97} on the NASA Ames Pleiades supercluster.  The assumed cosmology is $\Omega_m = 0.27$, $\Omega_\Lambda = 0.73$, $h = 0.7$, $n_s = 0.95$, and $\sigma = 0.82$, similar to WMAP7 results \citep{wmap7}; the effective force-softening length is $1$ kpc $h^{-1}$, and the particle mass is $1.36\times 10^8 \Msun$ $h^{-1}$.

We also use four simulations of different sizes from the LasDamas project (McBride et al, in preparation).\footnote{{\tt http://lss.phy.vanderbilt.edu/lasdamas/}}  These have $1120^3$ to $1400^3$ particles in comoving regions from 420 Mpc $h^{-1}$ to 2400 Mpc $h^{-1}$ on a side, and were run using the GADGET-2 code \citep{Springel05}.  Again, the assumed cosmology is similar to WMAP7, with $\Omega_m = 0.25$, $\Omega_\Lambda = 0.75$, $h = 0.7$, $n_s = 1.0$, and $\sigma = 0.8$.  The effective force-softening lengths range from 8 to 53 kpc $h^{-1}$, and the particle masses range from $1.87\times 10^9 \Msun$ $h^{-1}$ to $4.57\times 10^{11} h^{-1}\Msun$.  Details of all the simulations are summarized in Table \ref{t:sim_data}.

\subsection{Visual Demonstration}

\label{s:vis_demo}

In order to demonstrate how the algorithm performs on halos in a real simulation, Fig.\ \ref{f:vis_demo} shows an example of how particles are assigned to halos in a major merger; this example has been taken from the Bolshoi simulation at $z=0$.  From the top image in the figure, one might expect that two large halos separated by about 200 kpc $h^{-1}$ are merging together, but careful analysis reveals that the larger halo actually consists of \textit{another} major merger wherein the halo cores are separated by only 15 kpc $h^{-1}$.  As shown in the bottom-left panel of the figure, the existence of the third massive halo is visible at a moderate significance level in position space---however, a position-space halo finder would have no way to correctly assign particles beyond the immediate locality of the two cores. Yet, in the bottom-right panel, the separation of the two halo cores is clearly distinguishable in velocity space.  As such, not only can the distinct existence of the close-to-merging halos be reliably confirmed, but particle assignment to the two halos based on particle velocity coordinates can be reliably carried out as well.

We also remark that phase-space halo-finding helps improve the accuracy of subhalo shapes by removing the need to perform a position-space cut to distinguish host particles from substructure particles.  Satellite halos are usually offset in velocity space from their hosts, but just as importantly, they usually also have a much lower velocity dispersion than their hosts.  This implies that satellites may be reliably found even in the dense cores of halos---although the position space density is very high for host particles, the average velocity dispersion is large enough that the lower-dispersion subhalo particles can be reliably distinguished from host particles. Consequently, as shown in the middle-right panel of Fig.\ \ref{f:vis_demo}, satellites are picked out even in the dense halo centers without bias as regard to shape or size.

\begin{figure}
\plotgrace{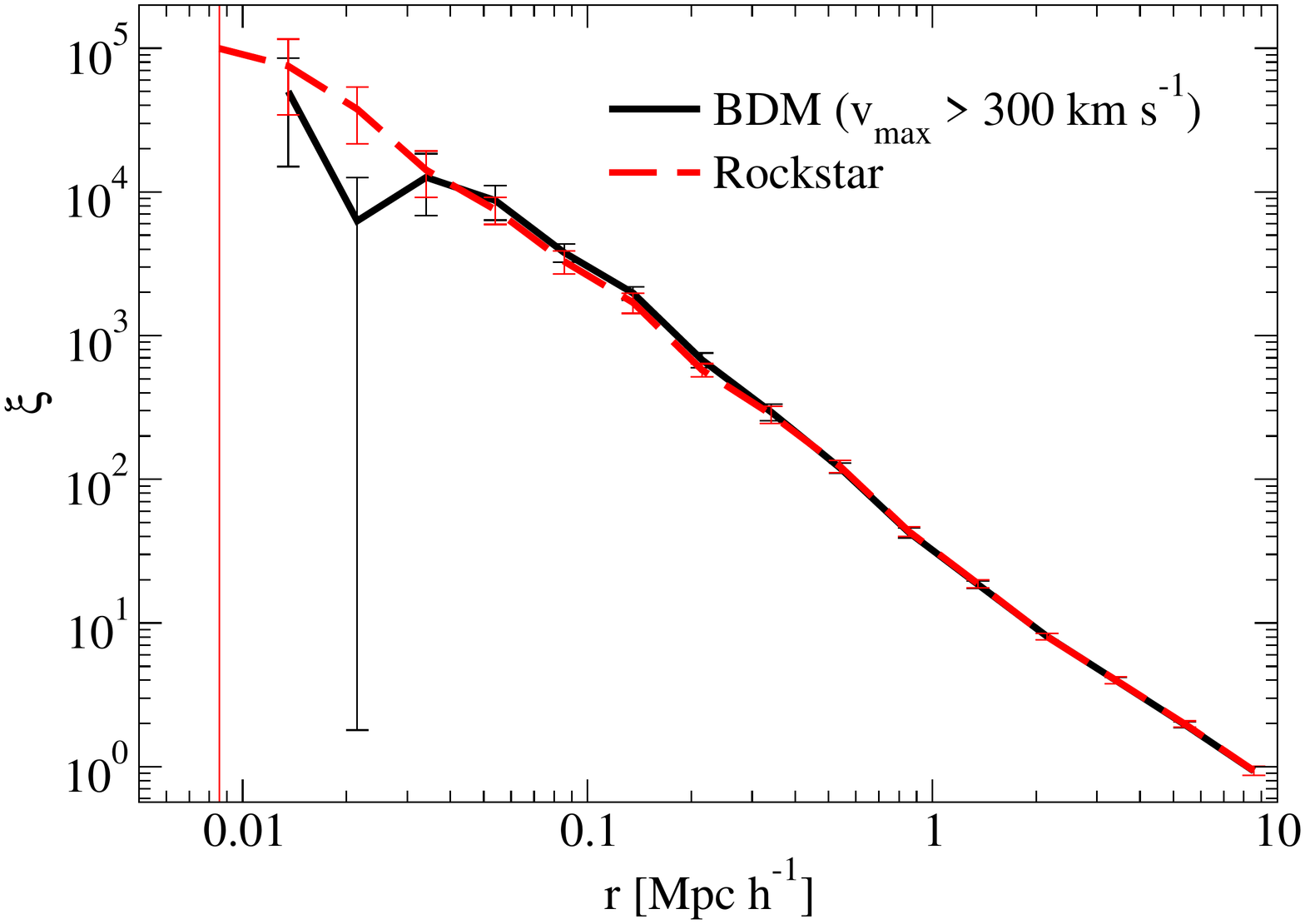}\\ \plotgrace{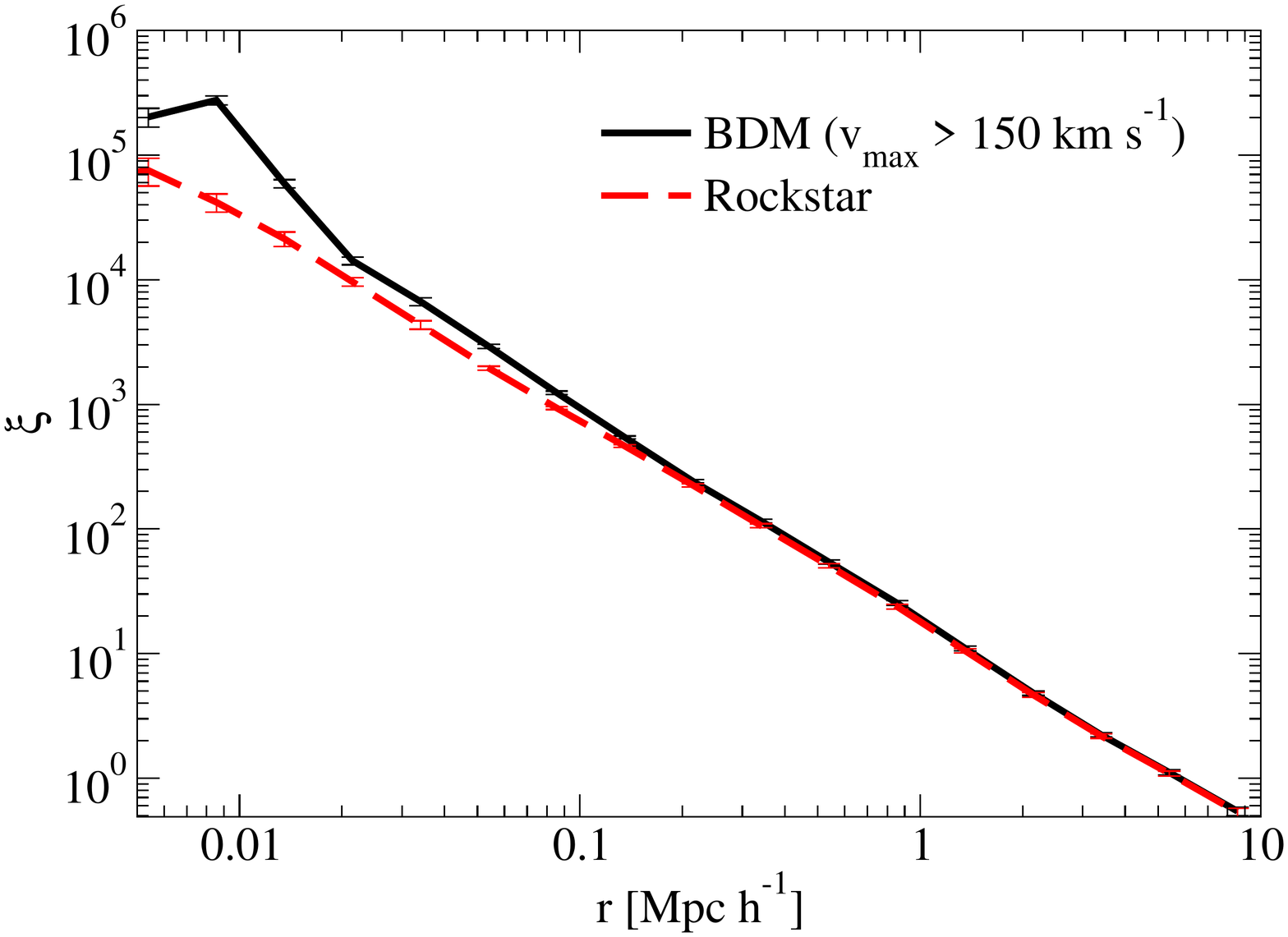}\\ \plotgrace{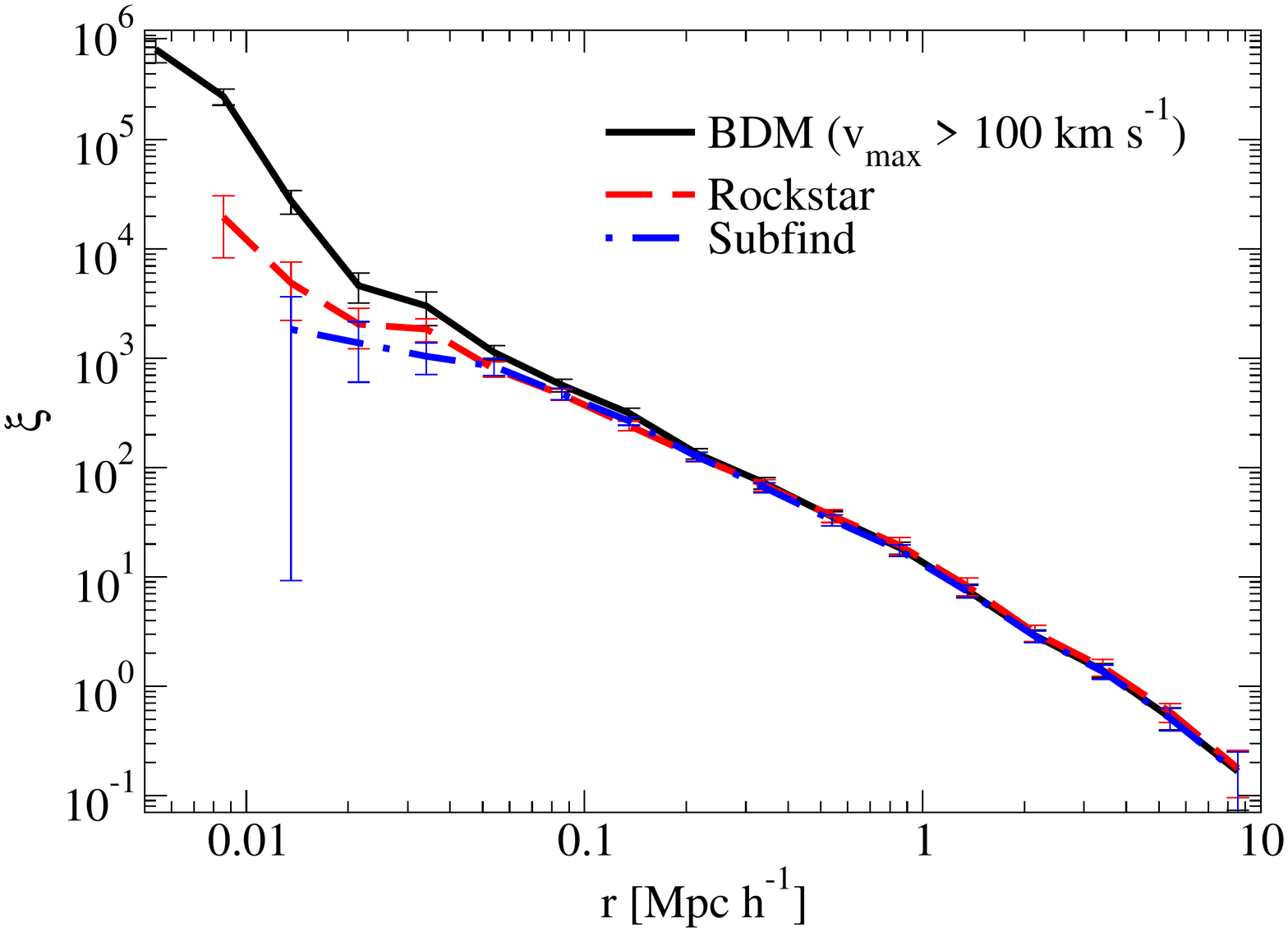}
  \caption{Two-point correlation functions are very similar to previously-published results except relatively close to the centers of halos.  This figure shows correlation functions for the \textsc{rockstar} halo finder and \textsc{bdm}  on the Bolshoi simulation (2048$^3$ particles, 250 Mpc $h^{-1}$).  The \textbf{top} panel shows the correlation function for $v_\mathrm{max} > 300$ km s$^{-1}$ (10,000 halos), the \textbf{middle} panel shows the correlation function for $v_\mathrm{max} > 150$ km $s^{-1}$ (100,000 halos), and the \textbf{bottom} panel shows the correlation function for $v_\mathrm{max} > 100$ km $s^{-1}$.  The bottom panel includes a comparison with Subfind. Subfind halos were only available for 1/125th of Bolshoi (a total of $\sim$ 1000 halos for this $v_\mathrm{max}$ threshold).  For a fair comparison, both \textsc{rockstar} and \textsc{bdm}  results for the bottom panel were computed on the same region of Bolshoi.}
\label{f:cf_comp}
\end{figure}

\begin{figure}
\includegraphics[width=\columnwidth]{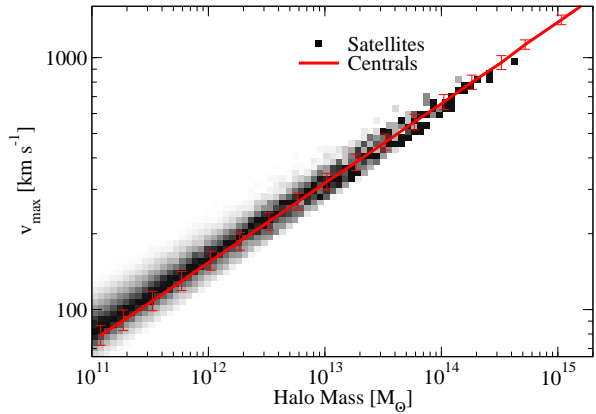}
\caption{Relationship between $M_\mathrm{vir}$ and $\vmax$ for satellite halos (conditional density plot) and centrals (red line; error bars show the 1-$\sigma$ scatter about the mean) at $z=0$ in Bolshoi for the \textsc{rockstar} halo finder.  While satellites have a very similar $M_\mathrm{vir}$-$\vmax$ relation to centrals, the relation for satellites has slightly more scatter and is slightly biased towards lower masses at fixed $\vmax$, a consequence of dynamical stripping.}
\label{f:sat_vmax}
\end{figure}

\begin{figure}
\hspace{2.5ex}\plotgrace{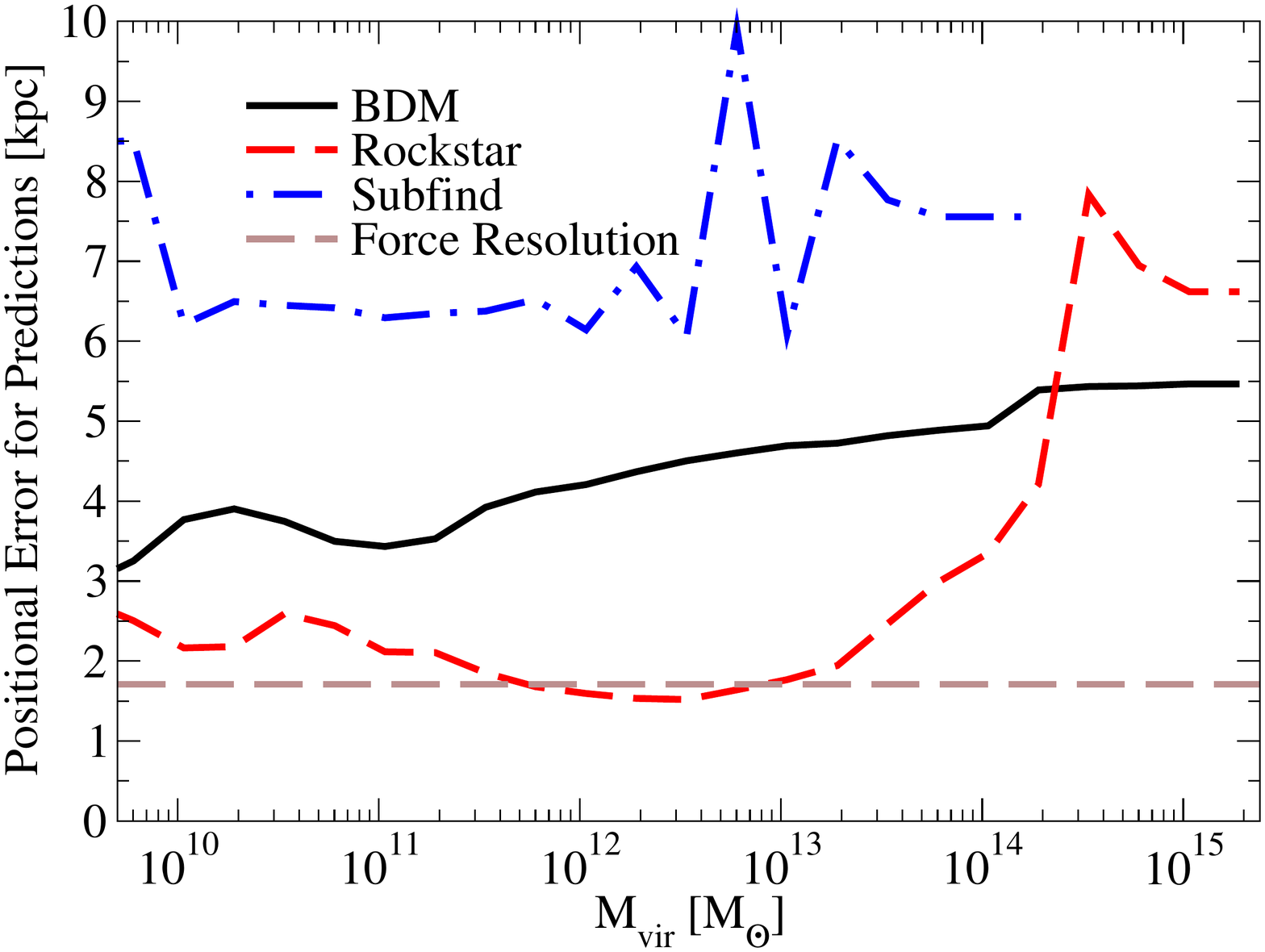}\\[1ex]
\plotgrace{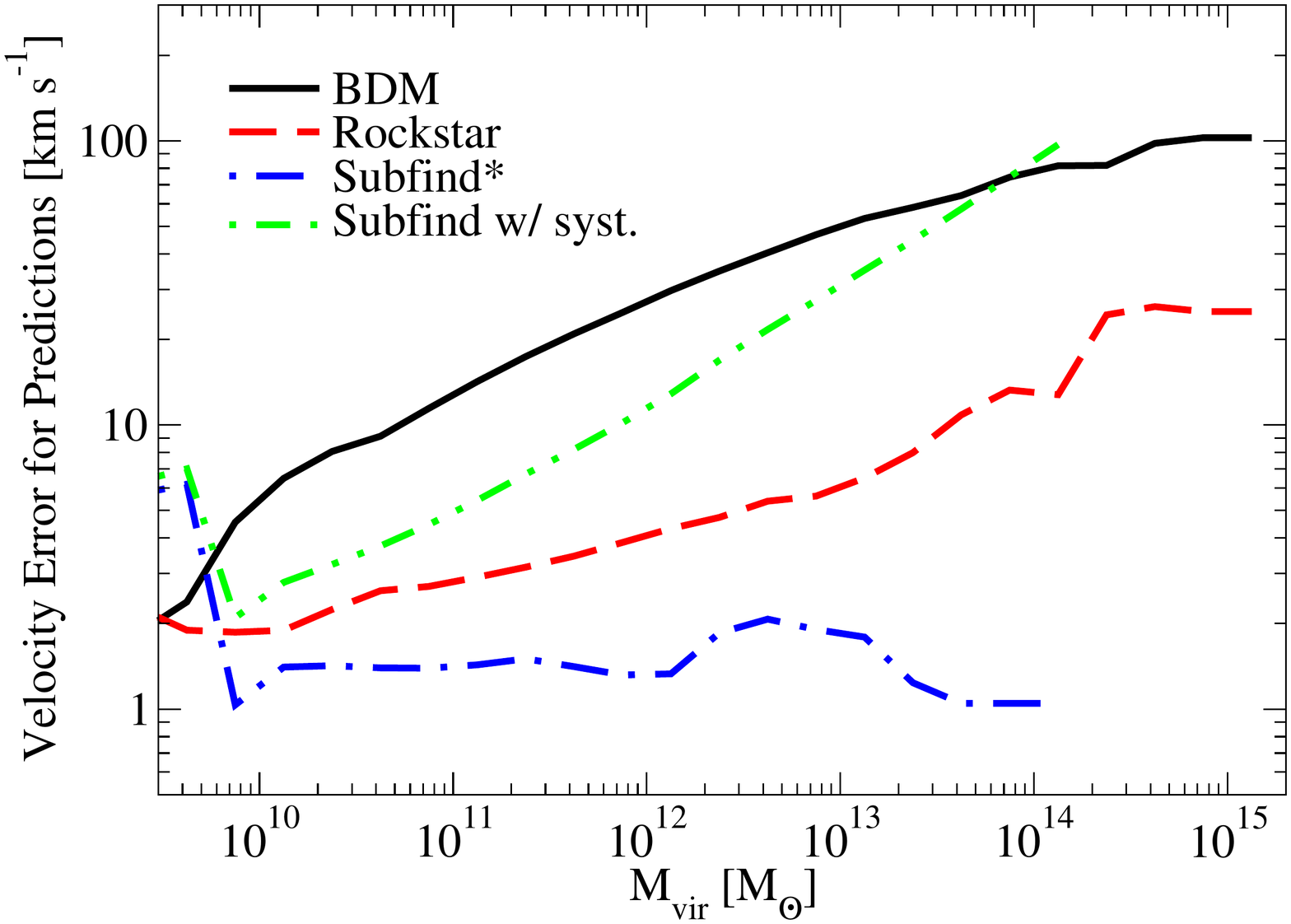}
\caption{\textsc{rockstar} shows superior recovery of halo positions and velocities for cosmological halos as compared to both the \textsc{bdm}  and Subfind halo finders.  This figure shows a comparison of the joint consistency of halo position and velocity for two timesteps of the Bolshoi simulation at $z=0$ separated by 40 Myr, including both distinct halos and subhalos.  By comparing the evolution of halo positions and velocities across timesteps to the values predicted by the laws of inertia and gravity, one may obtain an estimate of the reliability of halo property recovery (see text, also \citealt{BehrooziTree}).  The Y-axis shows the difference between the predicted and measured position (\textbf{top}) and velocity (\textbf{bottom}).   \textsc{rockstar} offers excellent performance in locating halo position centers, in all cases very close to the force resolution of the simulation; \textsc{bdm}  performs almost as well, but Subfind performs somewhat worse on account of using a different position estimator.  \textsc{rockstar} offers significantly better performance than \textsc{bdm}  for estimating halo velocities.  Subfind's halo velocities appear more consistent than \textsc{rockstar}, but because they represent the bulk velocity of the halo as opposed to the core velocity (which would correspond to measurable galaxy properties), they suffer from large systematic errors which make them significantly less accurate than \textsc{rockstar} (see \S \ref{s:cores}).  As such, the direct Subfind results are marked with an \textbf{asterisk} to signify caution in their interpretation.  For reference, the systematic errors calculated in \S \ref{s:cores} are included for Subfind in the green dash-double-dotted line (``Subfind w/ syst.'').  Subfind halos were only available for a small region of Bolshoi (1/125th of the total volume), and so error measurements do not extend to halo masses above $10^{14}\Msun$.}
\label{f:dyn_comp}
\end{figure}

\begin{figure*}
\begin{center}
\vspace{-6ex}
\begin{tabular}{@{\hspace{-6ex}}c@{\hspace{-3ex}}c@{\hspace{-3ex}}c}
\plotsmallgrace{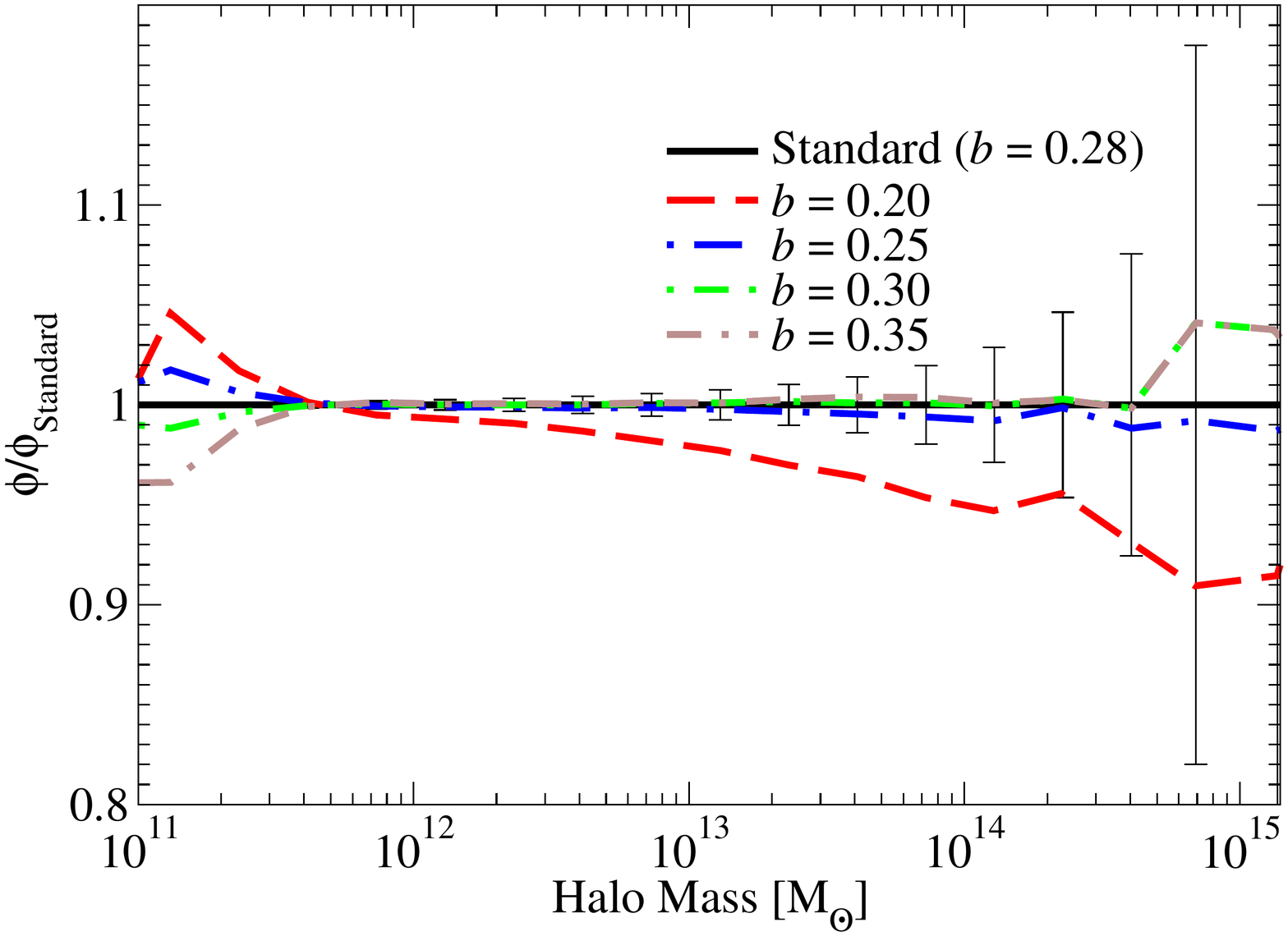} & \plotsmallgrace{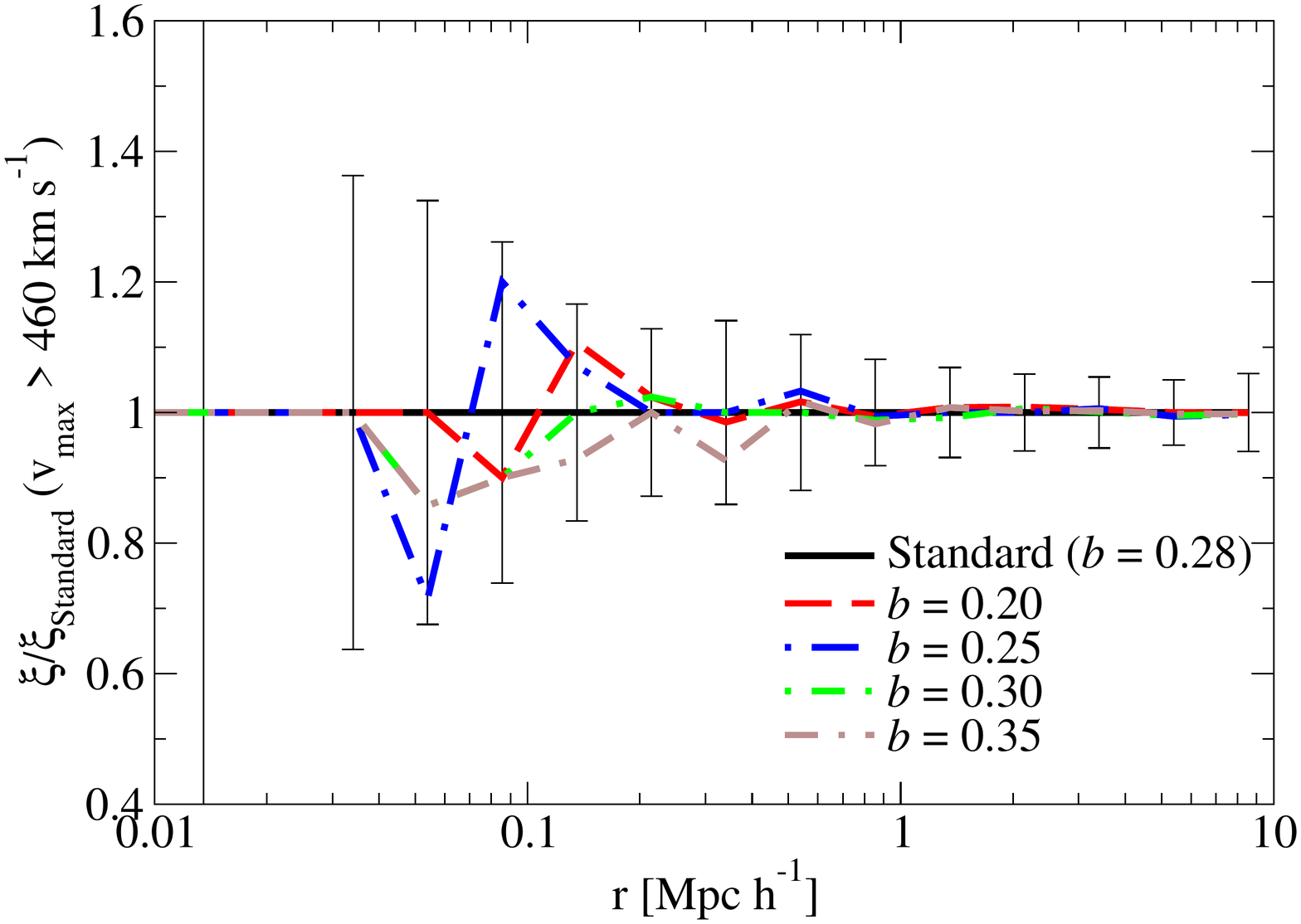} &
\plotsmallgrace{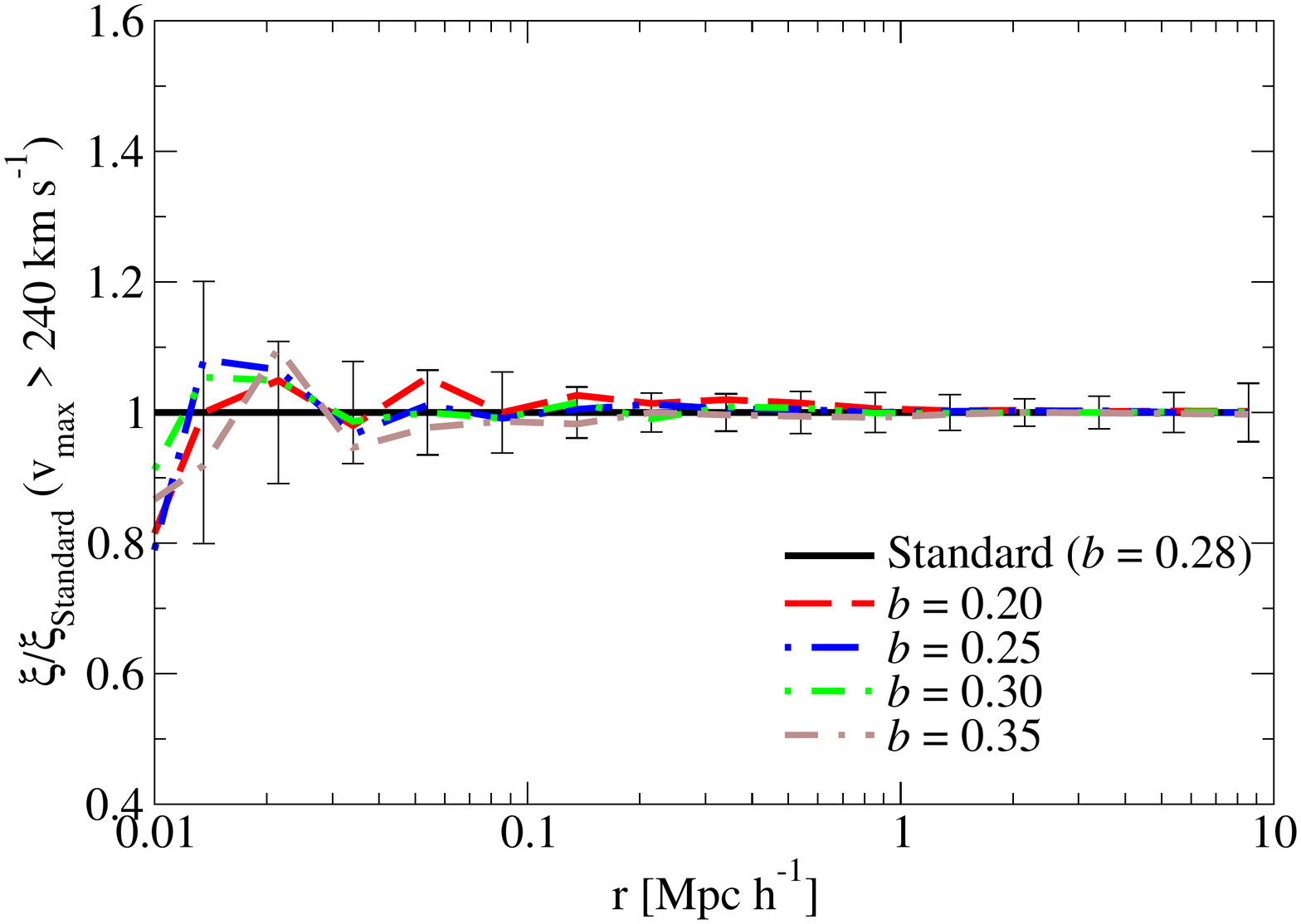} \\[-5ex]
\plotsmallgrace{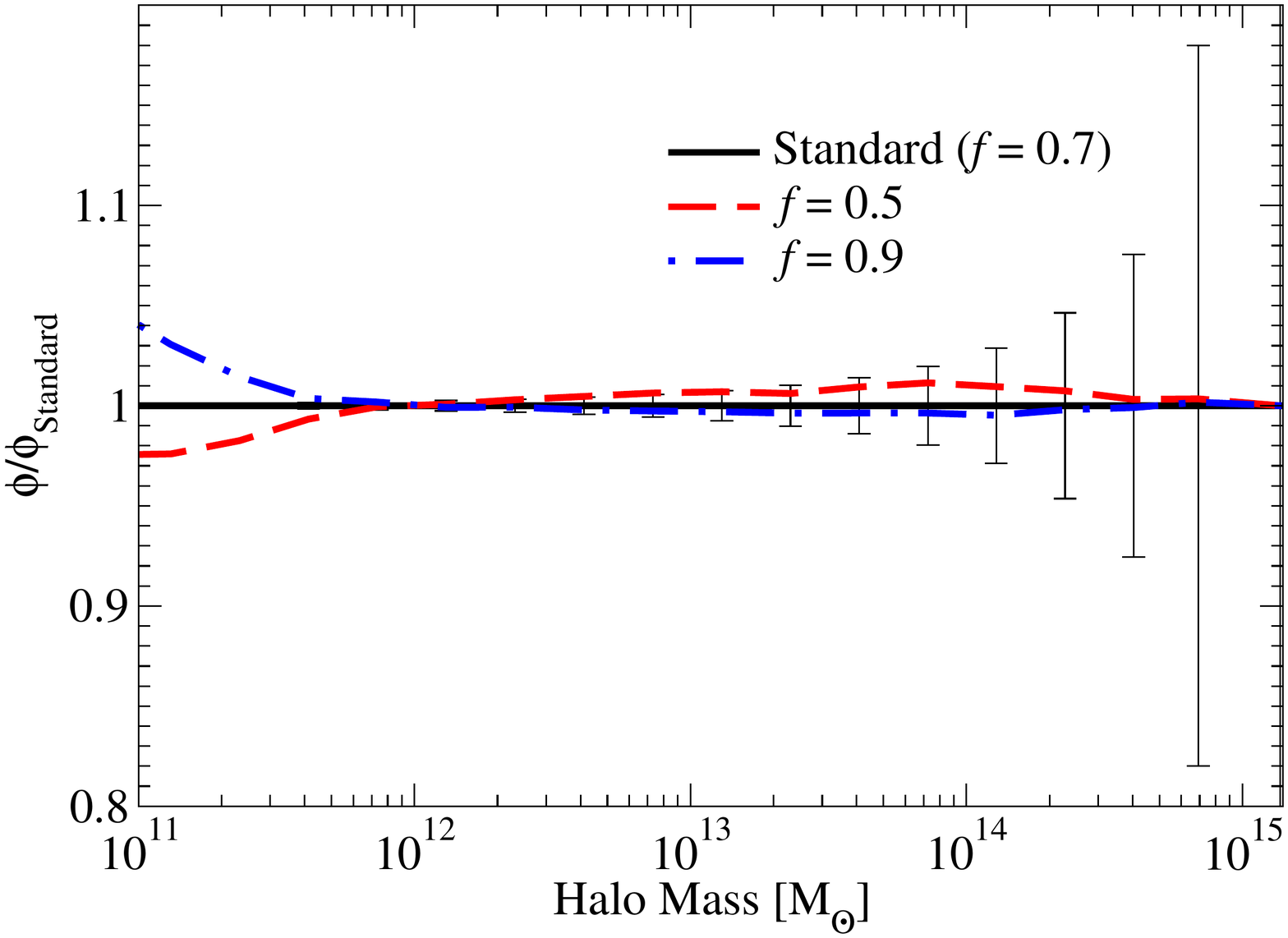} &
\plotsmallgrace{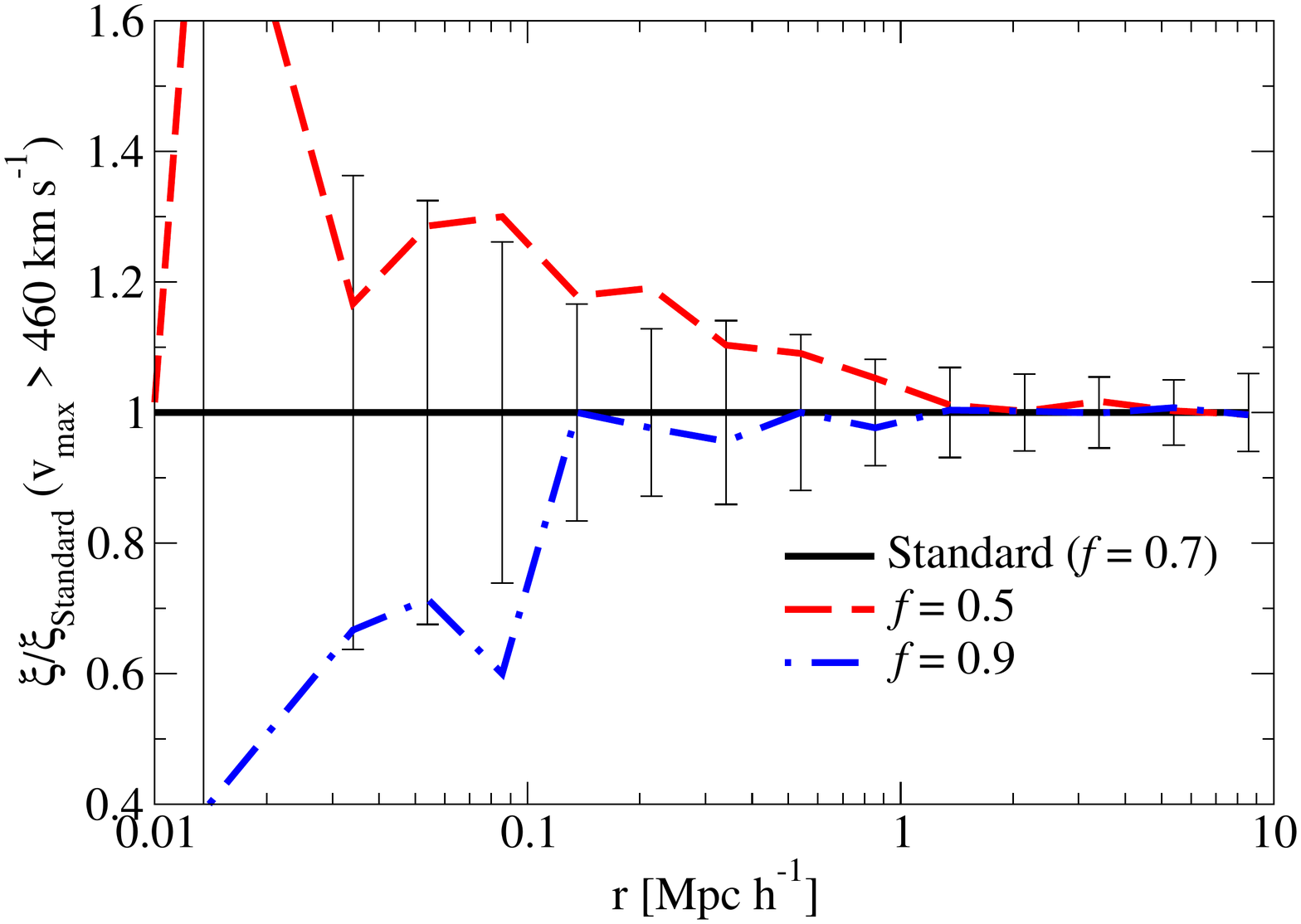} &
\plotsmallgrace{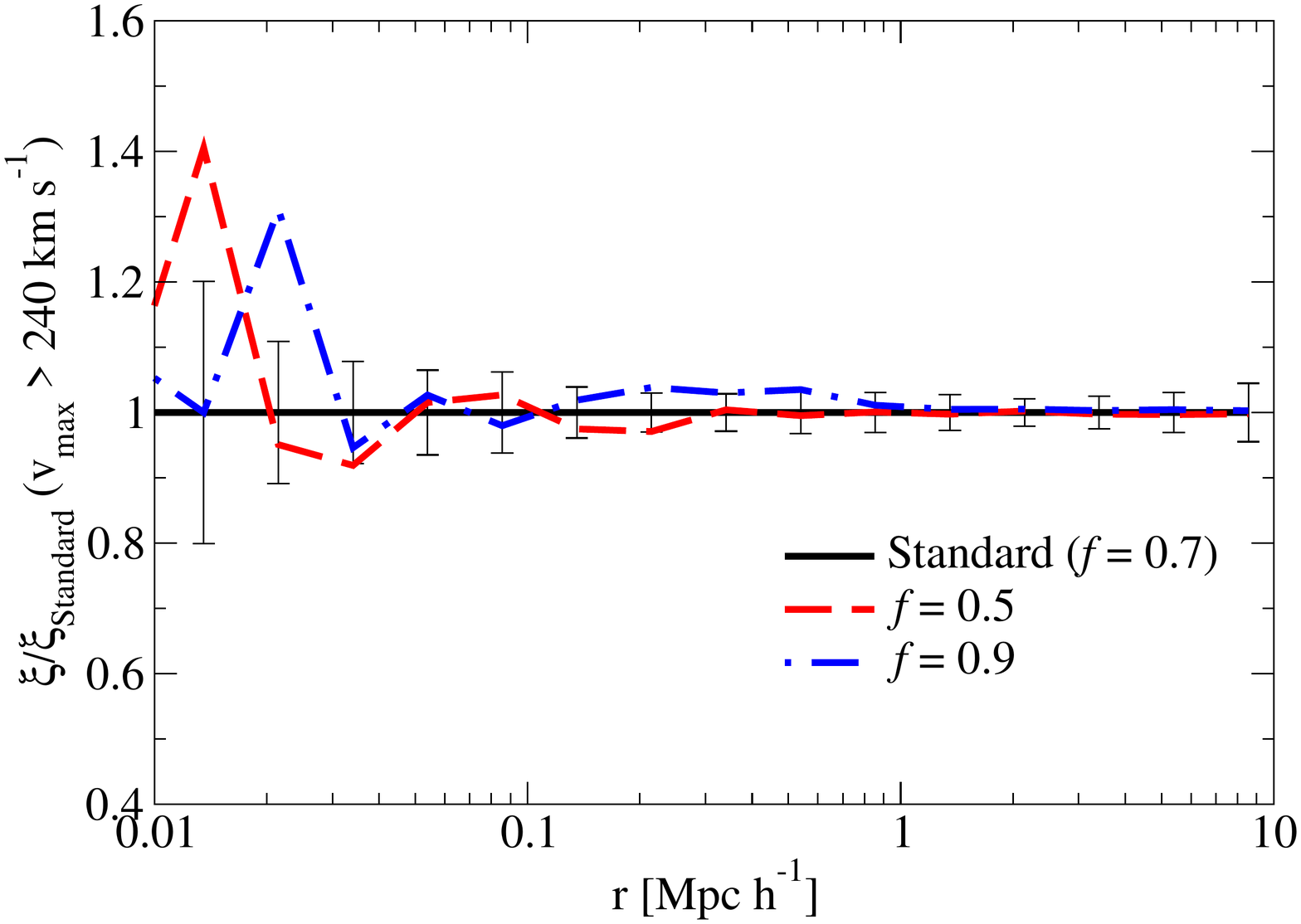}\\[-5ex]
\plotsmallgrace{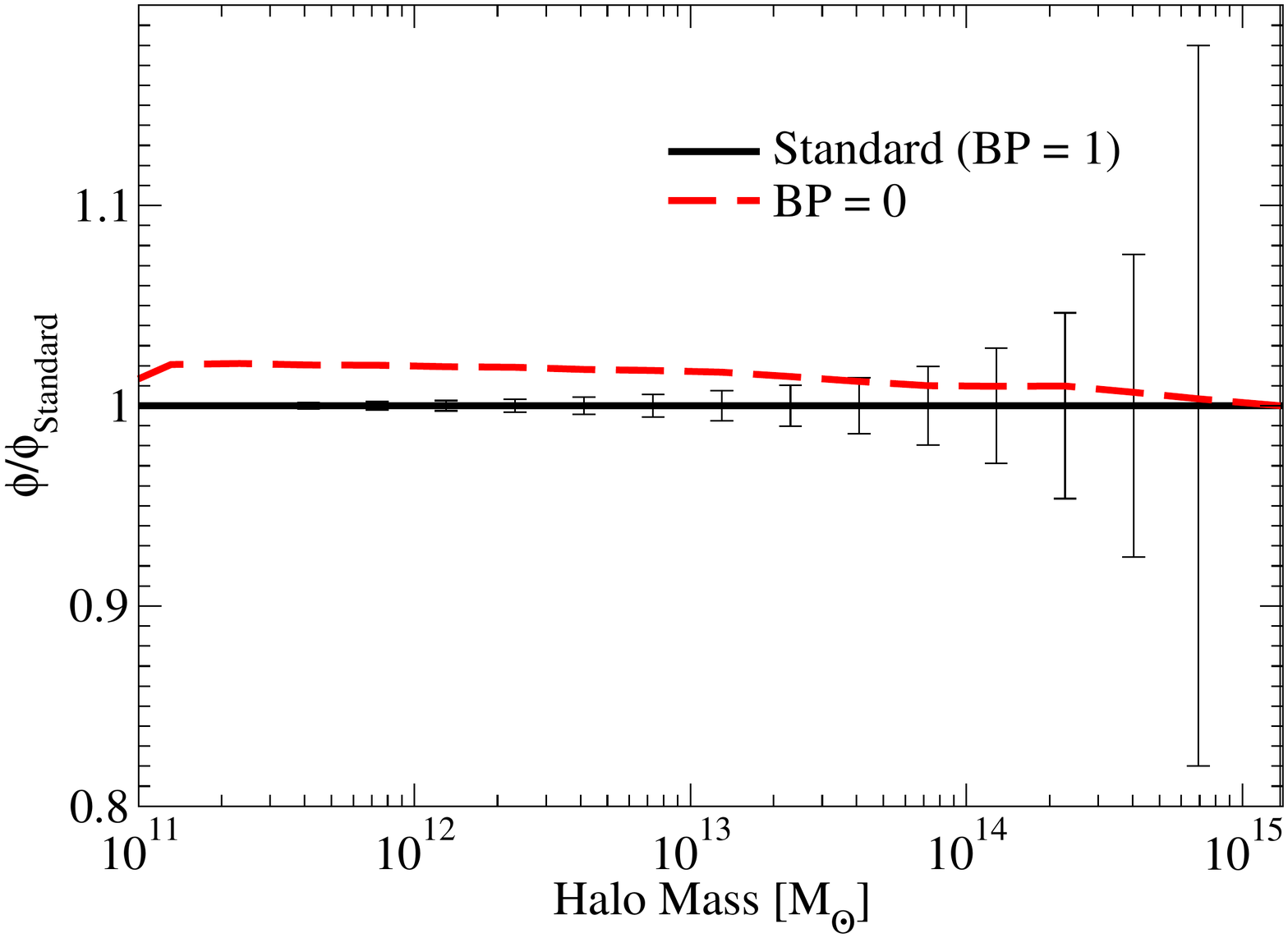} &
\plotsmallgrace{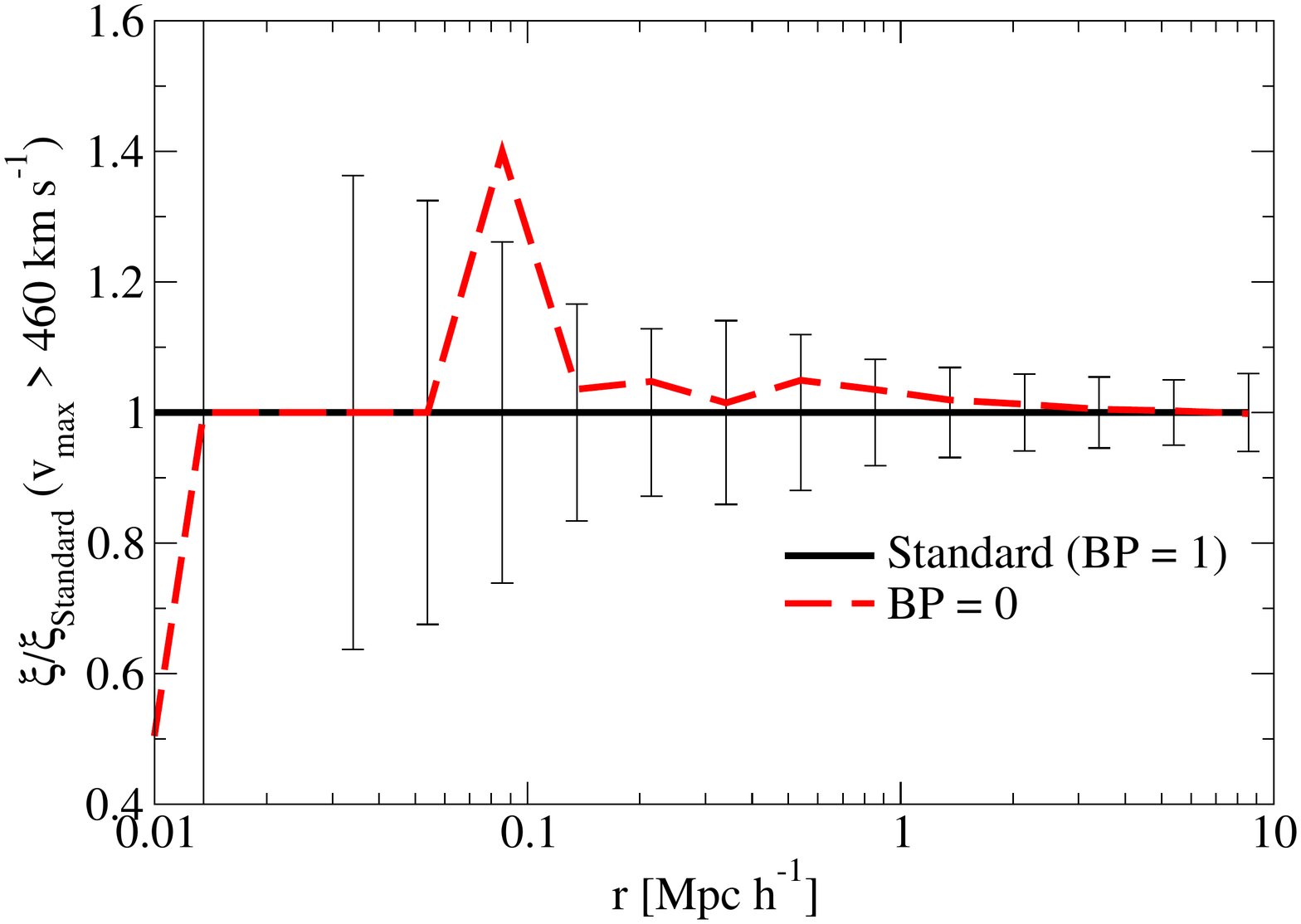} &
\plotsmallgrace{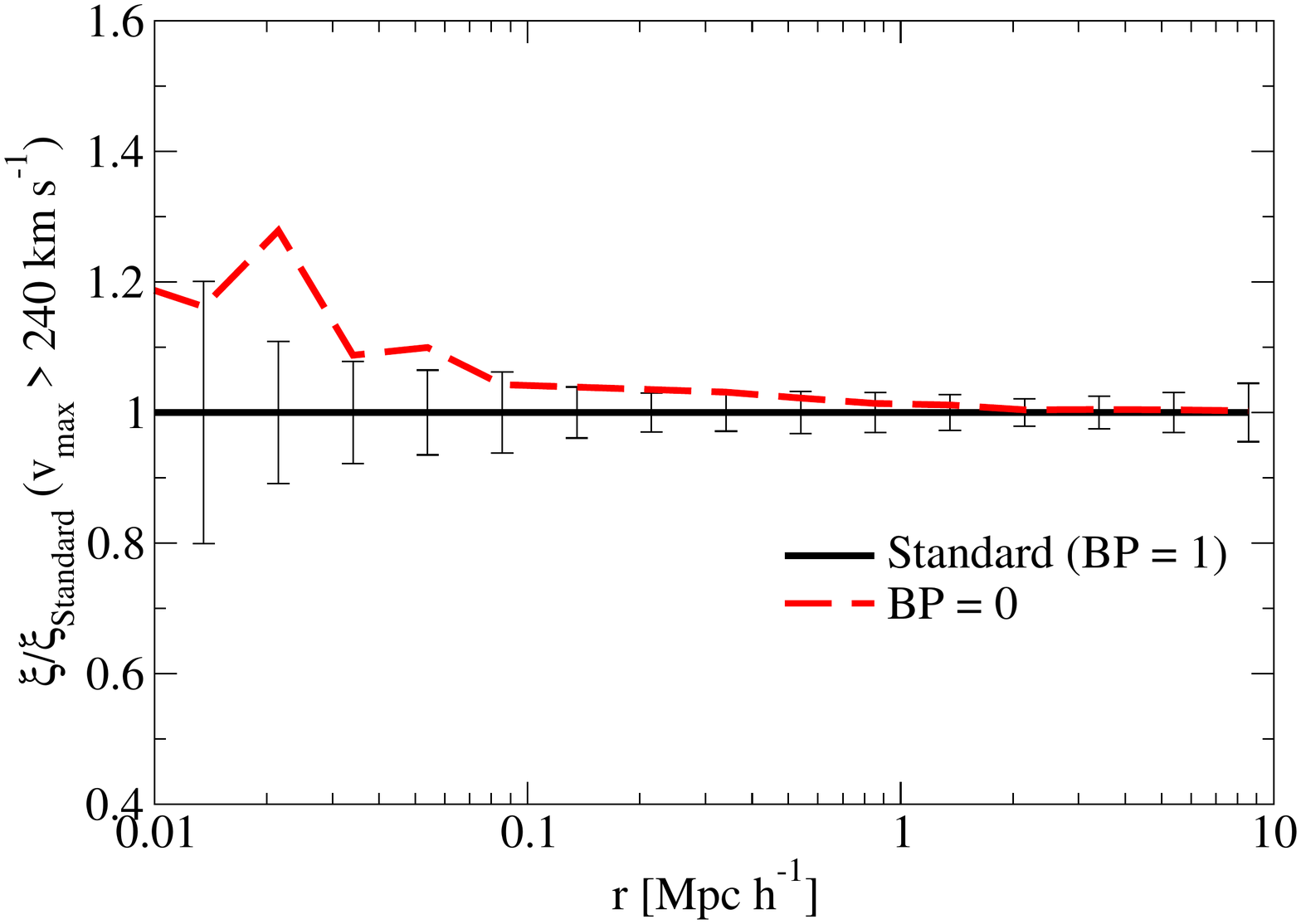}\\[-5ex]
\plotsmallgrace{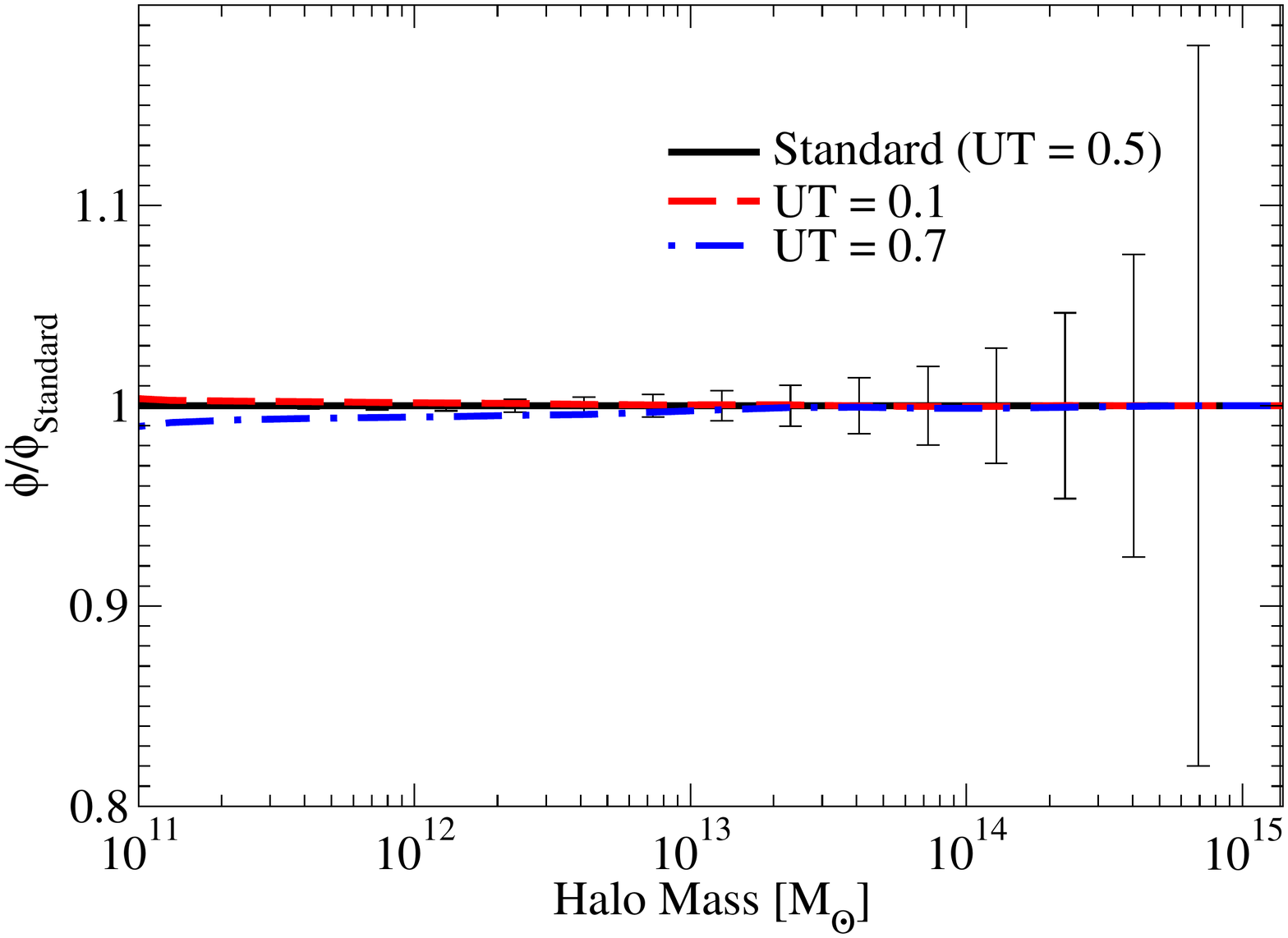} &
\plotsmallgrace{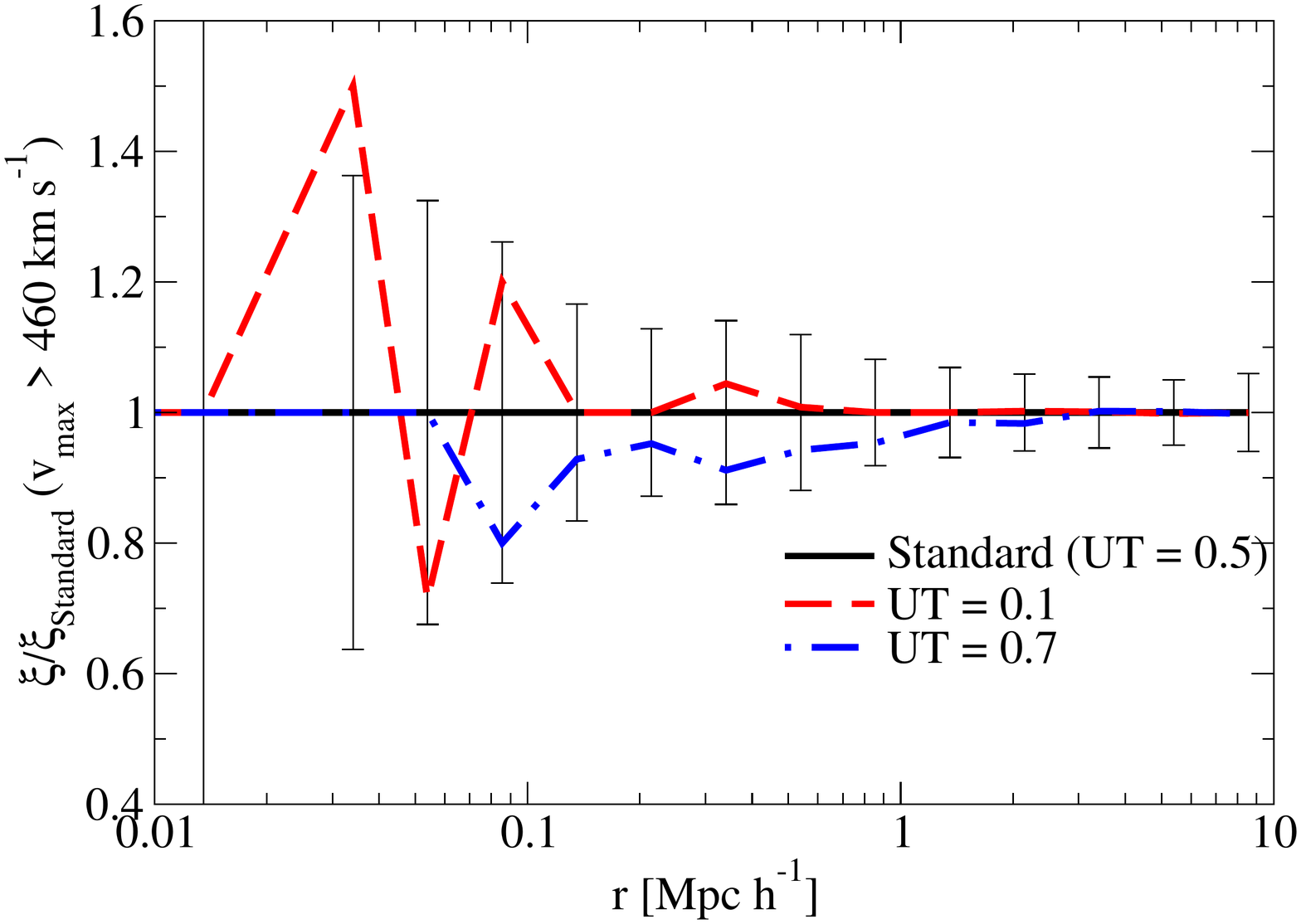} &
\plotsmallgrace{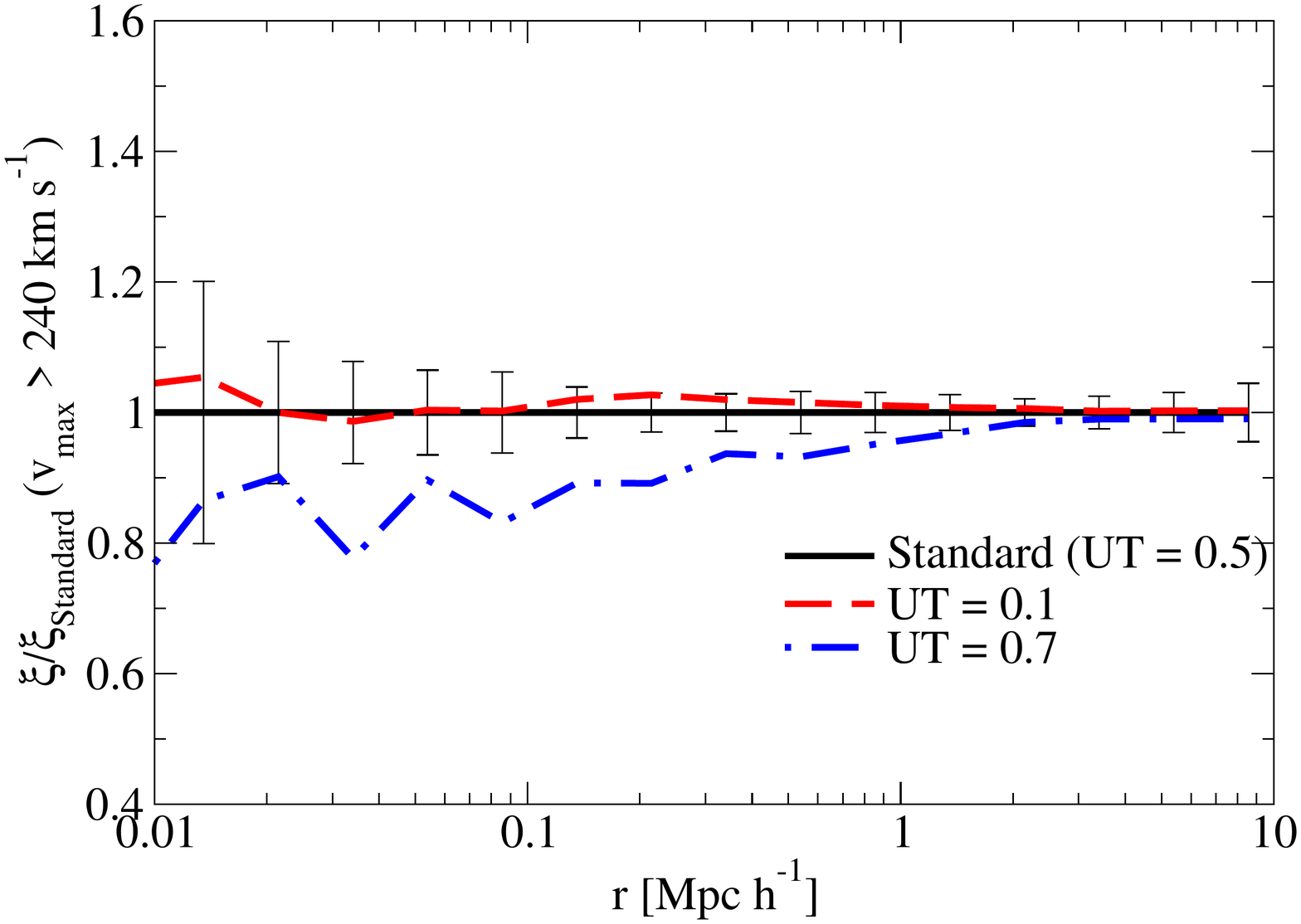}\\[-5ex]
\plotsmallgrace{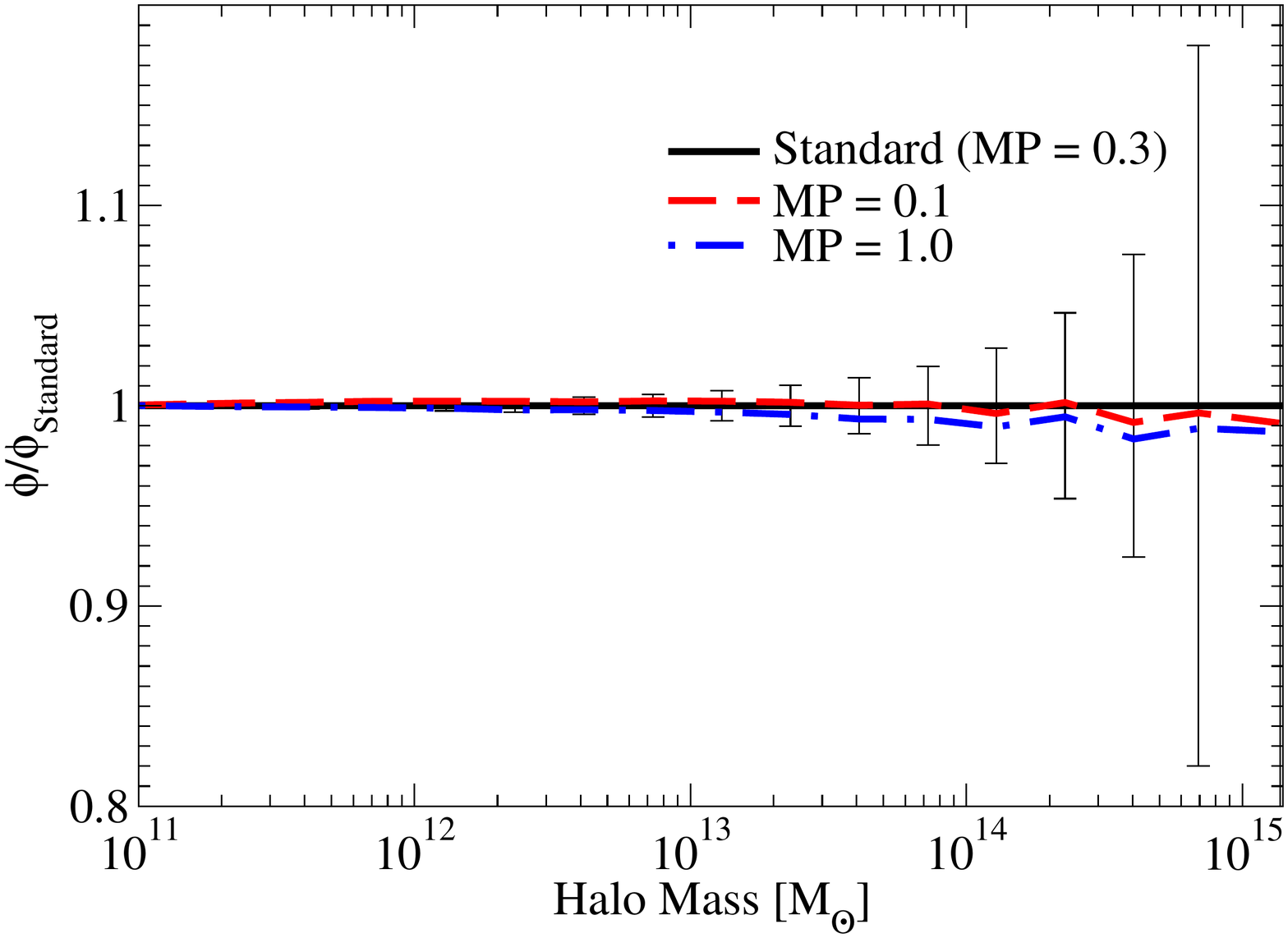} & \plotsmallgrace{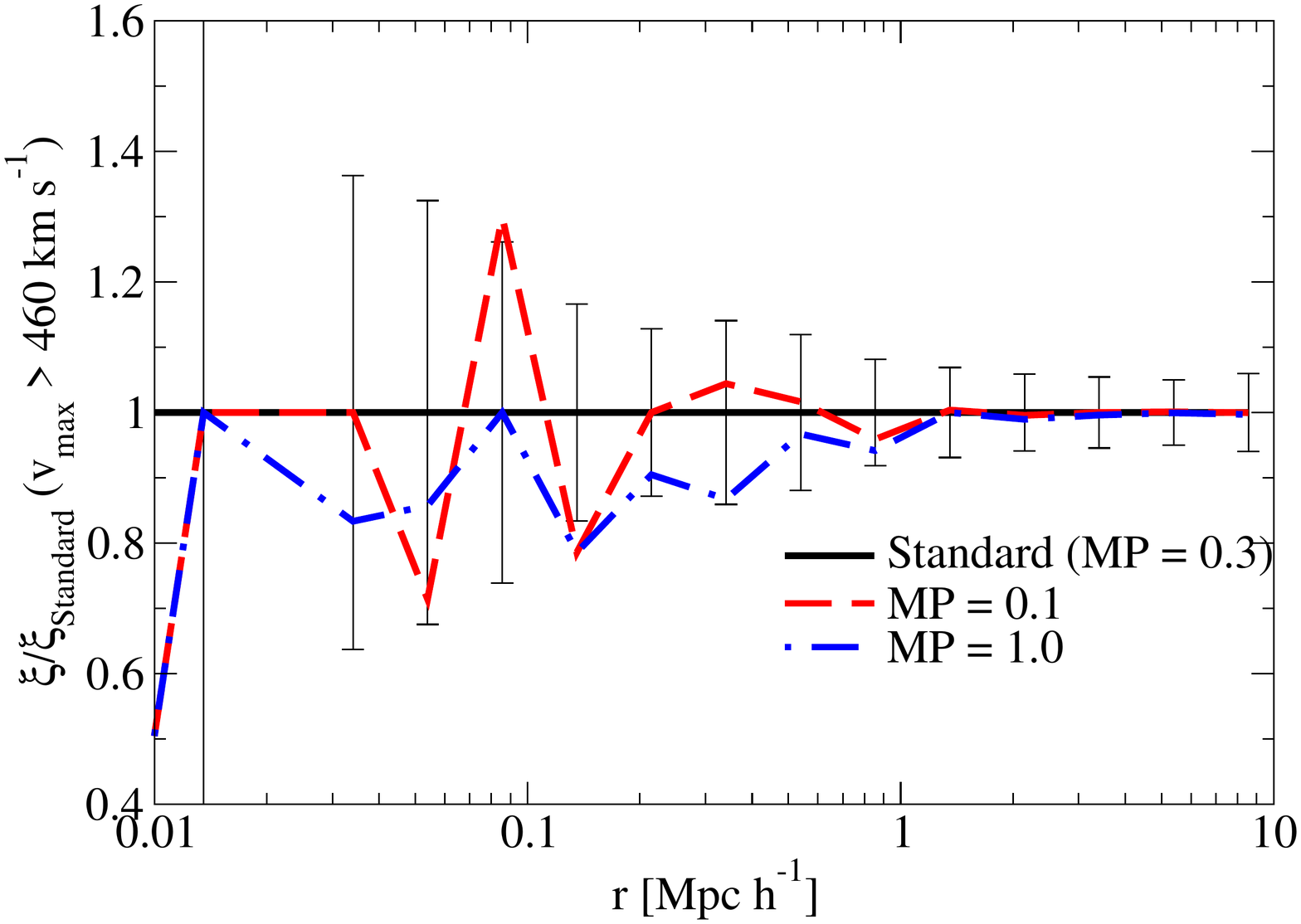} &\plotsmallgrace{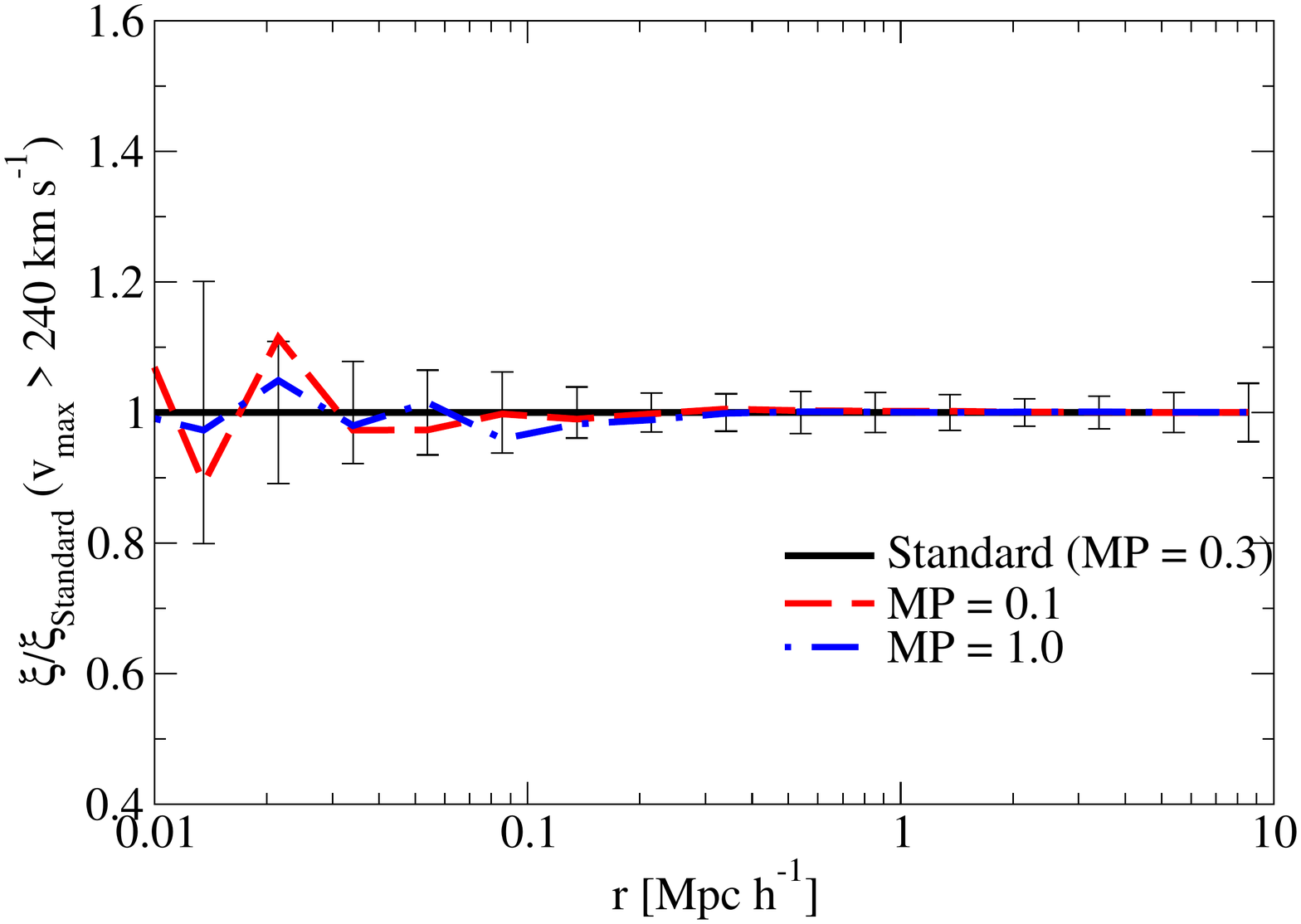}\\[-0.3ex]
\end{tabular}
\end{center}
\vspace{-5ex}
\caption{Dependence of the mass function and correlation functions ($v_{max} > 460$ km s$^{-1}$ and $v_{max} > 240$ km s$^{-1}$) for halos in Consuelo (1400$^3$ particles, 420 Mpc $h^{-1}$) on the parameter choices in the \textsc{rockstar} halo finder (Table \ref{t:params}).  Mass function error bars are Poisson, and a $10^{11} \Msun$ halo has approximately 30 particles; correlation function error bars are jackknife.  The \textbf{top row} shows how the mass function recovery is affected by changing the linking length used to preselect bound groups; clearly, $b=0.2$ is too small, but the default value $b=0.28$ gives results indistinguishable from larger values.  The \textbf{second row} shows how the mass function is affected when changing the fraction of particles retained in each successive 6D refinement level.  The default fraction (0.7) represents the optimal balance between recovery of small halos while at the same time being resilient to noise at low particle number or high particle density (see text).  The \textbf{third row} shows that roughly 1--2\% of halo particles are unbound; as shown in the correlation function plot on the right-hand side, satellite halos tend to have more unbound particles than centrals.  The \textbf{fourth row} shows that raising the threshold for the required fraction of bound particles mainly affects completeness for close halo pairs; otherwise, changing the threshold has no effect.  The \textbf{fifth row} shows almost negligible effects on the mass function and correlation function from including additional contributions to the gravitational potential; the main effect is better seen in merger trees.}
\label{f:param_comp}
\end{figure*}

\subsection{Mass and Correlation Function Comparisons}

\label{s:mf_comp}

An extensive comparison of the mass function and correlation function between \textsc{rockstar} and other halo finders has already been conducted in \cite{Knebe11}.  Our algorithm has changed somewhat since that paper was published, and the mass and $v_{max}$ function comparison in that paper did not separate out the effects of central halos from satellite halos.  In addition, the correlation function comparison only compared the outskirts of halos beyond $r = 2$ Mpc $h^{-1}$ and not smaller scales dominated by substructure.

In Fig.\ \ref{f:mf_comp}, we present a comparison of the mass function for central halos in the Bolshoi and Las Damas simulations.  The results from \textsc{rockstar} agree with those from \cite{tinker-umf} at all masses for halos with more than 100 particles on the level of a few percent, well within the calibration specification (5\%) except when Poisson errors dominate.  Detailed comparisons of the mass function from all of the Las Damas boxes, including fifty times the volume shown here, will be presented by McBride et al (in preparation).  In the Bolshoi simulation, the mass function for central halos returned by \textsc{rockstar} is virtually identical to \textsc{bdm}  \citep{Bolshoi}, differing by at most 5\% over the entire mass range until Poisson errors dominate; these differences stem mainly from different conventions for unbinding particles, especially in major mergers.

Figs.\ \ref{f:vf_comp} and \ref{f:cf_comp} show comparisons between the $v_\mathrm{max}$ function for all halos and for satellite halos and the correlation function for $v_\mathrm{max}$-selected halos between \textsc{rockstar} and the \textsc{bdm}  algorithm for the Bolshoi simulation \citep{Bolshoi}. The $v_\mathrm{max}$ function for all halos is also mostly identical, differing by only 5\% on average for halos above 100 km $s^{-1}$.  For satellites, the deviations are slightly more significant, especially for very massive halos.  Most notably, as a phase-space halo finder, \textsc{rockstar} is able to track subhalos into the extreme inner reaches of halos.  This enables it to track very massive subhalos even when their cores are very close to their hosts in position space, as long as they are sufficiently separated in velocity space.  This is evident both in the increased number density of subhalos with large $v_\mathrm{max}$ as compared to \textsc{bdm}  (Fig.\ \ref{f:vf_comp}), but also in the increased 1-halo contribution to the correlation function for massive halos (Fig.\ \ref{f:cf_comp}).  We also show comparisons for a small region of Bolshoi (1/125th of the total volume) where Subfind \citep{Springel01} halos were also available in Fig.\ \ref{f:cf_comp}.  For the halos considered in this latter comparison ($v_\mathrm{max}$ > 100 km s$^{-1}$), \textsc{bdm}  may overpredict the number of major mergers within 20 kpc $h^{-1}$, whereas Subfind may underpredict the number of major mergers within 30 kpc $h^{-1}$, given the deviations seen in the correlation function as compared to larger scales.

Fig.\ \ref{f:sat_vmax} shows the relationship between $M_\mathrm{vir}$ and $\vmax$ for both satellite halos and centrals using the \textsc{rockstar} halo finder on Bolshoi; as may be expected, satellites have a very similar relation as compared to centrals.  Due to dynamical stripping, however, the satellite relation has more scatter and is biased towards lower halo masses at fixed $\vmax$.

\subsection{Dynamical Tests}

\label{s:dyn_perf}

In \cite{Knebe11}, the authors performed a series of tests on mock halos in order to assess the accuracy of halo property recovery.  While \textsc{rockstar} performed extremely well in these tests, they nonetheless are not representative of the halos which one would expect to find in a real simulation (the tests considered only spherical, NFW/Plummer halos with very little substructure).  This is understandable---in a real simulation, there is no \textit{a priori} ``correct'' answer for the recovery of halo properties.  Nonetheless, it is still possible to assess the \textit{precision} of halo property recovery in a simulation by comparing halo properties between timesteps.

To this end, we have used code from \cite{BehrooziTree} to check the halo property consistency for two position-space halo finders (\textsc{bdm}  and Subfind: \citealt{Bolshoi,Springel01}) and \textsc{rockstar} on the Bolshoi simulation.  This code simulates the gravitational evolution of halos purely based on their recovered properties (mass, scale radius, position, and velocity).   Given the positions and velocities at one timestep, one can thus predict their positions and velocities at the next timestep with high accuracy; the difference between the predicted and actual positions and velocities is one measure of the uncertainty in halo property recovery for the halo finder.

Fig.\ \ref{f:dyn_comp} shows the results of this analysis.  Both \textsc{bdm}  and \textsc{rockstar} recover halo positions very precisely across timesteps; although \textsc{rockstar} recovers positions better by a factor of $\sim$ 2 for lower halo masses, both perform close to the force (softening) resolution of the simulation.  Subfind, where halo positions are averaged over all particles, performs poorly in comparison.  \textsc{rockstar} recovers halo velocities much better (2-3 times as precisely) as compared to \textsc{bdm} , due largely to the fact that \textsc{bdm}  uses only the innermost 100 particles to compute halo velocities \citep{Bolshoi}.  By comparison, Subfind appears to do extremely well; however, because it averages particle velocities over the entire halo instead of over the central region, it suffers from substantial systematic errors (as discussed in \S \ref{s:cores}) which make its actual performance in recovering estimates of galaxy velocities always worse than \textsc{rockstar}.

\subsection{Evaluation of Default Parameters}

\begin{table}
\caption{Summary of algorithm parameters}
\label{t:params}
\begin{center}
\begin{tabular}{rcp{130pt}l}
\hline\\[-1.7ex]
\hline
Variable & Default & Description & Section\\
\hline
$b$ & 0.28 & Friends-of-friends linking length. & \S \ref{s:fof}\\
$f$ & 0.7 & Fraction of particles in each hierarchical refinement level. & \S \ref{s:subfof}\\[3ex]
$\Delta(z)$ & virial & Spherical overdensity definition & \S \ref{s:halos} \\
 BP & 1 & (Bound Properties) Whether halo properties are calculated only using bound particles. & \S \ref{s:unbound}\\[5.5ex]
UT & 0.5 & (Unbound Threshold) The minimum ratio of bound to total mass required for halos. & \S \ref{s:unbound}\\[3ex]
MP & 0.3 & (Major-merger Potential) Mass ratio for mergers at which particles are evaluated for boundedness relative to the merging system, as opposed to individual halos/subhalos. & \S \ref{s:unbound}\\ 
\hline
\end{tabular}
\end{center}
\end{table}

\label{s:def_param}

In Fig.\ \ref{f:param_comp}, we show the residual mass and correlation functions for changes in the default parameter settings for \textsc{rockstar}.  All mass and correlation functions were calculated at $z=0$ for the Consuelo simulation.  A summary of the main tunable parameters is shown in Table \ref{t:params}.

In the top row of figures, it is evident that choosing a base 3D linking length of $b=0.2$ does not capture particles all the way out to the virial radius for massive halos.  Larger values of $b$ (0.25-0.35) result in very similar mass functions; the standard value of $b=0.28$ thus allows some degree of safety even for specific resimulations with unusual halo shapes.  No effect on $v_{max}$ results from choosing larger linking lengths, and thus the correlation function is unaffected.  For smaller values of $b$, FOF groups become more fragmented; at low particle numbers, this can result in a single halo at high $b$ being detected as multiple halos for a lower $b$.

In the second row, the effects of varying the refinement threshold for the 6D FOF hierarchy are shown.  The default behavior is to retain 70\% of particles from the next higher refinement level.  If one retains more particles, then one is more likely to find low-significance objects---however, one is also more likely to find coincidental particle groups which do not correspond to actual halos.  If one had a 90\% particle retainment threshold, then approximately 4\% of 30-particle halos ($10^{11} \Msun$ for Consuelo) would be false positives; this is exactly the residual difference between the mass functions for a 90\% threshold vs.\ the standard threshold in Fig.\ \ref{f:param_comp}.  However, a 50\% threshold is somewhat too low, as several percent fewer halos are detected as compared with the standard threshold.  For halos with masses above 100 particles, where percent-level comparisons are more trustworthy \citep{tinker-umf}, the mass functions are almost indistinguishable.  Some effect is seen in the autocorrelation function for $v_\mathrm{max} > 240$ km s$^{-1}$ halos within 40 kpc.  However, \S \ref{s:substructure} suggests that \textsc{rockstar} with default parameters is already substantially complete for major mergers down to 10 kpc; as such, the additional halos found with $f=0.9$ as opposed to the default $f=0.7$ may represent false detections.

In the third row, the effects of calculating halo properties using all particles (as opposed to the default behavior of only using bound particles) are shown.  When using all particles, halo masses increase by 1-2\% on average, implying that 98\% of the initially-assigned particles were bound.  Including unbound particles affects satellite halos more than centrals, as evidenced by an increase in the correlation function in the regime where the 1-halo term is dominant.

In the fourth row, the effects of changing the boundedness threshold requirement are shown.  Lowering the threshold substantially (to UT=0.1) does not result in any more halos being found; however, raising the threshold to UT=0.7 results in fewer satellite halos being found and a slight decrement in the mass function.

In the fifth row, the effects of changing the major-merger potential threshold are shown; this has extremely little (<1\%) effect on the mass function, and only a minor effect on the correlation function due to somewhat higher satellite masses for lower values of the threshold.

\begin{figure}
\includegraphics[width=\columnwidth]{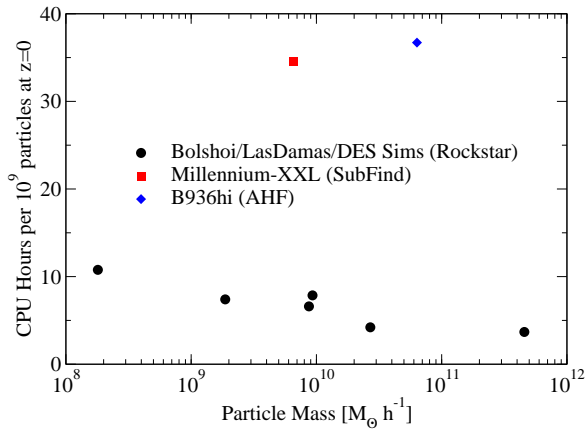}
\caption{\textsc{rockstar} demonstrates extremely efficient use of CPU time as compared to other halo finders with published results.  Figure shows how CPU time requirements per billion particles analyzed at $z=0$ depend on simulation particle mass for a number of simulations (Bolshoi and LasDamas simulations, as well as a lightcone for the DES Blind Cosmology Challenge).  \textsc{rockstar} tests were timed on four-year-old 2.3 GHz AMD Opteron 2376 cores.  Comparisons are included to two other halo finders \citep{Knollmann09,Springel10} with published performance data.  Note that these were not run on the same system; however, they most likely represent a lower bound to the run times if the calculations were repeated on our systems.  For \textsc{rockstar}, typical runtime depends on scale factor approximately as $T \propto a^{0.84}$; at earlier times, fewer particles are clustered and analysis is more efficient.}
\label{f:performance}
\end{figure}

\subsection{Performance}

\label{s:performance}

Figure \ref{f:performance} shows the excellent performance scaling of \textsc{rockstar}.  The costliest part of our algorithm is the process of calculating the hierarchical refinement levels, which takes more time for halos with more substructure.  For that reason, structure finding takes more time both for higher resolution simulations and for late redshifts (the runtime per snapshot scales approximately as $T \propto a^{0.84}$).  Nonetheless, the halo finder is still so efficient (5-10 CPU hours per billion particles at $z=0$) that high-throughput access to particle data is important to avoid wasting CPU time.  Indeed, despite its advanced capabilities, it is roughly 4-5 times as efficient as other halo finders that include substructure with published performance data \citep{Knollmann09,Springel10}, as shown in Fig.\ \ref{f:performance}.  Some caution is necessary in comparing these timings directly, as the systems used were not the same.  Nonetheless, the systems used for calculating \textsc{rockstar}'s performance (Sun X2200 systems with 2x2.3 GHz AMD Opteron 2376 processors, 32 GB of memory, and Cisco DDR HCA Infiniband adapters accessing a Lustre Filesystem) are over five years old, negating any advantage from newer technology.   The most modern friends-of-friends halo finders \citep[e.g.,][]{Woodring11} are faster by roughly a factor of two than \textsc{rockstar} on large datasets; however, these cannot identify substructure nor can they produce complete catalogs of halos with SO masses.

\section{Estimating Satellite Halo Completeness Limits}

\label{s:substructure}

For many decades, early computational simulations suffered from the so-called ``overmerging'' problem, where satellite halos were found to disappear almost as soon as they passed within the boundary of a larger cluster (see \citealt{Klypin99} for a discussion).  This situation has improved dramatically on account of increased mass and force resolution in simulations; however, even in modern simulations there is often a need to add ``orphan'' satellite galaxies (galaxies which exist without a corresponding subhalo) in order to match the small-scale clustering seen in observations \citep{MillOrphan}.  In detail, this depends on both simulation resolution and halo finding; for example, using the halo finder presented here, the clustering on scales larger than $\sim$ 100 kpc for galaxies in the Sloan Digital Sky Survey can be reproduced using only resolved halos and subhalos in the Bolshoi simulation \citep{Reddick12}.  With recent observations pushing the previous small-scale limit of clustering measurements from 100 kpc to 10 kpc \citep{Tinker11,Jiang11}, and with increased accuracy needs for modeling galaxy populations in clusters, it is important to understand the limitations of cosmological simulations' abilities to recover substructure at this level.

\begin{figure}
\plotgrace{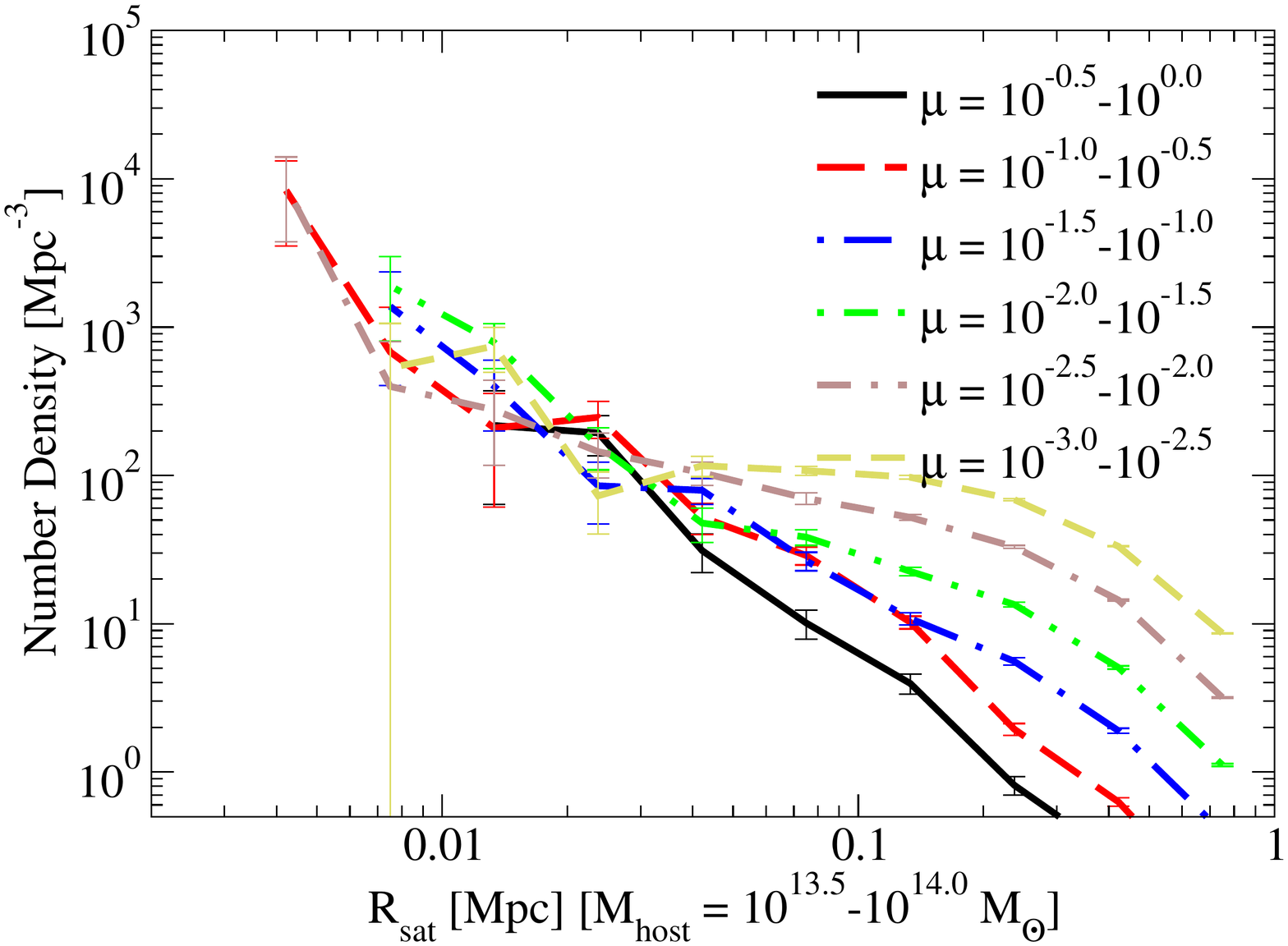}\\
\plotgrace{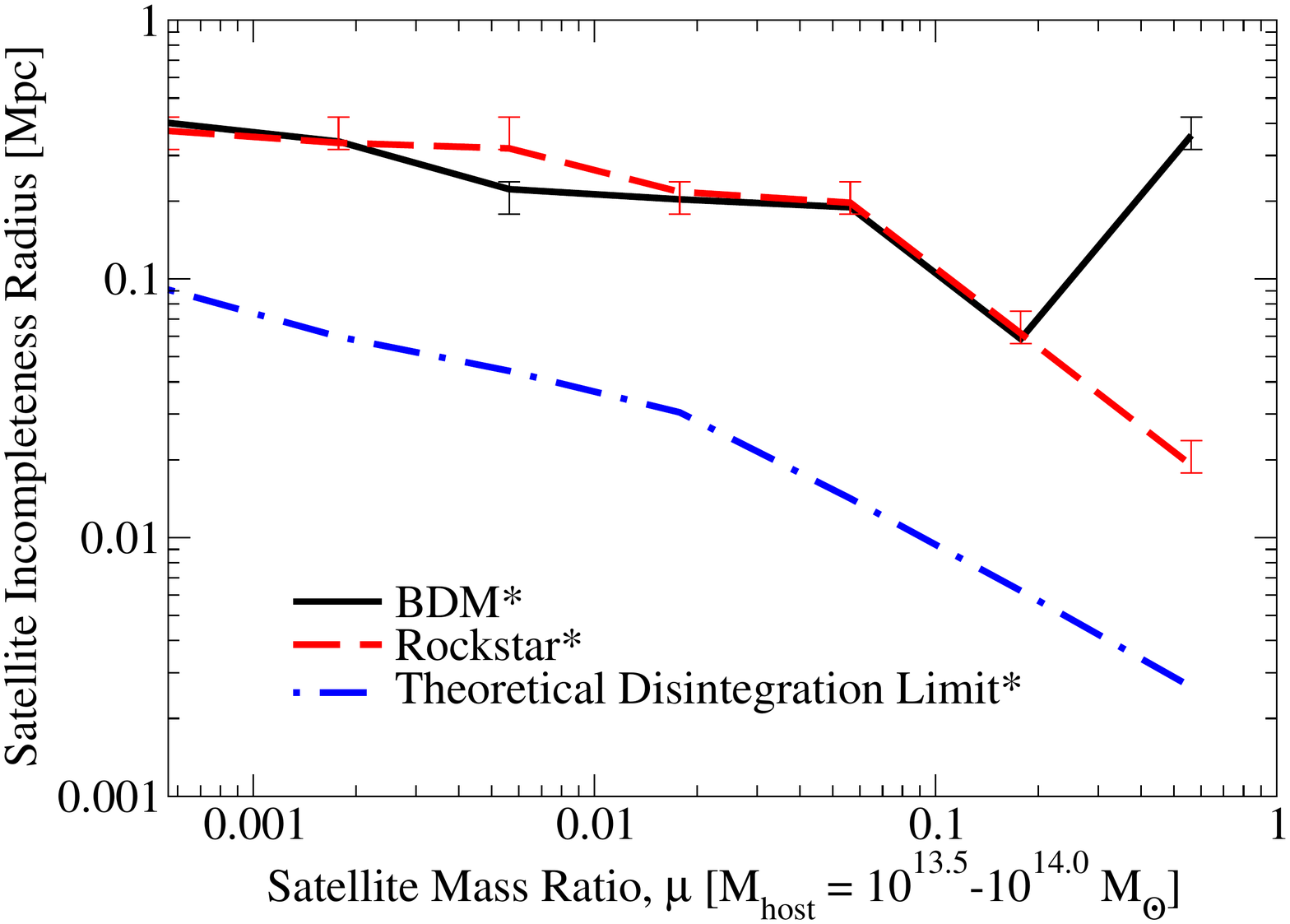}
\caption{\textbf{Top} panel: A plot showing the radial distribution of existing satellite halos at $z=0$ in cluster-scale halos as a function of the satellite mass ratio ($\mu$, the ratio between the peak satellite mass and the host mass) for halos found with \textsc{rockstar} in the Bolshoi simulation. \textbf{Bottom} panel: the radial completeness limit as a function of satellite mass ratio for cluster-scale halos in Bolshoi for both \textsc{rockstar} and the \textsc{bdm}  halo finder, defined as the radius at which the logarithmic slope of the satellite distribution reaches -1.5 (see text).  Error bars include Poisson uncertainties as well as uncertainties in this cutoff slope, which could reasonably be as high as -1.7 \citep{Tinker11}.  The upturn in radial incompleteness for major mergers in \textsc{bdm}  is a conscious choice of \textsc{bdm} 's author.  A theoretical limit is also shown, but the calculation is valid only for the assumption of circular orbits and spherical NFW profiles, and thus carries large systematic errors.  The \textbf{asterisks} thus denote that some caution is necessary in interpreting these results.  (See discussion in \S \ref{s:substructure}).}
\label{f:existing_subs}
\end{figure}

In observations, the distributions of satellite galaxies around such massive hosts is completely self-similar (density proportional to a power law of radius, with an exponent between $-1.7$ and $-1.5$, \citealt{Tinker11}).  However, Figure \ref{f:existing_subs} acutely shows the tension between this observational result and what is found in the Bolshoi simulation for cluster-sized halos ($R_\mathrm{vir} \sim 1$ Mpc); in the Bolshoi simulation, there is a clear radial incompleteness scale in the recovered volume density for merger ratios below 1:30.  

Making a quantitative estimate of the radial incompleteness scale is difficult because the true satellite halo distribution, $\rho(r)$, is unknown.  We can nonetheless connect the luminosity constraints in \cite{Tinker11} to constraints on dark matter halos via the assumption that galaxy luminosity is tightly correlated with halo peak mass (i.e., the largest mass that progenitors of a given halo or subhalo have ever had).  This assumption has been tested in a large number of studies and been found to accurately reproduce both the halo mass to stellar mass relation and the luminosity-dependent clustering of galaxies \citep{BWC12,Reddick12,Behroozi10,moster-09,Guo-09,cw-08,Wang09,conroy:06}.  As concerns halos, this assumption would imply that the radial dependence of satellite halos with a given peak mass should also follow a power law with an exponent between $-1.7$ and $-1.5$.

We thus conservatively define the satellite halo incompleteness scale as the radius at which the logarithmic slope of the satellite halo density $\rho(r)$ becomes shallower than $-1.5$; steeper numbers could overestimate the true logarithmic slope of the satellite halo radial distribution.  However, we include the possibility that the true cutoff slope should be $-1.7$ in our treatment of the errors.  We show the results for the radial completeness scale as a function of satellite mass ratio (satellite halo peak mass compared to host mass) for massive hosts ($10^{13.5}\Msun < M < 10^{14}\Msun$) in Fig.\ \ref{f:existing_subs}.  We also include results for the \textsc{bdm}  halo finder in Fig.\ \ref{f:existing_subs} to show the comparison with non-phase-space halo finding.  The benefit of phase-space finding is striking for satellite halo mass ratios above 1:30, with \textsc{rockstar} able to find major mergers consistently down to a tiny fraction of the virial radius.  However, for satellites with smaller mass ratios, \textsc{rockstar} is unable to perform any better than \textsc{bdm} .  This finding is especially unexpected considering the results in \cite{Knebe11}, wherein \textsc{rockstar} was the only halo finder able to accurately recover all the halo properties of a satellite (mass ratio 1:100) placed directly at the center of a million-particle $10^{14}\Msun$ halo.

In order to explain such a drastic reduction in ability to find satellites, it is instructive to consider the effects of tidal stripping on a satellite in a simulation with limited force and mass resolution.  In particular, due to gravitational softening, the maximum density of a halo cannot continue increasing indefinitely towards its center; instead, it will threshold at some distance $f_{res}$ from the center.  For halos with a low enough mass, there may not be any particles within $f_{res}$ of the center; for these halos, the effective maximum density is degraded even further.  Due to this effect, satellites are vulnerable to tidal disruption much earlier for larger values of $f_{res}$ and for larger particle masses.  In particular, the fluid Roche limit (under the assumption of a significantly more massive host) is given by 
\begin{equation}
\label{e:roche}
d_\mathrm{Roche} \approx 2.4 R_H \left(\frac{\rho_H}{\rho_s}\right)^{\frac{1}{3}}
\end{equation}
where $d_\mathrm{Roche}$ is the radius within which tidal disruption occurs, $R_H$ is the host radius, $\rho_H$ is the enclosed averaged host density, and $\rho_s$ is the average satellite density.  In terms of halos, this means that tidal disruption will occur at a distance $R$ from the host center where the enclosed host halo average density is $2.4^{-3} \sim 7$\% of the satellite halo peak density $\rho_s$.

A naive application of the Roche limit under the assumption of a spherical NFW profile, circular satellite halo orbits, the mean mass-concentration relation from Bolshoi, as well as the mass and force resolutions of Bolshoi yields an estimate of the theoretical disintegration limit shown in Figure \ref{f:existing_subs}.  A number of significant limitations apply to this calculation, because it does not include any mass stripping for satellites after they enter the host, and it does not include any effects from eccentric orbits, where satellites at a given radius would be expected to have passed through closer radii many times.  The latter bias is stronger for low-mass satellites, which experience less dynamical friction and so are expected to orbit many more times before destruction than larger halos; as such, low-mass satellites at a given radius are on average more stripped and have had more exposure to high tidal fields than high-mass satellites at the same radius.  This calculation also does not include numerical simulation artifacts wherein forces on one part of a satellite may be calculated differently from forces on another part (e.g., direct summation versus multipole expansion).  Nonetheless, many of these biases go in the direction of earlier destruction of satellites; as such, the calculation represents a useful lower limit on the radius where satellites may exist in massive clusters.  For a much more detailed assessment of subhalo completeness, we refer the reader to \cite{Wu11}.

This calculation and the results in Fig.\ \ref{f:existing_subs} would suggest that, regardless of the halo finding technique used, simulations of clusters still suffer from a overmerging at the very innermost radii due to limited force and mass resolution.  Depending on the scientific questions being addressed, these incomplete halos may be more or less relevant.  For a high resolution simulation like Bolshoi, the incompleteness does not impact the correlation function on scales larger than $\sim$ 100 kpc, and is present only in the inner regions of massive halos (see e.g. \citealt{Reddick12}).  In general, when determining the desired mass and force resolution to required, the requirements to resolve satellites in the inner regions of massive clusters will be much more stringent.  In these regions, tracking galaxies after their halos are destroyed with ``orphans'' may represent an attractive solution to reduce computational requirements, but the need for these galaxies is somewhat mitigated with increased force resolution and with the effective halo finding in inner regions available with \textsc{rockstar}.

\section{Halo Core Velocity Offsets}

\label{s:core_offsets}

\begin{figure}
\includegraphics[width=\columnwidth]{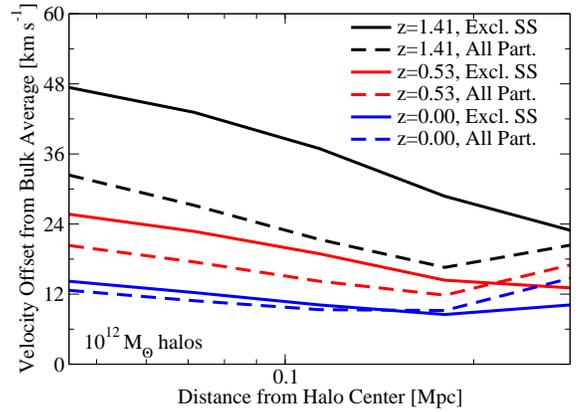}\\
\includegraphics[width=\columnwidth]{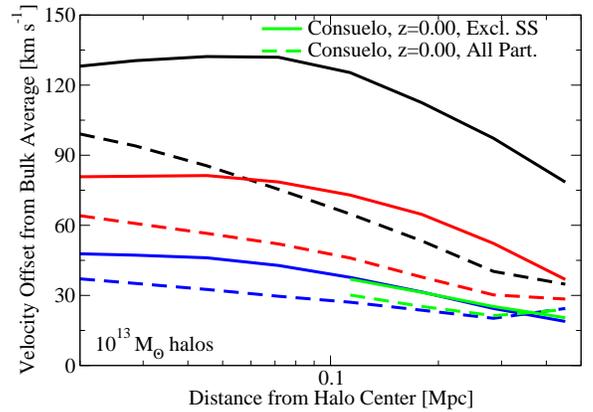}\\
\includegraphics[width=\columnwidth]{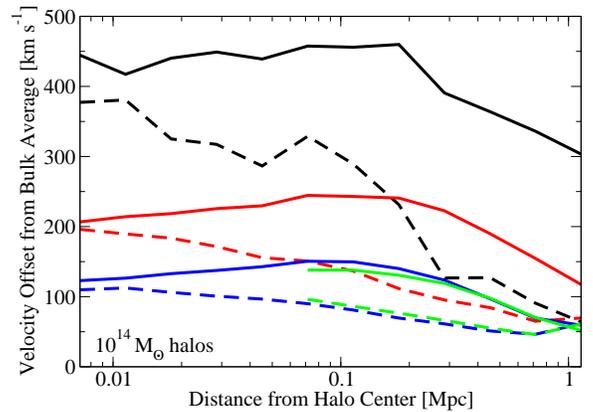}
\caption{
Significant differences are seen between the halo bulk velocity and particle velocities especially within 10\% of the virial radius.  Panels show comparisons between velocities averaged in radial bins (i.e., spherical shells) and the total average bulk velocity of the halo as a function of radius.  Results are plotted in terms of the median magnitude of the velocity difference at each radius for halos in several mass and redshift bins.  Calculations including substructure (``\textbf{All Part.}'') as well as excluding particles in substructure (``\textbf{Excl. SS}'') are shown.  All panels show central halos from the Bolshoi simulation; the middle and lower panels also show central halos from the Consuelo simulation.  In all cases, each radial bin contains at least 400 particles so as to minimize the effects of Poisson noise.  For the lower panel, the numbers of halos used in Bolshoi were 18, 226, and 484 at redshifts $z=1.41$, $0.53$, and $0.00$, respectively.  All other panels (as well as Consuelo for $z=0.00$) averaged results from more than 2000 halos.  Each panel shows halos in a mass bin of 0.48 dex ($1$ to $3$ times $10^{12}\Msun$, $10^{13}\Msun$, and $10^{14}\Msun$, respectively); the respective \textbf{virial radii} are $\sim 0.3$ \hmpc, $\sim 0.6$ \hmpc, and $\sim 1.3$ \hmpc.}
\label{f:core_highz}
\end{figure}

\begin{figure}
\includegraphics[width=\columnwidth]{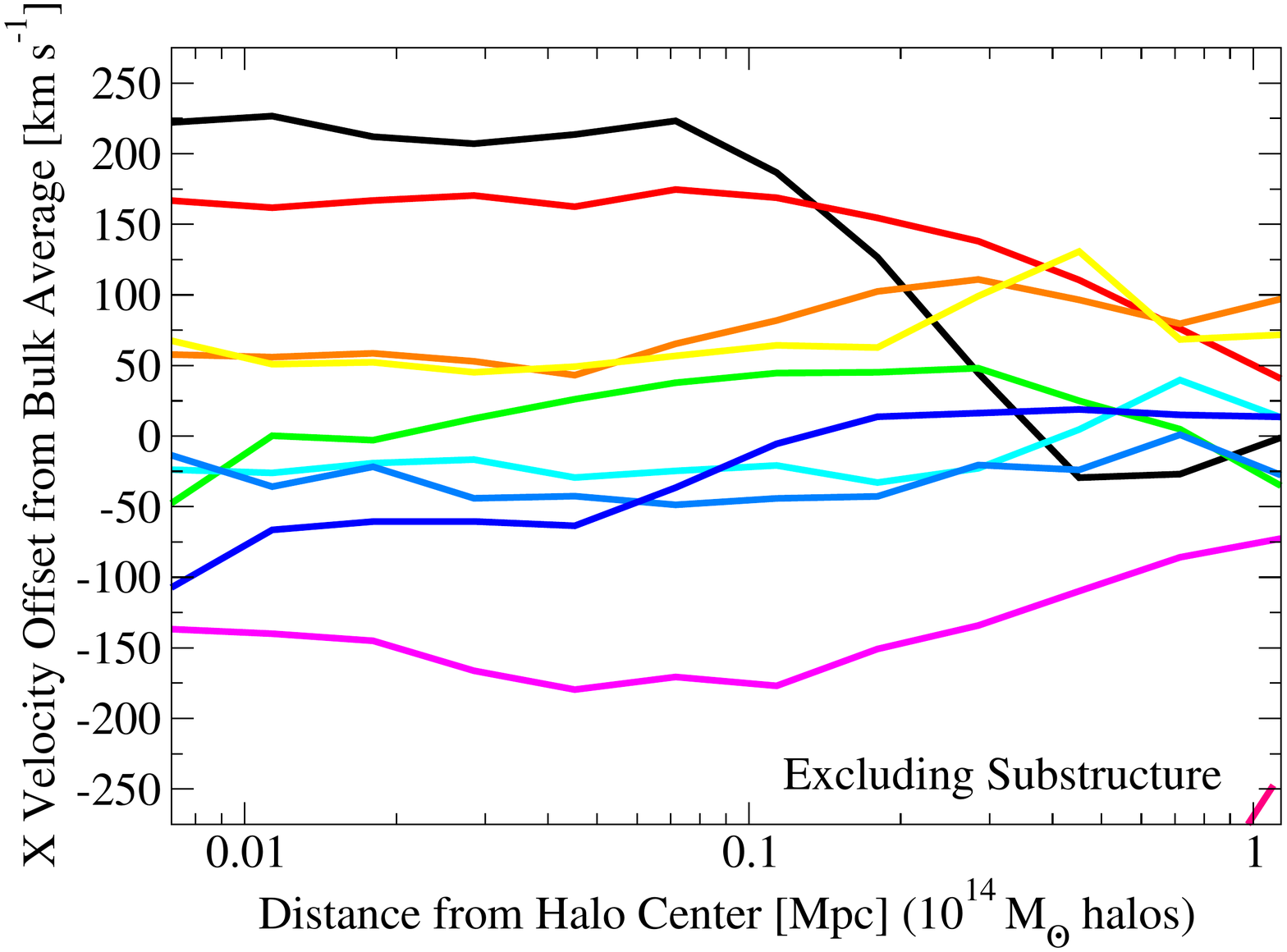}\\
\includegraphics[width=\columnwidth]{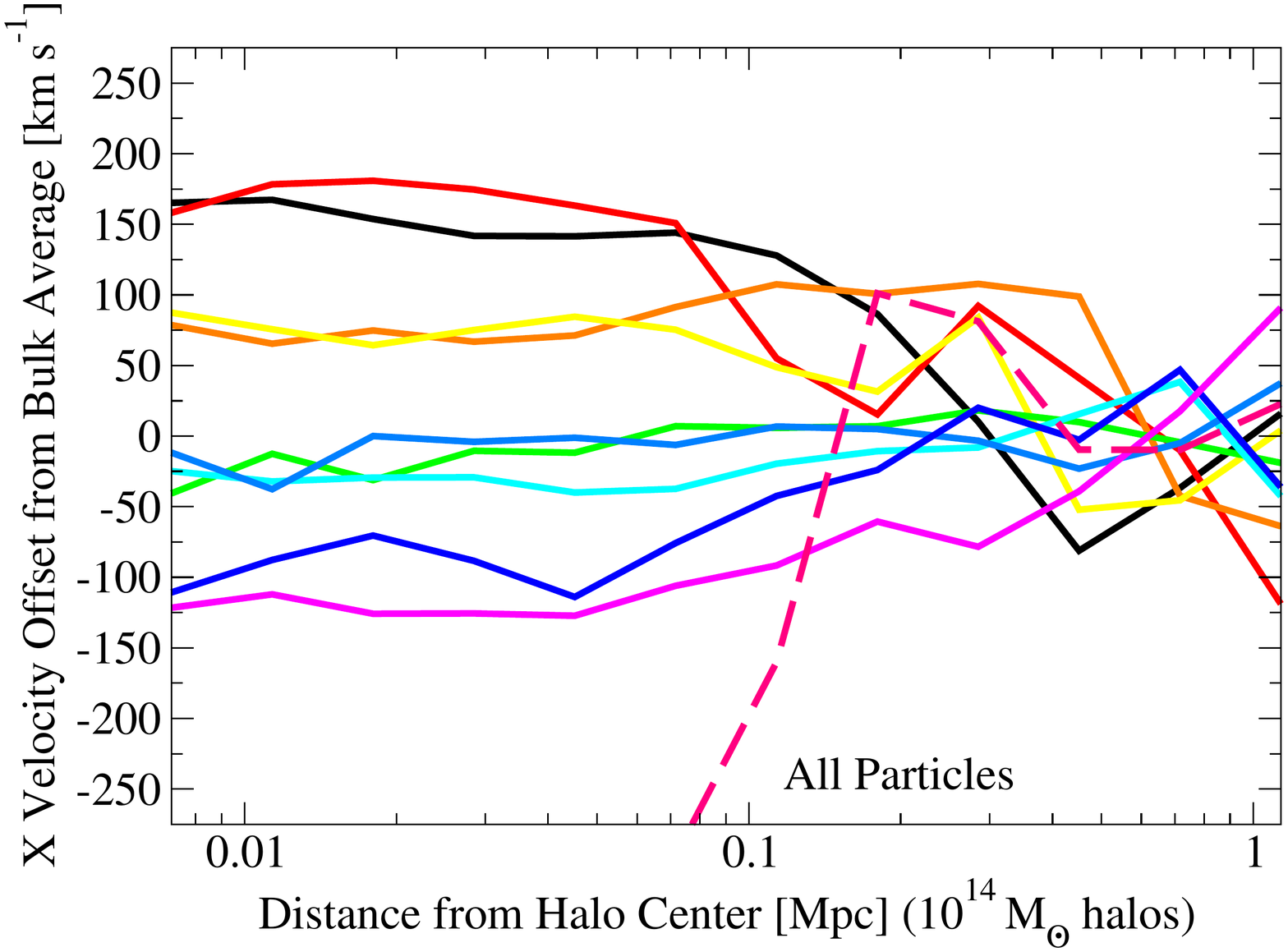}\\
\includegraphics[width=\columnwidth]{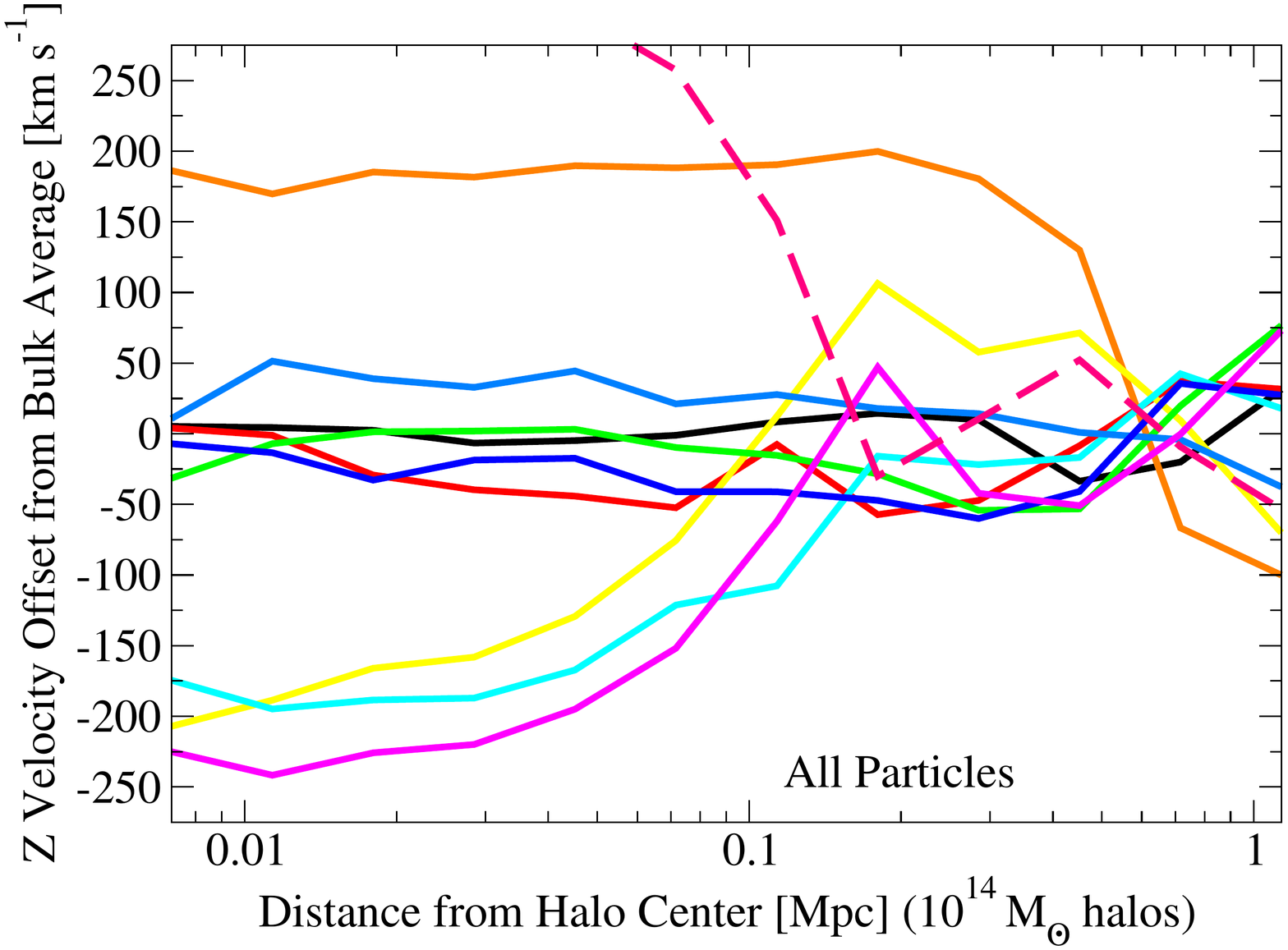}
\caption{Velocity offsets of the particle velocities at a given radius from the halo bulk velocity.  There is no
distance at which they match consistently.  Nonetheless, inner radial bins (excluding substructure) often have consistent velocities, suggesting that the halo core velocity is both well-defined and physical.  The \textbf{top two} panels show the $x$-velocity offsets between the halo bulk velocity and radially-binned halo particle velocity for 10 central halos with masses between $10^{14}$ and $3\times 10^{14} \Msun$ randomly selected from the Bolshoi simulation at $z=0$.  The bottom panel shows $z$-velocity offsets for the same halos.  The \textbf{top} panel shows calculations excluding substructure; the \textbf{middle} and \textbf{bottom} panels include all particles.  Matching colors between the panels correspond to the same halo.  Radial bins are evenly spaced in $\log(r)$, and as such contain significantly more particles at large radii.  The red-violet halo (\textbf{dashed line}) is undergoing a major merger in its outskirts, as such, the halo bulk velocity is over 400 km s$^{-1}$ offset along the $x$-direction from the halo core velocity.}
\label{f:core_examples}
\end{figure}

\begin{figure}
\includegraphics[width=\columnwidth]{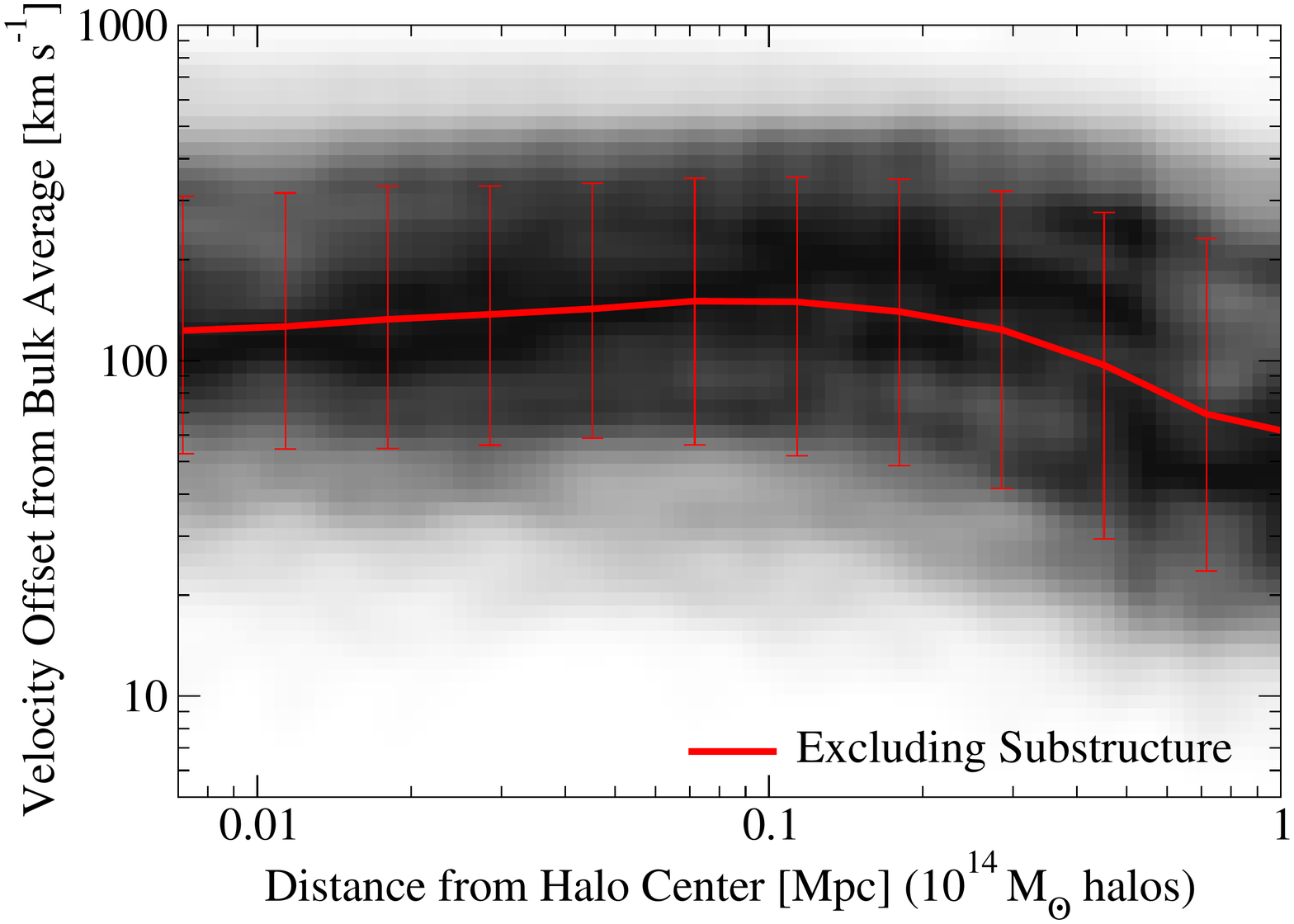}\\
\includegraphics[width=\columnwidth]{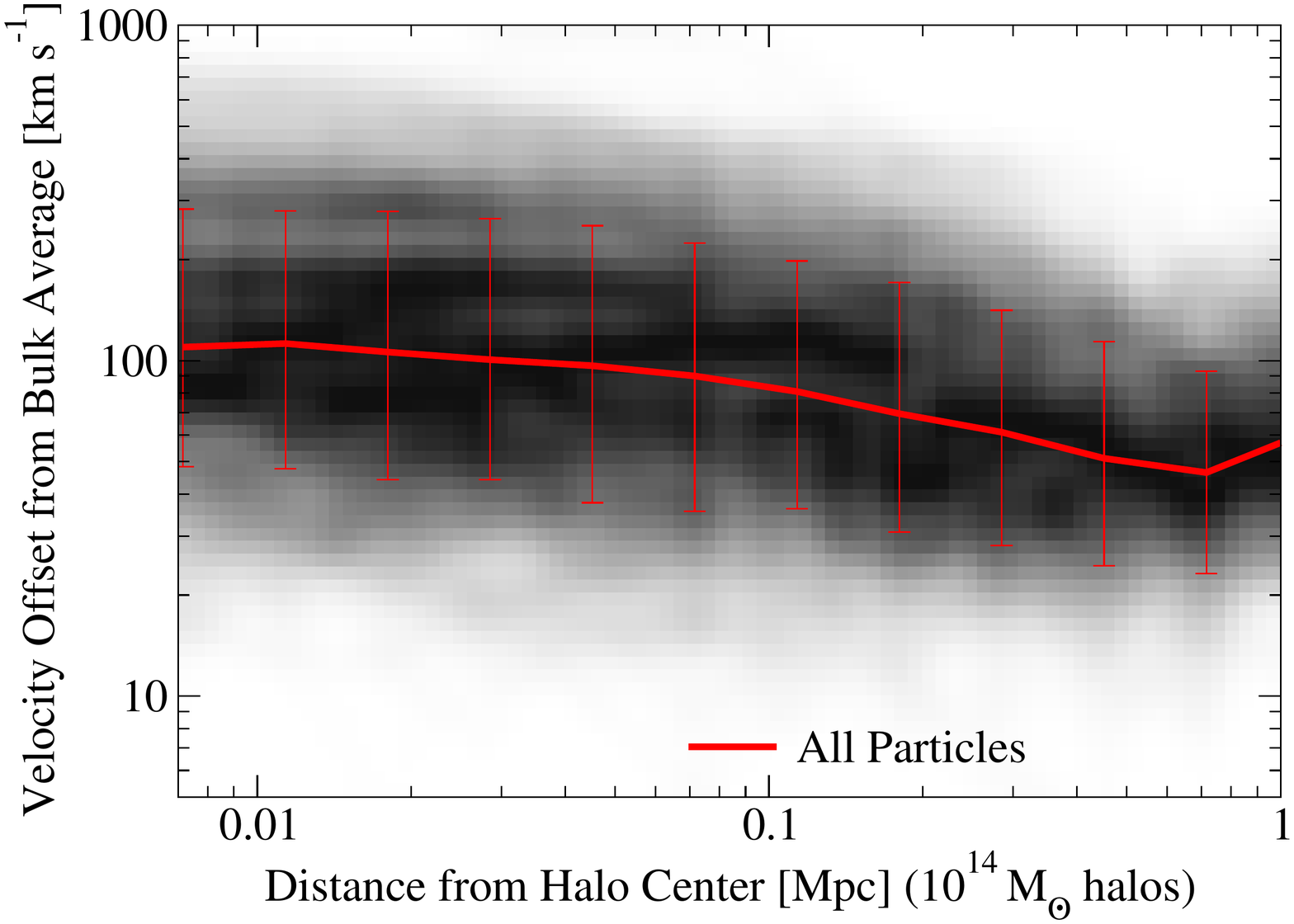}
\caption{There is a large intrinsic scatter in the core-bulk velocity offset, at least $\pm$0.3 dex among halos at a given mass.  These panels show the conditional probability density of offsets between the halo bulk velocity and radially-binned halo particle velocity for 484 central halos with masses between $10^{14}$ and $3\times 10^{14} \Msun$ from the Bolshoi simulation at $z=0$.  The red line shows the median of the offsets, and the error bars encompass $\pm$ 1$\sigma$ of the distribution.}
\label{f:core_scatter}
\end{figure}

\label{s:cores}
\begin{figure}
\plotgrace{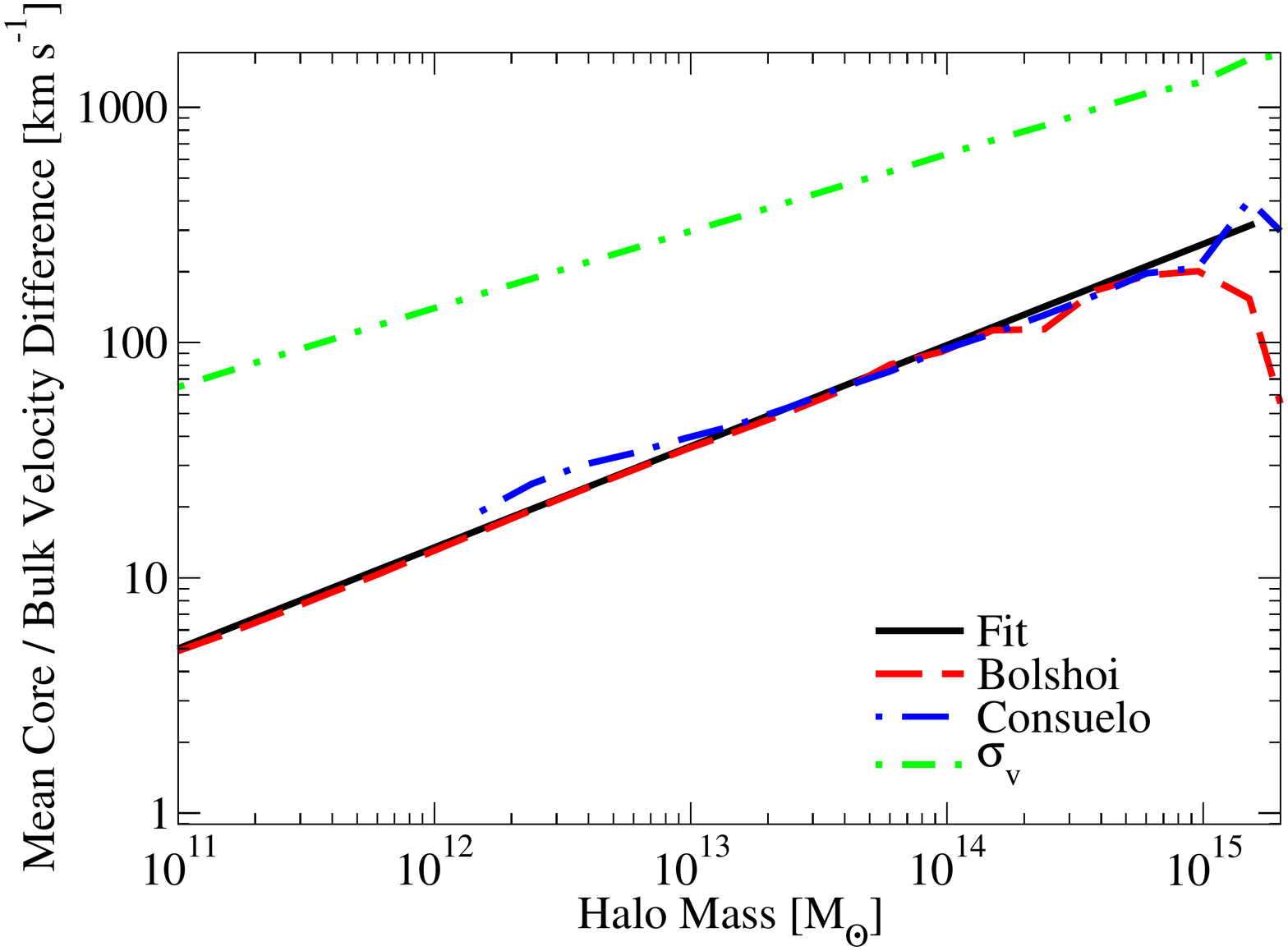}
\caption{The difference between halo core velocity (averaged within $0.1R_\mathrm{vir}$) and the bulk velocity of all halo particles within $R_\mathrm{vir}$ as a function of halo mass at $z=0$.  This is also compared to the halo particle velocity dispersion, $\sigma_v$.}
\label{f:core_bulk}
\end{figure}

In the $\Lambda$CDM paradigm, the main method by which halos reach a relaxed state is dynamical friction; namely, energy in bulk motions relative to the halo center of gravity is transformed into increased halo velocity dispersion.  Dynamical friction depends not only on the background density that a satellite halo is passing through, but also on the satellite mass and velocity.  For a massive host halo, the high background density means that incoming satellites will initially transfer momentum to the host with high efficiency.  However, massive host halos also have very strong tidal fields, and once satellites are disrupted into cold velocity streams, dynamical friction becomes much less effective at transferring momentum into the inner halo regions.  This combination of dynamical friction and tidal forces suggests that momentum transfer may be more efficient in the outer regions of a host halo and less efficient in the inner regions, leading to an offset between the mean velocity of the central density peak and the bulk velocity of the halo.

We examine the presence of this effect by calculating radially-averaged halo particle velocities (i.e., particle velocities averaged in spherical shells) and comparing to the halo bulk velocity (averaged over all halo particles) for a large number of central halos in both the Bolshoi and Consuelo simulations over a range of halo masses ($10^{12}$ to $10^{14} \Msun$) and redshifts ($z=0$ to $1.4$).  We additionally consider the effects of excluding particles belonging to substructure from calculations of the radially-binned velocities; results for all of the median offsets (both including and excluding substructure) are shown in Fig.\ \ref{f:core_highz}.

Figure \ref{f:core_highz} demonstrates that the core-bulk velocity difference can be quite dramatic at high redshift for massive halos: halos with $10^{14} \Msun < M < 3\times 10^{14}\Msun$ at $z = 1.4$ have a median offset of over 450 km s$^{-1}$ between the velocity averaged within $0.1 R_\mathrm{vir}$ (excluding substructure) and the bulk velocity of the halo.  This effect is lower for both smaller halos and later times, presumably due to lower velocity offsets in incoming substructure for smaller halos and a reduced merger rate at later times.  Nonetheless, a clear difference between the inner and bulk velocities is present in all halos tested even down to $z=0$.  In all cases, there is a definite radial trend, with the velocity in the inner region of the halo being farther away from the bulk velocity than the halo outskirts, consistent with the hypothesis of reduced efficiency of momentum transfer in the inner regions of the halo.

In the calculations which exclude substructure, a clear plateau is evident below $0.1-0.2 R_\mathrm{vir}$ where the core velocity offset stabilizes, at least for halos with $M>10^{13}\Msun$; we have insufficient mass resolution to test this effect robustly for smaller halos.  This transition is extremely consistent with expectations for where most of the satellite mass will be stripped in such halos.  For a satellite halo with a concentration of $c=10$ falling into a massive host with $c=5$, Eq.\ \ref{e:roche} would suggest that 90\% of the satellite mass (and momentum) will be stripped from the main satellite density peak at radii greater than $0.3 R_\mathrm{vir,host}$, with another 9\% stripped by $0.1 R_\mathrm{vir, host}$.  As tidal forces increase dramatically towards the center of a halo, they should disrupt efficient momentum transfer within $0.1 R_\mathrm{vir, host}$ even for satellites on highly radial orbits.

In the calculations which include substructure, the median velocity offsets are less than the results which exclude substructure because substructure is included when calculating the bulk halo velocity.  In the very innermost regions of halos, however, both radial velocity averages converge, suggesting that our results should be robust to the subhalo finding technique used.  On the outskirts of halos, the presence of non-disrupted subhalos results in an upturn in the offset between the radially-averaged velocity (including substructure) and the bulk halo velocity.

Due to significantly higher mass resolution, the Bolshoi simulation is able to probe the velocity offsets to significantly smaller radii as compared to the Consuelo simulation.  Furthermore, the Bolshoi simulation is able to better resolve substructure than Consuelo; as such, the difference between radial velocity offset profiles which include and exclude substructure in Consuelo is less than the difference for Bolshoi.  Except for this caveat, the median velocity offsets are in excellent agreement between the two simulations.

One more interesting feature of Fig.\ \ref{f:core_highz} is that the median velocity offset never reaches zero for any of the halos in either simulation.  That is to say, there is \textit{no single radius} where the bulk-averaged halo velocity corresponds to the actual average motion of particles in a radial shell.  We examine this issue further by considering ten halos randomly drawn from the Bolshoi simulation at $z=0$ in the mass range $10^{14}$ to $3\times 10^{14}\Msun$.  As before, we compute the average velocity in radial bins for each halo, but instead of the absolute offset from the bulk velocity, we show just the offsets between the $x$-components of the two velocities in Fig.\ \ref{f:core_examples}.  

The results for individual halos in Fig.\ \ref{f:core_examples} are just as striking as for the median offsets in Fig.\ \ref{f:core_highz}.  Almost every halo in Fig.\ \ref{f:core_highz} exhibits the same velocity offset plateau within $0.1 R_\mathrm{vir}$, suggesting that the velocity within $0.1 R_\mathrm{vir}$ has a physical interpretation corresponding to the actual bulk motion of those particles.  If substructure is excluded, the radially-averaged $x$-velocity often never matches the bulk $x$-velocity, implying an overall offset between the main halo motion and satellites.  If substructure is included, the radially-averaged $x$-velocity can briefly match the bulk $x$-velocity (and it must do so, by the mean value theorem), but there is no reason why the radially averaged velocities along other axes must match the bulk velocity at the same time; in fact, as shown for the $z$-velocities in the bottom panel of Fig.\ \ref{f:core_examples}, they do not.  As such, the bulk velocity of the particles in the halos shown here does not actually correspond to the velocity averaged over any of the radial shells.

To investigate this latter point in more detail, we show conditional density plots of the absolute offsets between radially-averaged velocities and the bulk velocity for the complete sample of $10^{14}$ to $3\times 10^{14}\Msun$ (again at $z=0$ in Bolshoi) in Figure \ref{f:core_scatter}.  While there is significant scatter in the absolute offsets (on the order of 0.3 dex, after correcting for the variation over the bin width), only a tiny fraction ($\ll1$\%) of the halos ever have a radial bin in which the averaged velocity matches the bulk velocity to better than 20 km s$^{-1}$, and most have significant offsets ($\sim 100$ km s$^{-1}$) at all radii from the halo center.

These results support the use of the Poisson minimization method of \S \ref{s:posvel} as a velocity estimator for \textsc{rockstar}, as the method primarily probes the inner regions of halos with velocity profiles like those shown in Fig.\ \ref{f:core_examples}.  Indeed, for both the Bolshoi and Consuelo simulations, we find an extremely consistent mean velocity offset between the core velocity (as discussed in \S \ref{s:posvel}) and the halo bulk velocity, as shown in Fig.\ \ref{f:core_bulk}.  Significantly, these offsets are much larger than either the direct Poisson error estimates in determining velocities (\S \ref{s:posvel}) or the cross-timestep velocity error estimates in Fig.\ \ref{f:dyn_comp} (\S \ref{s:dyn_perf}).  The main differences between Bolshoi and Consuelo come from reduced mass resolution and substructure resolution in the latter simulation, resulting in a velocity estimate slightly closer to the bulk average for Consuelo as compared to Bolshoi.

These velocity offsets display a power-law-like behavior across more than four orders of magnitude in halo mass; however, this power law is not the same as the power law for the average velocity dispersion $\sigma_v$ (which is proportional to $M^{1/3}$); instead, the velocity offset relative to the velocity dispersion increases for increasing halo masses.  At $z=0$, the average core-velocity offset may be fit by the empirical relation:
\begin{equation}
\sqrt{\langle \Delta V^2 \rangle} = 5\; \mathrm{km}\; \mathrm{s}^{-1} \left(\frac{M}{10^{11}\Msun}\right)^{0.43}
\end{equation}
This relation implies that, for the most massive $M \sim 2\times 10^{15}\Msun$ clusters at $z=0$, the average core-bulk velocity offset is of order 350 km s$^{-1}$.

We note that \cite{Gao06} have previously studied the question of core-bulk velocity offsets in Milky-Way to cluster-size halos in the Millennium Simulation.  In their work, the core velocity is defined as the average velocity of the most-bound 100-1000 particles as opposed to being defined directly by the halo finder.  As may be expected from our finding that the core-bulk velocity offset is relatively constant within 0.1 $R_\mathrm{vir}$, the choice of the number of particles to use in the definition of core velocity only impacts their results for the smallest halos in their analysis, which have only $\sim 2000$ to $4000$ particles to begin with.  They find that the core-bulk velocity offsets are approximately 10 to 15 percent of $V_{200}$ (defined as $\sqrt{G M_{200} / r_{200}}$), increasing with increasing halo mass.  As $V_{200}$ and $\sigma_v$ both scale as $M^{1/3}$ and are approximately the same magnitude across a wide range of halo masses at $z=0$, this compares well with our result in Fig.\ \ref{f:core_bulk} that the core-bulk offsets are on the order of 10-20\% of $\sigma_v$, also increasing with increasing halo mass.

Interestingly, \cite{Gao06} also explore the accretion histories of halo cores; they find that 75\% of cluster-scale halos have accreted less than 10\% of their core particles since $z=0.5$.  By contrast, cluster-scale halos have on average accreted more than 40\% of their \textit{bulk mass} in that same time period \citep{Wechsler02}.  While this does not by itself support our hypothesis that momentum transfer to the core becomes inefficient due to tidal stripping, it leads to the same conclusion that accreted material makes its way only very slowly to the halo core.

\section{Conclusions}

\label{s:conclusions}
We have presented a novel method of finding halos in phase space, the \textsc{rockstar} halo finder. This method has several key strengths which, in combination, 
improve on previous approaches to robustly identifying halos and their substructures:
\begin{enumerate}
\item high accuracy of recovered halo properties (as discussed in \citealt{Knebe11}),
\item the ability to consistently and usefully define a subhalo mass (\S \ref{s:hmass}),
\item the ability to reliably track major mergers down to the inner host halo core (\S \ref{s:tests} and \S \ref{s:substructure}),
\item explicit grid-independence, orientation-independence, and halo shape-independence for recovered halos (\S \ref{s:methodology}),
\item calibrated estimates of the accuracy of recovered halo properties in full cosmological simulations (\S \ref{s:dyn_perf}),
\item extremely efficient resource consumption both in CPU time and memory (\S \ref{s:performance}),
\item massively parallel implementation, supporting large future simulations (Appendix \ref{a:load_balancing}), and
\item a publicly available codebase:\\ http://code.google.com/p/rockstar .
\end{enumerate}

In addition, the halo finder complements the method of \cite{BehrooziTree} for creating merger trees, making it part of the first halo finding system to recover halos using seven dimensions of information (phase space plus time).  These properties make it ideal for a large number of scientific applications, especially for studies of subhalos and subhalo particle distributions.  We have demonstrated the ability to probe substructure into the very inner regions of halos, at least as far as simulations are capable of resolving.  This is critical for comparison with galaxy statistics in dense regimes, and paves the way for direct comparisons of clustering between simulations and observations in the regions where the unknown effects of baryons are the most significant.

We have also investigated the radial dependence of halo velocity.  We find that there exists a well-defined core velocity (for $r < 0.1 R_\mathrm{vir}$) which is always distinct from the bulk-averaged halo velocity.  Using the bulk velocity to estimate galaxy peculiar velocities thus carries significant systematic error, which can be up to 350 km s$^{-1}$ in the most massive clusters at $z=0$ and is larger at higher redshifts.  Furthermore (as discussed in \S \ref{s:core_offsets}), the halo velocity averaged in radial bins never matches the bulk average velocity; as such, the bulk average velocity is only an ensemble property of the entire halo, and the local dynamics have a much more complicated and interesting structure than can be described by the bulk velocity and the velocity dispersion alone.

\acknowledgements

We would like to thank Anatoly Klypin and Joel Primack for providing access to the Bolshoi simulation, which was run on the NASA Advanced Supercomputing Pleiades computer at NASA Ames Research Center, Michael Busha for running Subfind on the Bolshoi simulation, and Michael Busha, Cameron McBride, and the rest of the LasDamas collaboration for providing access to the Las Damas simulations, which were run on the Orange cluster at SLAC and on the NSF Teragrid machine Ranger (PI: Andreas Berlind).  We are grateful to Anatoly Klypin, Joel Primack, Michael Busha, Houjun Mo, Alexie Leauthaud, Andrey Kravtsov, James Bullock, Alex Knebe, Rachel Reddick, Yu Lu, Oliver Hahn, Matt Becker, Louie Strigari, Jeremy Tinker, Tom Abel, Steven Rieder, Markus Haider, Shea Garrison-Kimmel, Gus Evrard, Marcelo Alvarez, Joanne Cohn, Martin White, Simon White, Manodeep Sinha, and Annika Peters for interesting discussions, helpful suggestions, and testing. We gratefully acknowledge the support of Stuart Marshall and the SLAC computational team, as well as the computational resources at SLAC.  Finally, we would like to thank our anonymous referee for a very careful and thorough review which resulted in a great many improvements to the clarity of this paper.  PSB and RHW received support from NASA HST Theory grant HST-AR-12159.01-A; RHW was also supported by the National Science Foundation under grant NSF AST-0908883.  This work was additionally supported by the U.S. Department of Energy under contract number DE-AC02-76SF00515.

\newpage 

\bibliography{rockstar}

\begin{thebibliography}{76}
\expandafter\ifx\csname natexlab\endcsname\relax\def\natexlab#1{#1}\fi

\bibitem[{{Allgood} {et~al.}(2006){Allgood}, {Flores}, {Primack}, {Kravtsov},
  {Wechsler}, {Faltenbacher}, \& {Bullock}}]{Allgood06}
{Allgood}, B., {Flores}, R.~A., {Primack}, J.~R., {Kravtsov}, A.~V.,
  {Wechsler}, R.~H., {Faltenbacher}, A., \& {Bullock}, J.~S. 2006, \mnras, 367,
  1781

\bibitem[{{Ascasibar et al.}(2010)}]{Ascasibar10}
{Ascasibar et al.} 2010, In prep.

\bibitem[{{Barnes} \& {Hut}(1986)}]{BarnesHut86}
{Barnes}, J., \& {Hut}, P. 1986, \nat, 324, 446

\bibitem[{{Behroozi} {et~al.}(2010){Behroozi}, {Conroy}, \&
  {Wechsler}}]{Behroozi10}
{Behroozi}, P.~S., {Conroy}, C., \& {Wechsler}, R.~H. 2010, \apj, 717, 379

\bibitem[{{Behroozi} {et~al.}(2012{\natexlab{a}}){Behroozi}, {Loeb}, \&
  {Wechsler}}]{BehrooziUnbound}
{Behroozi}, P.~S., {Loeb}, A., \& {Wechsler}, R.~H. 2012{\natexlab{a}},
  arXiv:1208.0334

\bibitem[{{Behroozi} {et~al.}(2012{\natexlab{b}}){Behroozi}, {Wechsler}, \&
  {Conroy}}]{BWC12}
{Behroozi}, P.~S., {Wechsler}, R.~H., \& {Conroy}, C. 2012{\natexlab{b}},
  arXiv:1207.6105

\bibitem[{{Behroozi} {et~al.}(2013){Behroozi}, {Wechsler}, {Wu}, {Busha},
  {Klypin}, \& {Primack}}]{BehrooziTree}
{Behroozi}, P.~S., {Wechsler}, R.~H., {Wu}, H.-Y., {Busha}, M.~T., {Klypin},
  A.~A., \& {Primack}, J.~R. 2013, \apj, 763, 18

\bibitem[{{Bryan} \& {Norman}(1998)}]{mvir_conv}
{Bryan}, G.~L., \& {Norman}, M.~L. 1998, \apj, 495, 80

\bibitem[{{Bullock} {et~al.}(2001){Bullock}, {Dekel}, {Kolatt}, {Kravtsov},
  {Klypin}, {Porciani}, \& {Primack}}]{Bullock01}
{Bullock}, J.~S., {Dekel}, A., {Kolatt}, T.~S., {Kravtsov}, A.~V., {Klypin},
  A.~A., {Porciani}, C., \& {Primack}, J.~R. 2001, \apj, 555, 240

\bibitem[{{Chambers}(2006)}]{Pan-STARRS}
{Chambers}, K.~C. 2006, {Pan-STARRS Mission Concept Statement for PS1,
  23000200}

\bibitem[{{Conroy} \& {Wechsler}(2009)}]{cw-08}
{Conroy}, C., \& {Wechsler}, R.~H. 2009, \apj, 696, 620

\bibitem[{Conroy {et~al.}(2006)Conroy, Wechsler, \& Kravtsov}]{conroy:06}
Conroy, C., Wechsler, R.~H., \& Kravtsov, A.~V. 2006, \apj, 647, 201

\bibitem[{{Cunha} \& {Evrard}(2010)}]{Cunha10}
{Cunha}, C.~E., \& {Evrard}, A.~E. 2010, \prd, 81, 083509

\bibitem[{{Davis} {et~al.}(1985){Davis}, {Efstathiou}, {Frenk}, \&
  {White}}]{Davis85}
{Davis}, M., {Efstathiou}, G., {Frenk}, C.~S., \& {White}, S.~D.~M. 1985, \apj,
  292, 371

\bibitem[{{Diemand} {et~al.}(2006){Diemand}, {Kuhlen}, \& {Madau}}]{Diemand06}
{Diemand}, J., {Kuhlen}, M., \& {Madau}, P. 2006, \apj, 649, 1

\bibitem[{{Eisenstein} \& {Hut}(1998)}]{HOP}
{Eisenstein}, D.~J., \& {Hut}, P. 1998, \apj, 498, 137

\bibitem[{{Evrard} {et~al.}(2002){Evrard}, {MacFarland}, {Couchman}, {Colberg},
  {Yoshida}, {White}, {Jenkins}, {Frenk}, {Pearce}, {Peacock}, \&
  {Thomas}}]{Evrard02}
{Evrard}, A.~E., {et~al.} 2002, \apj, 573, 7

\bibitem[{{Faltenbacher} {et~al.}(2005){Faltenbacher}, {Allgood},
  {Gottl{\"o}ber}, {Yepes}, \& {Hoffman}}]{Faltenbacher05}
{Faltenbacher}, A., {Allgood}, B., {Gottl{\"o}ber}, S., {Yepes}, G., \&
  {Hoffman}, Y. 2005, \mnras, 362, 1099

\bibitem[{{Gao} \& {White}(2006)}]{Gao06}
{Gao}, L., \& {White}, S.~D.~M. 2006, \mnras, 373, 65

\bibitem[{{Gardner} {et~al.}(2007){Gardner}, {Connolly}, \&
  {McBride}}]{Gardner07}
{Gardner}, J.~P., {Connolly}, A., \& {McBride}, C. 2007, Accepted to Challenges
  of Large Applications in Distributed Environments (CLADE) 2007

\bibitem[{{Gardner} {et~al.}(2006){Gardner}, {Mather}, {Clampin}, {Doyon},
  {Greenhouse}, {Hammel}, {Hutchings}, {Jakobsen}, {Lilly}, {Long}, {Lunine},
  {McCaughrean}, {Mountain}, {Nella}, {Rieke}, {Rieke}, {Rix}, {Smith},
  {Sonneborn}, {Stiavelli}, {Stockman}, {Windhorst}, \& {Wright}}]{JWST}
{Gardner}, J.~P., {et~al.} 2006, \ssr, 123, 485

\bibitem[{{Giocoli} {et~al.}(2010){Giocoli}, {Tormen}, {Sheth}, \& {van den
  Bosch}}]{Giocoli10}
{Giocoli}, C., {Tormen}, G., {Sheth}, R.~K., \& {van den Bosch}, F.~C. 2010,
  \mnras, 404, 502

\bibitem[{Gottl\"ober(1998)}]{Stefan98}
Gottl\"ober, S. 1998, in Large Scale Structure: Tracks and Traces, ed.
  {V.~M\"uller, S.~Gottl\"ober, J.-P.~M\"ucket, and J.~Wambsganss}, World
  Scientific, 43--46

\bibitem[{{Guo} {et~al.}(2010){Guo}, {White}, {Li}, \&
  {Boylan-Kolchin}}]{Guo-09}
{Guo}, Q., {White}, S., {Li}, C., \& {Boylan-Kolchin}, M. 2010, \mnras, 404,
  1111

\bibitem[{{Habib} {et~al.}(2009){Habib}, {Pope}, {Luki{\'c}}, {Daniel},
  {Fasel}, {Desai}, {Heitmann}, {Hsu}, {Ankeny}, {Mark}, {Bhattacharya}, \&
  {Ahrens}}]{Habib09}
{Habib}, S., {et~al.} 2009, Journal of Physics Conference Series, 180, 012019

\bibitem[{{Han} {et~al.}(2011){Han}, {Jing}, {Wang}, \& {Wang}}]{Han11}
{Han}, J., {Jing}, Y.~P., {Wang}, H., \& {Wang}, W. 2011, arXiv:1103.2099

\bibitem[{{Harker} {et~al.}(2006){Harker}, {Cole}, {Helly}, {Frenk}, \&
  {Jenkins}}]{Harker06}
{Harker}, G., {Cole}, S., {Helly}, J., {Frenk}, C., \& {Jenkins}, A. 2006,
  \mnras, 367, 1039

\bibitem[{{Jenkins} {et~al.}(2001)}]{Jenkins01}
{Jenkins}, A., {et~al.} 2001, \mnras, 321, 372

\bibitem[{{Jiang} {et~al.}(2011){Jiang}, {Hogg}, \& {Blanton}}]{Jiang11}
{Jiang}, T., {Hogg}, D.~W., \& {Blanton}, M.~R. 2011, arXiv:1104.5483

\bibitem[{{Kitzbichler} \& {White}(2008)}]{MillOrphan}
{Kitzbichler}, M.~G., \& {White}, S.~D.~M. 2008, \mnras, 391, 1489

\bibitem[{{Klypin} {et~al.}(1999){Klypin}, {Gottl{\"o}ber}, {Kravtsov}, \&
  {Khokhlov}}]{Klypin99}
{Klypin}, A., {Gottl{\"o}ber}, S., {Kravtsov}, A.~V., \& {Khokhlov}, A.~M.
  1999, \apj, 516, 530

\bibitem[{{Klypin} {et~al.}(2011){Klypin}, {Trujillo-Gomez}, \&
  {Primack}}]{Bolshoi}
{Klypin}, A.~A., {Trujillo-Gomez}, S., \& {Primack}, J. 2011, \apj, 740, 102

\bibitem[{{Knebe} {et~al.}(2011){Knebe}, {Knollmann}, {Muldrew}, {Pearce},
  {Aragon-Calvo}, {Ascasibar}, {Behroozi}, {Ceverino}, {Colombi}, {Diemand},
  {Dolag}, {Falck}, {Fasel}, {Gardner}, {Gottl{\"o}ber}, {Hsu}, {Iannuzzi},
  {Klypin}, {Luki{\'c}}, {Maciejewski}, {McBride}, {Neyrinck}, {Planelles},
  {Potter}, {Quilis}, {Rasera}, {Read}, {Ricker}, {Roy}, {Springel}, {Stadel},
  {Stinson}, {Sutter}, {Turchaninov}, {Tweed}, {Yepes}, \& {Zemp}}]{Knebe11}
{Knebe}, A., {et~al.} 2011, \mnras, 415, 2293

\bibitem[{{Knollmann} \& {Knebe}(2009)}]{Knollmann09}
{Knollmann}, S.~R., \& {Knebe}, A. 2009, \apjs, 182, 608

\bibitem[{{Komatsu} {et~al.}(2011){Komatsu}, {Smith}, {Dunkley}, {Bennett},
  {Gold}, {Hinshaw}, {Jarosik}, {Larson}, {Nolta}, {Page}, {Spergel},
  {Halpern}, {Hill}, {Kogut}, {Limon}, {Meyer}, {Odegard}, {Tucker}, {Weiland},
  {Wollack}, \& {Wright}}]{wmap7}
{Komatsu}, E., {et~al.} 2011, \apjs, 192, 18

\bibitem[{Kravtsov {et~al.}(1997)Kravtsov, Klypin, \&
  Khokhlov}]{kravtsov_etal:97}
Kravtsov, A.~V., Klypin, A.~A., \& Khokhlov, A.~M. 1997, \apj, 111, 73

\bibitem[{{Lacey} \& {Cole}(1994)}]{Cole94}
{Lacey}, C., \& {Cole}, S. 1994, \mnras, 271, 676

\bibitem[{{LSST Science Collaborations} {et~al.}(2009){LSST Science
  Collaborations}, {Abell}, {Allison}, {Anderson}, {Andrew}, {Angel}, {Armus},
  {Arnett}, {Asztalos}, {Axelrod}, \& et~al.}]{LSST}
{LSST Science Collaborations} {et~al.} 2009, arXiv:0912.0201

\bibitem[{{Maciejewski} {et~al.}(2009){Maciejewski}, {Colombi}, {Springel},
  {Alard}, \& {Bouchet}}]{Maciejewski09}
{Maciejewski}, M., {Colombi}, S., {Springel}, V., {Alard}, C., \& {Bouchet},
  F.~R. 2009, \mnras, 396, 1329

\bibitem[{{More} {et~al.}(2011){More}, {Kravtsov}, {Dalal}, \&
  {Gottl{\"o}ber}}]{More11}
{More}, S., {Kravtsov}, A.~V., {Dalal}, N., \& {Gottl{\"o}ber}, S. 2011, \apjs,
  195, 4

\bibitem[{{Moster} {et~al.}(2010){Moster}, {Somerville}, {Maulbetsch}, {van den
  Bosch}, {Macci{\`o}}, {Naab}, \& {Oser}}]{moster-09}
{Moster}, B.~P., {Somerville}, R.~S., {Maulbetsch}, C., {van den Bosch}, F.~C.,
  {Macci{\`o}}, A.~V., {Naab}, T., \& {Oser}, L. 2010, \apj, 710, 903

\bibitem[{{Navarro} {et~al.}(1997){Navarro}, {Frenk}, \& {White}}]{NFW97}
{Navarro}, J.~F., {Frenk}, C.~S., \& {White}, S.~D.~M. 1997, \apj, 490, 493

\bibitem[{{Neto} {et~al.}(2007){Neto}, {Gao}, {Bett}, {Cole}, {Navarro},
  {Frenk}, {White}, {Springel}, \& {Jenkins}}]{Neto07}
{Neto}, A.~F., {et~al.} 2007, \mnras, 381, 1450

\bibitem[{{Neyrinck} {et~al.}(2005){Neyrinck}, {Gnedin}, \&
  {Hamilton}}]{Neyrinck05}
{Neyrinck}, M.~C., {Gnedin}, N.~Y., \& {Hamilton}, A.~J.~S. 2005, \mnras, 356,
  1222

\bibitem[{{Peebles}(1969)}]{Peebles69}
{Peebles}, P.~J.~E. 1969, \apj, 155, 393

\bibitem[{{Pilbratt} {et~al.}(2010){Pilbratt}, {Riedinger}, {Passvogel},
  {Crone}, {Doyle}, {Gageur}, {Heras}, {Jewell}, {Metcalfe}, {Ott}, \&
  {Schmidt}}]{Herschel}
{Pilbratt}, G.~L., {et~al.} 2010, \aap, 518, L1

\bibitem[{{Planelles} \& {Quilis}(2010)}]{Planelles10}
{Planelles}, S., \& {Quilis}, V. 2010, \aap, 519, A94+

\bibitem[{{Predehl} {et~al.}(2010){Predehl}, {Andritschke}, {B{\"o}hringer},
  {Bornemann}, {Br{\"a}uninger}, {Brunner}, {Brusa}, {Burkert}, {Burwitz},
  {Cappelluti}, {Churazov}, {Dennerl}, {Eder}, {Elbs}, {Freyberg}, {Friedrich},
  {F{\"u}rmetz}, {Gaida}, {H{\"a}lker}, {Hartner}, {Hasinger}, {Hermann},
  {Huber}, {Kendziorra}, {von Kienlin}, {Kink}, {Kreykenbohm}, {Lamer},
  {Lapchov}, {Lehmann}, {Meidinger}, {Mican}, {Mohr}, {M{\"u}hlegger},
  {M{\"u}ller}, {Nandra}, {Pavlinsky}, {Pfeffermann}, {Reiprich}, {Robrade},
  {Roh{\'e}}, {Santangelo}, {Sch{\"a}chner}, {Schanz}, {Schmid}, {Schmitt},
  {Schreib}, {Schrey}, {Schwope}, {Steinmetz}, {Str{\"u}der}, {Sunyaev},
  {Tenzer}, {Tiedemann}, {Vongehr}, \& {Wilms}}]{eROSITA}
{Predehl}, P., {et~al.} 2010, in Society of Photo-Optical Instrumentation
  Engineers (SPIE) Conference Series, Vol. 7732, Society of Photo-Optical
  Instrumentation Engineers (SPIE) Conference Series

\bibitem[{{Rasera} {et~al.}(2010){Rasera}, {Alimi}, {Courtin}, {Roy},
  {Corasaniti}, {F{\"u}zfa}, \& {Boucher}}]{Rasera10}
{Rasera}, Y., {Alimi}, J., {Courtin}, J., {Roy}, F., {Corasaniti}, P.,
  {F{\"u}zfa}, A., \& {Boucher}, V. 2010, in American Institute of Physics
  Conference Series, Vol. 1241, American Institute of Physics Conference
  Series, ed. {J.-M.~Alimi \& A.~Fu{\"o}zfa}, 1134--1139

\bibitem[{{Reddick} {et~al.}(2012){Reddick}, {Wechsler}, {Tinker}, \&
  {Behroozi}}]{Reddick12}
{Reddick}, R.~M., {Wechsler}, R.~H., {Tinker}, J.~L., \& {Behroozi}, P.~S.
  2012, arXiv:1207.2160

\bibitem[{{Rudd} {et~al.}(2008){Rudd}, {Zentner}, \& {Kravtsov}}]{Rudd08}
{Rudd}, D.~H., {Zentner}, A.~R., \& {Kravtsov}, A.~V. 2008, \apj, 672, 19

\bibitem[{{Salmon} \& {Warren}(1994)}]{Salmon94}
{Salmon}, J.~K., \& {Warren}, M.~S. 1994, Journal of Computational Physics,
  111, 136

\bibitem[{{Schlegel} {et~al.}(2011){Schlegel}, {Abdalla}, {Abraham}, {Ahn},
  {Allende Prieto}, {Annis}, {Aubourg}, {Azzaro}, {Baltay}, {Baugh}, {Bebek},
  {Becerril}, {Blanton}, {Bolton}, {Bromley}, {Cahn}, {Carton},
  {Cervantes-Cota}, {Chu}, {Cortes}, {Dawson}, {Dey}, {Dickinson}, {Diehl},
  {Doel}, {Ealet}, {Edelstein}, {Eppelle}, {Escoffier}, {Evrard}, {Faccioli},
  {Frenk}, {Geha}, {Gerdes}, {Gondolo}, {Gonzalez-Arroyo}, {Grossan},
  {Heckman}, {Heetderks}, {Ho}, {Honscheid}, {Huterer}, {Ilbert}, {Ivans},
  {Jelinsky}, {Jing}, {Joyce}, {Kennedy}, {Kent}, {Kieda}, {Kim}, {Kim},
  {Kneib}, {Kong}, {Kosowsky}, {Krishnan}, {Lahav}, {Lampton}, {LeBohec}, {Le
  Brun}, {Levi}, {Li}, {Liang}, {Lim}, {Lin}, {Linder}, {Lorenzon}, {de la
  Macorra}, {Magneville}, {Malina}, {Marinoni}, {Martinez}, {Majewski},
  {Matheson}, {McCloskey}, {McDonald}, {McKay}, {McMahon}, {Menard},
  {Miralda-Escude}, {Modjaz}, {Montero-Dorta}, {Morales}, {Mostek}, {Newman},
  {Nichol}, {Nugent}, {Olsen}, {Padmanabhan}, {Palanque-Delabrouille}, {Park},
  {Peacock}, {Percival}, {Perlmutter}, {Peroux}, {Petitjean}, {Prada},
  {Prieto}, {Prochaska}, {Reil}, {Rockosi}, {Roe}, {Rollinde}, {Roodman},
  {Ross}, {Rudnick}, {Ruhlmann-Kleider}, {Sanchez}, {Sawyer}, {Schimd},
  {Schubnell}, {Scoccimaro}, {Seljak}, {Seo}, {Sheldon}, {Sholl},
  {Shulte-Ladbeck}, {Slosar}, {Smith}, {Smoot}, {Springer}, {Stril}, {Szalay},
  {Tao}, {Tarle}, {Taylor}, {Tilquin}, {Tinker}, {Valdes}, {Wang}, {Wang},
  {Weaver}, {Weinberg}, {White}, {Wood-Vasey}, {Yang}, {Yeche}, {Zakamska},
  {Zentner}, {Zhai}, \& {Zhang}}]{BigBOSS}
{Schlegel}, D., {et~al.} 2011, arXiv:1106.1706

\bibitem[{{Schlegel} {et~al.}(2009){Schlegel}, {White}, \& {Eisenstein}}]{BOSS}
{Schlegel}, D., {White}, M., \& {Eisenstein}, D. 2009, in ArXiv Astrophysics
  e-prints, Vol. 2010, astro2010: The Astronomy and Astrophysics Decadal
  Survey, 314

\bibitem[{{Springel}(2005)}]{Springel05}
{Springel}, V. 2005, \mnras, 364, 1105

\bibitem[{{Springel} {et~al.}(2010){Springel}, {Angulo}, {White}, {Jenkins},
  {Frenk}, {Baugh}, \& {Cole}}]{Springel10}
{Springel}, V., {Angulo}, R., {White}, S. D.~M., {Jenkins}, A., {Frenk}, C.~S.,
  {Baugh}, C., \& {Cole}, S. 2010, inSiDE, 8, 20

\bibitem[{{Springel} {et~al.}(2008){Springel}, {Wang}, {Vogelsberger},
  {Ludlow}, {Jenkins}, {Helmi}, {Navarro}, {Frenk}, \& {White}}]{Aquarius}
{Springel}, V., {et~al.} 2008, \mnras, 391, 1685

\bibitem[{{Springel} {et~al.}(2005){Springel}, {White}, {Jenkins}, {Frenk},
  {Yoshida}, {Gao}, {Navarro}, {Thacker}, {Croton}, {Helly}, {Peacock}, {Cole},
  {Thomas}, {Couchman}, {Evrard}, {Colberg}, \& {Pearce}}]{Springel05Mill}
---. 2005, \nat, 435, 629

\bibitem[{{Springel} {et~al.}(2001){Springel}, {White}, {Tormen}, \&
  {Kauffmann}}]{Springel01}
{Springel}, V., {White}, S.~D.~M., {Tormen}, G., \& {Kauffmann}, G. 2001,
  \mnras, 328, 726

\bibitem[{{Stadel}(2001)}]{Stadel01}
{Stadel}, J.~G. 2001, PhD thesis, University of Washington

\bibitem[{{Stanek} {et~al.}(2009){Stanek}, {Rudd}, \& {Evrard}}]{Stanek09}
{Stanek}, R., {Rudd}, D., \& {Evrard}, A.~E. 2009, \mnras, 394, L11

\bibitem[{{Sutter} \& {Ricker}(2010)}]{Sutter10}
{Sutter}, P.~M., \& {Ricker}, P.~M. 2010, \apj, 723, 1308

\bibitem[{{The Dark Energy Survey Collaboration}(2005)}]{DES}
{The Dark Energy Survey Collaboration}. 2005, arXiv:astro-ph/0510346

\bibitem[{{The Planck Collaboration}(2006)}]{Planck}
{The Planck Collaboration}. 2006, arXiv:astro-ph/0604069

\bibitem[{{Tinker} {et~al.}(2008){Tinker}, {Kravtsov}, {Klypin}, {Abazajian},
  {Warren}, {Yepes}, {Gottl{\"o}ber}, \& {Holz}}]{tinker-umf}
{Tinker}, J., {Kravtsov}, A.~V., {Klypin}, A., {Abazajian}, K., {Warren}, M.,
  {Yepes}, G., {Gottl{\"o}ber}, S., \& {Holz}, D.~E. 2008, \apj, 688, 709

\bibitem[{{Tinker} {et~al.}(2012){Tinker}, {Sheldon}, {Wechsler}, {Becker},
  {Rozo}, {Zu}, {Weinberg}, {Zehavi}, {Blanton}, {Busha}, \&
  {Koester}}]{Tinker11}
{Tinker}, J.~L., {et~al.} 2012, \apj, 745, 16

\bibitem[{{Tweed} {et~al.}(2009){Tweed}, {Devriendt}, {Blaizot}, {Colombi}, \&
  {Slyz}}]{Tweed09}
{Tweed}, D., {Devriendt}, J., {Blaizot}, J., {Colombi}, S., \& {Slyz}, A. 2009,
  \aap, 506, 647

\bibitem[{{Wang} \& {Jing}(2010)}]{Wang09}
{Wang}, L., \& {Jing}, Y.~P. 2010, \mnras, 402, 1796

\bibitem[{{Wechsler} {et~al.}(2002){Wechsler}, {Bullock}, {Primack},
  {Kravtsov}, \& {Dekel}}]{Wechsler02}
{Wechsler}, R.~H., {Bullock}, J.~S., {Primack}, J.~R., {Kravtsov}, A.~V., \&
  {Dekel}, A. 2002, \apj, 568, 52

\bibitem[{{Wetzel} {et~al.}(2009){Wetzel}, {Cohn}, \& {White}}]{Wetzel09}
{Wetzel}, A.~R., {Cohn}, J.~D., \& {White}, M. 2009, \mnras, 395, 1376

\bibitem[{{Woodring} {et~al.}(2011){Woodring}, {Heitmann}, {Ahrens}, {Fasel},
  {Hsu}, {Habib}, \& {Pope}}]{Woodring11}
{Woodring}, J., {Heitmann}, K., {Ahrens}, J., {Fasel}, P., {Hsu}, C.-H.,
  {Habib}, S., \& {Pope}, A. 2011, \apjs, 195, 11

\bibitem[{{Wu} {et~al.}(2011){Wu}, {Hahn}, {Wechsler}, \& {Behroozi}}]{Wu11}
{Wu}, H., {Hahn}, O., {Wechsler}, R.~H., \& {Behroozi}, P.~S. 2011, in prep.

\bibitem[{{Wu} {et~al.}(2010){Wu}, {Zentner}, \& {Wechsler}}]{Wu10}
{Wu}, H., {Zentner}, A.~R., \& {Wechsler}, R.~H. 2010, \apj, 713, 856

\bibitem[{{Wu} {et~al.}(2012{\natexlab{a}}){Wu}, {Hahn}, {Wechsler},
  {Behroozi}, \& {Mao}}]{Wu12b}
{Wu}, H.-Y., {Hahn}, O., {Wechsler}, R.~H., {Behroozi}, P.~S., \& {Mao}, Y.-Y.
  2012{\natexlab{a}}, arXiv:1210.6358

\bibitem[{{Wu} {et~al.}(2012{\natexlab{b}}){Wu}, {Hahn}, {Wechsler}, {Mao}, \&
  {Behroozi}}]{Wu12a}
{Wu}, H.-Y., {Hahn}, O., {Wechsler}, R.~H., {Mao}, Y.-Y., \& {Behroozi}, P.~S.
  2012{\natexlab{b}}, arXiv:1209.3309

\bibitem[{{Zemp} {et~al.}(2011){Zemp}, {Gnedin}, {Gnedin}, \&
  {Kravtsov}}]{Zemp11}
{Zemp}, M., {Gnedin}, O.~Y., {Gnedin}, N.~Y., \& {Kravtsov}, A.~V. 2011, \apjs,
  197, 30

\end{thebibliography}

\appendix

\section{Load Balancing}

The simplest approach to load balancing---equal volumes---works well on scales where the universe is homogenous, namely, when individual analysis regions are at least 100 Mpc on a side.  However, cosmic variance on smaller scales requires a more sophisticated approach for algorithms designed to run on small simulations or very large numbers of processors.

The approach we use is to divide the simulation volume into chunks requiring approximately equal memory.  We divide the number of available processors, $N$, into three factors ($N_x$, $N_y$, and $N_z$), which control the number of divisions along each of the principal axes of the simulation.  Naturally, $N_x$, $N_y$, and $N_z$ are chosen to be as close as possible to $N^{1/3}$ so as to minimize the surface-to-volume ratio of the divisions.  Initially, particles are binned into $B=10^4$ bins according to their $x$-coordinate, and the simulation is divided into $N_x$ regions of equal particle count along the $x$-axis.  Then, within each of these regions, particles are binned according to their $y$-coordinate, and each region is split accordingly along the $y$-axis into $N_y$ subregions of equal particle count.  Finally, each subregion is split in the same way into $N_z$ sub-subregions along the $z$-axis.  This process is accurate and efficient, requiring a fixed total data transfer size of ${\cal{O}}(BN^{2/3})$, an amount which is completely independent of the total number of particles.

3D FOF group calculation is performed in parallel on these regions without any inter-process communication.  However, for phase-space analysis, the number of recovered groups can vary considerably depending on the overall density of the analysis region (e.g., voids will have lower fractions of bound particles).  For that reason, FOFs are distributed to processors in small work units; as soon as a processor finishes one work unit, it receives another to work on (possibly from a completely different part of the simulation), so that maximum concurrency can be maintained at all times.  However, a single FOF group is never (in the current implementation) split up for analysis among multiple processors.  Thus, the analysis time for the largest FOF group sets the lower limit for the wall clock time taken to analyze the simulation snapshot, regardless of the number of processors available for use.

\label{a:load_balancing}

\section{Unbinding}

\label{a:unbinding}

To calculate which particles are bound to halos, we use a modified Barnes-Hut algorithm \citep{BarnesHut86} to calculate particle potential energies with a binary space partitioning (BSP) tree.\footnote{Several other halo finders, such as \textsc{bdm}  and AHF, use the assumption of spherical symmetry to calculate particle potentials; we find that, while this works well for the vast majority of halos, the potential calculated in this way can be dramatically wrong for halos undergoing major mergers, especially if the halo center is incorrectly identified.}  As noted in \cite{Salmon94}, the cell refinement criterion suggested by Barnes and Hut gives unacceptably high \textit{maximum} errors (as opposed to average errors); in addition, our use of a BSP tree (which gives fewer wasted refinement levels than an octree) renders the Barnes and Hut criterion somewhat inapplicable.  Instead, given a cell containing $n$ particles with mass center $x_0$, we calculate a threshold monopole approximation distance---that is, the distance from $x_0$ at which a monopole would be an appropriate approximation of the potential.  Following \cite{Salmon94}, we can estimate the relative error $\theta$ in the calculated potential as a function of distance $d$ from the center $x_0$ as being bounded by:
\begin{eqnarray}
\theta(d) & \le & \frac{1}{1-\frac{r_{max}}{d}} \frac{\sigma_x^2}{d^2}\\
\sigma_x^2 & = & \frac{\sum_{i=1}^n |m_i| | x_i - x_0 |^2}{\sum_{i=1}^n |m_i|},
\end{eqnarray}
where $x_i$ and $m_i$ are the locations and masses of the $n$ particles in the cell, and $r_{max}$ is the maximum distance between $x_0$ and any of the particles in the cell.  Thus, for a given choice of the relative error $\theta$, this equation may be inverted to give a minimum distance for acceptable use of the monopole approximation:
\begin{equation}
d(\theta) \ge \frac{r_{max}}{2} + \sqrt{\left(\frac{r_{max}}{2}\right)^2 + \frac{\sigma_x^2}{\theta}},
\end{equation}
This equation is valid regardless of the boundary sizes of the cell; hence it is appropriate for use even with a BSP tree.  We choose $\theta$ to give potentials accurate to $4\%$; from our tests, this is sufficient to correctly estimate the boundedness of $99.8\%$ of all halo particles.

\end{document}